\documentclass[aps,prd,showpacs,nofootinbib,floats,floatfix,preprintnumbers,groupedaddress,twocolumn]{revtex4-1}
\usepackage{graphicx,epsfig}
\usepackage{dcolumn}
\usepackage{bm}
\usepackage{latexsym}
\usepackage{color}
\usepackage{tabularx}
\usepackage{ulem}
\usepackage{hyperref}
\usepackage{float}
\usepackage{tabularx}
\usepackage{color}
\usepackage{comment}
\usepackage{physics}
\usepackage{caption}
\usepackage{subcaption}

\usepackage{float}
\usepackage{tikz}
\usepackage{hyperref}
\usepackage{colortbl}
\usepackage{listings}
\usepackage{amsmath,amsfonts,amssymb}
\usepackage{fancyhdr}
\usepackage{hyperref}
\usepackage{natbib}
\hypersetup{
	colorlinks   = true, 
	urlcolor     = blue, 
	linkcolor    = blue, 
	citecolor    = red 
}

\hypersetup{
	colorlinks   = true, 
	urlcolor     = blue, 
	linkcolor    = blue, 
	citecolor   = red 
}

\begin{document}
\title{Quantum corrections enhance chaos: study of particle motion near a generalized Schwarzschild black hole}
\author{Avijit Bera}
\email{Avijit.Bera@utdallas.edu}
\affiliation{Department of Physics, The University of Texas at Dallas, Richardson, TX75080, USA}
\author{Surojit Dalui}
\email{suroj176121013@iitg.ac.in}
\affiliation{Department of Physics, Indian Institute of Technology Guwahati, Assam, India}
\author{Subir Ghosh}
\email{subirghosh20@gmail.com}
\affiliation{Physics and Applied Mathematics Unit, Indian Statistical Institute Kolkata, India}
\author{Elias C. Vagenas}
\email{elias.vagenas@ku.edu.kw}
\affiliation{Theoretical Physics Group, Department of Physics Kuwait University, P.O. Box 5969, Safat 13060, Kuwait}

\date{\today}

\begin{abstract}
\par\noindent
The paper is devoted to a detailed study  of the effects of quantum corrections on the chaotic behavior in the dynamics of a (massless) probe particle near the horizon of a generalized Schwarzschild black hole. Two possible origins inducing the modification  of black hole metric are considered separately; the noncommutative geometry inspired metric (suggested by   Nicolini,  Smailagic and  Spallucci) and the metric with  quantum field theoretic corrections (derived by Donoghue). Our results clearly show that in both cases, the metric extensions favour chaotic behavior, namely chaos is attained for relatively lower particle energy. This is demonstrated numerically by exhibiting the breaking of the KAM tori in Poincar\'e sections of particle trajectories and also via explicit computation of the (positive) Lyapunov exponents of the trajectories.
\end{abstract}

\maketitle

%
%
%
%
\section{Introduction}
%
%
%
%
%
\par\noindent
A universal upper bound on chaos in quantum field theory at temperature $T$ has been discovered by Maldacena, Shenker and Stanford \cite{Maldacena:2015waa}, in terms of the  Lyapunov exponent $\lambda_L$ of out-of-time-ordered correlators (OTOC)
\begin{equation}
\lambda_L \leq 2\pi T/\hbar ~.
 \label{m1}   
\end{equation}
The bound is saturated for the Sachdev-Ye-Kitaev models \cite{sy,k}. This can be exploited to study the effect of temperature on gravity since black hole thermodynamics connects black hole surface gravity $\kappa$  to Hawking temperature
\begin{equation}
\kappa = 2\pi T/\hbar ~.
 \label{m2}   
\end{equation}
Now, the Lyapunov exponent can be calculated independently for a black hole from the study of chaotic   behavior in particle motion near   a black hole horizon \cite{Hashimoto:2016dfz,Hashimoto:2018fkb}. Thus, the universality of the bound given by Eq.  (\ref{m1}) can be established in a different setting. This has triggered an immediate interest in the particle motion near a black hole horizon. Analysis of shock waves near black hole
horizons \cite{3} and AdS/CFT correspondence \cite{10} originally provided $T$-dependence of $\lambda$. Of topical interest is the possibility that signatures of chaotic behavior around
black holes  on gravitational waves emitted from it \cite{49} might be observed in the recently realized  LIGO experiment \cite{52}.
Recently it has been shown (in works involving one of us) that in the presence of the Rindler horizon where the intrinsic curvature of the space-time is absent, unlike the case of black holes, the particle dynamics becomes chaotic in the near-horizon region \cite{Dalui:2018qqv,Dalui:2019umw}. 

Studies on the influence of the black hole horizon in inducing chaos on particle motion near it have a long history  \cite{Bombelli:1991eg,Sota:1995ms,Vieira:1996zf,  Suzuki:1996gm,Cornish:1996ri,deMoura:1999wf,Hartl:2002ig,Lei:2020clg,Han:2008zzf,Takahashi:2008zh,Hashimoto:2016dfz,Li:2018wtz,Hashimoto:2018fkb,Dalui:2018qqv,Colangelo:2020tpr}. This feature has led many researchers  to the study of near-horizon chaotic dynamics both in the classical as well as in the quantum regime. In the former, it has been shown that in the presence of event horizon for different kinds of black holes, either  static spherically symmetric \cite{Sota:1995ms,Suzuki:1996gm} or rotational \cite{Hartl:2002ig}, or magnetized \cite{Li:2018wtz}, the particle dynamics becomes chaotic in the vicinity  of the horizon.  In all those cases, the considered test particle was either massive, charged \cite{Lei:2020clg}, spinning \cite{Han:2008zzf,Takahashi:2008zh} or  massless. This shows the fact that in the classical scenario, the horizon has an inherent property of inducing chaos in a system \cite{Bombelli:1991eg,Suzuki:1996gm,Cornish:1996ri,deMoura:1999wf,Takahashi:2008zh}. Exploration of chaotic dynamics in the latter, namely in the context of quantum chaos, the characteristics OTOC has been mainly studied \cite{Maldacena:2015waa,Hashimoto:2017oit}. The exponential growth of OTOC is the main signature of  quantum chaos \cite{Maldacena:2015waa,Hashimoto:2017oit}. Therefore, all these analyses indicate towards one conclusion that \textit{a horizon is the nest of chaos.}

Let us come to our work in this perspective. In the extreme environment near the black hole horizon, it is expected that quantum gravity will play a decisive role. In the absence of  a fundamental theory one considers physically motivated and viable models that are extensions of conventional theories that incorporate quantum gravity effects in a phenomenological way. One such extension is the Generalized Uncertainty Principle (GUP) \cite{kemp} that takes into account the existence of  a minimal length scale, which is advocated, through diverse models, as a characteristic feature   of quantum gravity scenario. The GUP framework has been very productive in generating a plethora of quantum gravity signatures in conventional physics (see for example \cite{rev1,rev2,rev3}). In the context of the theory of gravitation, a significant application of GUP effect is a modified black hole metric, derived by Nicolini, Smailagic and  Spallucci \cite{nic}. This metric appears as a solution of Einstein field equations where the matter density is given by a distribution with an inbuilt minimal length (of quantum gravity origin). This metric has the cherished feature that, on one hand, the essential black hole singularity is removed, whereas on the other hand, at large distance the GUP effects weaken and eventually the standard black hole metric is recovered (see \cite{Nicolini:2008aj} for a review). Because of its connection to quantum gravity inspired Non-Commutative (NC) geometry the above metric \cite{nic} is also referred to as NC-inspired metric.

In an alternative scheme, quantum field theoretic effects in general relativity have also been considered by Donoghue \cite{don} that are manifested in a generalized form of black hole metric, that is distinct from \cite{nic}. We will refer to this deformed metric as quantum-corrected metric.

In the present work we will study the near-horizon chaotic behavior  of probe particles in the background of the two types of generalized metrics mentioned above \cite{nic,don}. In a nutshell, our results indicate that both NC and quantum effects tend to increase the chaotic nature of the particle motion, namely chaos appears at a lower particle energy if NC or quantum effects are present.  In fact it is natural to ask whether these two deformations can be related. An interesting option is to exploit the approach in the work (involving one of us) \cite{elias} that estimated the NC parameter in generalized uncertainty principle by comparing the two results: NC parameter corrected Hawking temperature for a black hole on one hand and the same temperature computed from  Newtonian dynamics for an effective potential arising from a deformed black hole metric that incorporated quantum field theoretic corrections \cite{don}. We will come back to this question at the end.

The paper is organized as follows: In Section II A, the NC-extended metric, derived in \cite{nic}, together with the Hamiltonian equations of motion for the probe particle is given. In Section II B, the Poincar\'e sections are plotted numerically to reveal the near-horizon chaotic behavior. In Section II C, an approximate analytic form of the metric, derived in  \cite{nic}, is introduced which induces qualitatively similar chaotic behavior similar to the exact metric, as demonstrated  numerically. Subsequently, the Lyapunov exponents for the chaotic paths are derived to reveal the chaotic features quantitatively. In Section III, the metric with quantum field theoretic corrections, derived in \cite{don} is considered and the corresponding chaotic trajectories are studied via Poincar\'e sections (subsection IIIA) and via Lyapunov exponents (subsection IIIB). The paper concludes in Section IV with a discussion of results and future directions of research.
%
%
%
%
\section{Near horizon chaos for  NC-inspired Schwarzschild metric\label{NC SCH}}
%
%
%
%
%
\subsection{Exact form of NC-deformed metric}
\par\noindent
In this section we shall start with the NC-inspired Schwarzschild metric given in the form \cite{Nicolini:2008aj} 
\begin{eqnarray}
ds^2 = -f_{nc}(r)dt^2 + f_{nc}(r)^{-1} dr^2 + r^2 d\Omega^2\label{Nicolini metric}
\end{eqnarray}
where $f_{nc}(r)=\left(1-(4 M/r \sqrt{\pi})\gamma(3/2,r^2/4 \theta_{nc})\right)$ and 
$\gamma(3/2, r^2/4\theta_{nc}) = \int_{0}^{r^2/4\theta_{nc}}x^{1/2} e^{-x} dx $. The dimensional constant ${\sqrt{\theta_{nc}}}\sim length $ corresponds to the NC parameter and fixes the scale for NC effects to be appreciable. For $r>>{\sqrt{\theta_{nc}}}$, $f_{nc}$ reduces to the standard Schwarzschild form $f_{nc}\approx 1-2M/r$. The horizon $r=r_{H}$ is determined by $f_{nc}(r=r_{H})=0$ and $d\Omega^{2}=(d\theta^{2}+\sin^{2}\theta d\phi^{2})$. This line element is a solution of the Einstein equation where the energy density distribution of a static, spherically symmetric,  particle-like gravitational source is  diffused due to the presence of the inherent length scale in $\theta_{nc}$.  The generalized density is given by \cite{Nicolini:2008aj}
\begin{eqnarray}
\rho_{\theta_{nc}}(r)=\frac{M}{(4\pi\theta_{nc})^{3/2}}\exp\left(-r^{2}/4\theta_{nc}\right)~.\label{energy density}
\end{eqnarray}	
%
%
%
\par\noindent
Since we are interested in the near-horizon dynamics of the probe particle and noncommutativity obviously affects the horizon, it is important  to  note that there is an interplay between the black hole mass and numerical value of the NC parameter $\theta_{nc}$. This is discussed in detail in \cite{Nicolini:2008aj}. From the NC-corrected metric structure (\ref{Nicolini metric}), there is a critical value of the black hole mass $M=M_{0}\thickapprox1.90\sqrt{\theta_{{nc}_0}}$ below which horizon will not form. In fact, for $M>M_{0}$ there are two horizons that coalesce to a single horizon at $M=M_{0}$ which disappears for $M<M_{0}$ \cite{Nicolini:2008aj}. Thus, corresponding to a given value of the black hole mass, there is an upper bound for $\theta_{nc}=\theta_{{nc}_0}$.
\par\noindent
It should be  pointed out  that as we decrease the value of $\theta_{nc}$, the diffusive nature of  energy density  gets reduced so that the results tend towards  the standard Schwarzschild black hole (without noncommutative correction) \cite{Dalui:2018qqv}. As we decrease the value of $\theta_{nc}$, the expression of $\rho_{\theta_{nc}}(r)$ becomes the energy density distribution of a $\delta$-functional like that of a point  gravitational source \cite{Nicolini:2008aj}. So, it is evident that with the decrease of $\theta_{nc}$, the result will tend towards the results of the standard Schwarzschild black hole. Profile of the NC-modified metric is shown in Fig. 1. Here all the graphs are for different $\theta_{nc}$ but fixed black hole mass $M$. It is interesting to note that there is a critical black hole mass (function of $\theta_{nc}$) below which there is no horizon and above which there are two horizons, as is the present case. The two values coalesce at the critical mass. For still larger $M$ the horizons move apart until in the limit  the inner horizons shrinks to zero and the outer one equals the Schwarzschild horizon $2M$. Within the inner horizon the original singularity is replaced by  a de Sitter core of constant curvature (for details see \cite{nic,Nicolini:2008aj}) as a result of noncommutative effects. In our work, we will always consider the region near and outside  of the outer horizon since it is a closer analogue of the Schwarzschild horizon. 
\par\noindent
Similar to standard Schwarzschild metric, the NC-corrected metric as given by (\ref{Nicolini metric}) has a coordinate singularity at the horizon $r=r_{H}$. To remove this, we shall adopt the Painleve coordinate transformation \cite{Painleve:1921,Parikh:1999mf}
\begin{eqnarray}
dt\rightarrow dt-\frac{\sqrt{1-f_{nc}(r)}}{f_{nc}(r)}dr~.
\end{eqnarray} 
Implementing this  transformation, the metric (\ref{Nicolini metric}) takes the following form
\begin{eqnarray}
ds^{2}=-f_{nc}(r)dt^{2}+2\sqrt{1-f_{nc}(r)}dtdr+dr^{2}+r^{2}d\Omega^{2}~.~~~~\label{Nicolini painleve metric}
\end{eqnarray}
The above metric has a timelike Killing vector $\chi^{a}=(1,0,0,0)$. With the help of this Killing vector $\chi^{a}$, we can define the energy of the particle, moving under this background as $E=-\chi^{a}p_{a}=-p_{t}$ where $p_{a}$ is the four momentum vector, i.e., $p_{a}=(p_{t},p_{r},p_{\theta},p_{\phi})$. In the present context, our aim is to study the particle motion near the horizon. To do that first we need to formulate the expression of the particle energy in the background (\ref{Nicolini painleve metric}). With the help of the dispersion relation $g^{ab}p_{a}p_{b}=-m^{2}$,  $m$ being the mass of the particle, we obtain
\begin{eqnarray}
E^{2}+2\sqrt{1-f_{nc}(r)}p_{r}E-\left(f_{nc}(r)p_{r}^{2}+\frac{p_{\theta}^{2}}{r^{2}}\right)=m^{2}~.
\end{eqnarray}   
At this point it should be noted that we have considered  the motion of the particle in  the poloidal plane, i.e., in the $r~ - ~\theta$ plane with $p_{\phi}=0$. Our entire calculation will be done for the case of a massless particle, i.e., $m=0$, and with this, we obtain the two solutions of energy
\begin{eqnarray}
E=-\sqrt{1-f_{nc}(r)}p_{r}\pm\sqrt{p_{r}^{2}+\frac{p_{\theta}^{2}}{r^{2}}}~
\end{eqnarray}
where the positive sign corresponds to the outgoing particle, and the negative sign corresponds to the ingoing one. In the present instance, we shall be considering only the case of  the outgoing particle having the positive sign solution. 

Our next task is to find out the trajectory of the outgoing particle, and it will be computed using Hamilton's equation of motion. However, before that, let us concentrate for a moment on our particular system. In many contexts, it has been shown that the particle trajectory experiences instability in the near-horizon regions. 
In \cite{Dalui:2019esx,Dalui:2020qpt,Dalui:2021tvy}, it has been shown that through considering a model  in the presence of a static spherically symmetric black hole \cite{Dalui:2019esx,Dalui:2020qpt} or of a Kerr black hole \cite{Dalui:2021tvy}, an outgoing massless and chargeless particle experiences instability and the Hamiltonian of the system is found to be of the kind $H= xp$  which is an unstable one. In the quantum region, it has been shown that this instability leads to thermality in the system, which suggests the fact that instability and thermality have some intimate connection in the context of the horizon. On a similar note in \cite{Hegde:2018xub},  it has been shown that the effective motion of the particle near the black hole horizon is equivalent to the motion in an inverted harmonic potential (IHO), and such IHO gives rise to temperature under quantization which is proportional to the instability factor of the system. \\
Now, coming back to the chaotic motion of the particle in the near-horizon region, in different contexts it has been shown that for different values of parameters  this instability leads to chaotic motion of the particle \cite{Dalui:2018qqv,Dalui:2019umw}. Keeping in mind that the particle must not fall into the black holes, an external potential (like harmonic potential \cite{Dalui:2018qqv} or any other effective potential \cite{Hashimoto:2016dfz}) has been applied in order to keep the particle bounded in the near-horizon region. \\
%
%
%
For our case, we have chosen an external harmonic potential in the $(r,\theta)$ plane for the simplicity of the problem. The significance of considering the harmonic potential is that it is the simplest stable potential where the particle always finds its fixed point at the minima of the potential where it can bound itself. Now, by tuning the parameters of the spring constants of the harmonic potential we can bound the particle very easily in the near-horizon region. However, one may check the change in the dynamics of the particle trajectory by introducing any other arbitrary external potential in place of a harmonic potential. Interestingly, the motion along the radial direction remains unaffected as long as a massless particle is concerned (see Section III of \cite{Dalui:2018qqv}).\\
Another way of visualising the effect of the harmonic potential (introduced by hand) on top of the black hole potential is the following. Note that  by itself the  particle trapped in a harmonic potential is an  integrable system, devoid of chaos. When it is  placed in the vicinity  of the black hole horizon the system can become chaotic. The harmonic potential in no way {\it{introduces}} chaos. It only helps in visualizing and quantifying chaos by making the orbits bounded and not ending in the black hole singularity. We would like to see how the dynamics of the system changes as we change the parameters $E$ and $\theta_{nc}$ of the system. In Appendix \ref{App2}, we have plotted the orbits of the particle trapped in the harmonic potential in the near-horizon region for different values of the system energy $E$ but for a particular value of the NC parameter $(\theta_{nc})$. In these figures, i.e., Fig. \ref{fig:14} and Fig. \ref{fig:15}, we have plotted the orbits in the $(r,\theta)$ plane in order to understand the characteristics of the dynamics of the composite system in the near-horizon region.\\
Therefore, introducing the harmonic potentials into the picture, we obtain the total energy of the system as
\begin{eqnarray}
E = &&-\sqrt{(1-f_{nc}(r))}p_r +  \sqrt{p_r^2 + \frac{ p_\theta^2}{r^2}} + \frac{1}{2}K_r (r-r_c)^2 \nonumber\\
&&+ \frac{1}{2}K_\theta (y-y_c)^2\label{energy}
\end{eqnarray}  
where $y=r_{H}\theta,~K_{r}$ and $K_{\theta}$ are spring constants while $r_c$ and $y_c$ are the equilibrium position of the two harmonic potentials. Using  Hamilton's equations of motion, we obtain  the particle dynamics
\begin{eqnarray}
\dot{r}&=&\frac{\partial E}{\partial p_{r}}=-\sqrt{1-f_{nc}(r)}+\frac{p_{r}}{\sqrt{p_{r}^{2}+\frac{p_{\theta}^{2}}{r^{2}}}}\label{rdot}\\
\dot{p_r}&=&- \frac{\partial E}{\partial r} = -\frac{f_{nc}'(r)}{2 \sqrt{(1-f_{nc}(r))}}p_r + \frac{p_\theta ^2 /r^3}{\sqrt{p_r^2 + \frac{ p_\theta ^2}{r^2}}}\nonumber\\
&&- K_r(r-r_c)\\
\dot{\theta}&=&\frac{\partial E}{\partial p_\theta} = \frac{p_\theta /r^2}{\sqrt{p_r^2 + \frac{ p_\theta ^2}{r^2}}}\\\label{theta_dot}
\dot{p_\theta}&=& - \frac{\partial E}{\partial \theta} = -K_\theta r_H(y-y_c)\label{ptheta_dot}
\end{eqnarray} 
where the derivative is taken with respect to some affine parameter.
In the next subsection, we shall study these equations with the help of numerical analysis to reveal the characteristics of the particle motion. 
\subsection{Numerical analysis for NC-deformed metric}
\par
\begin{figure}[H]
	\centering
	\begin{subfigure}[b]{0.49\linewidth}
		\includegraphics[width=\linewidth]{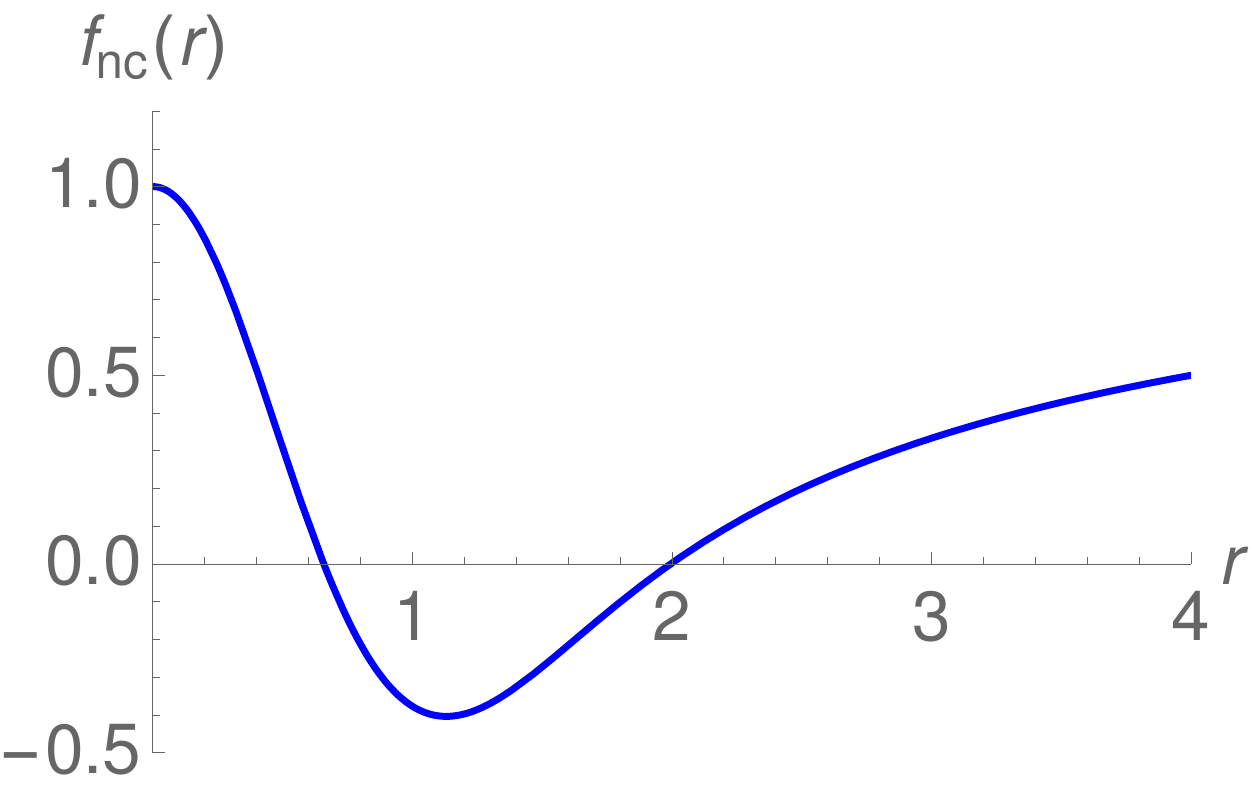}
		\caption{$\theta_{nc} = 0.14$}
	\end{subfigure}
	\begin{subfigure}[b]{0.49\linewidth}
		\includegraphics[width=\linewidth]{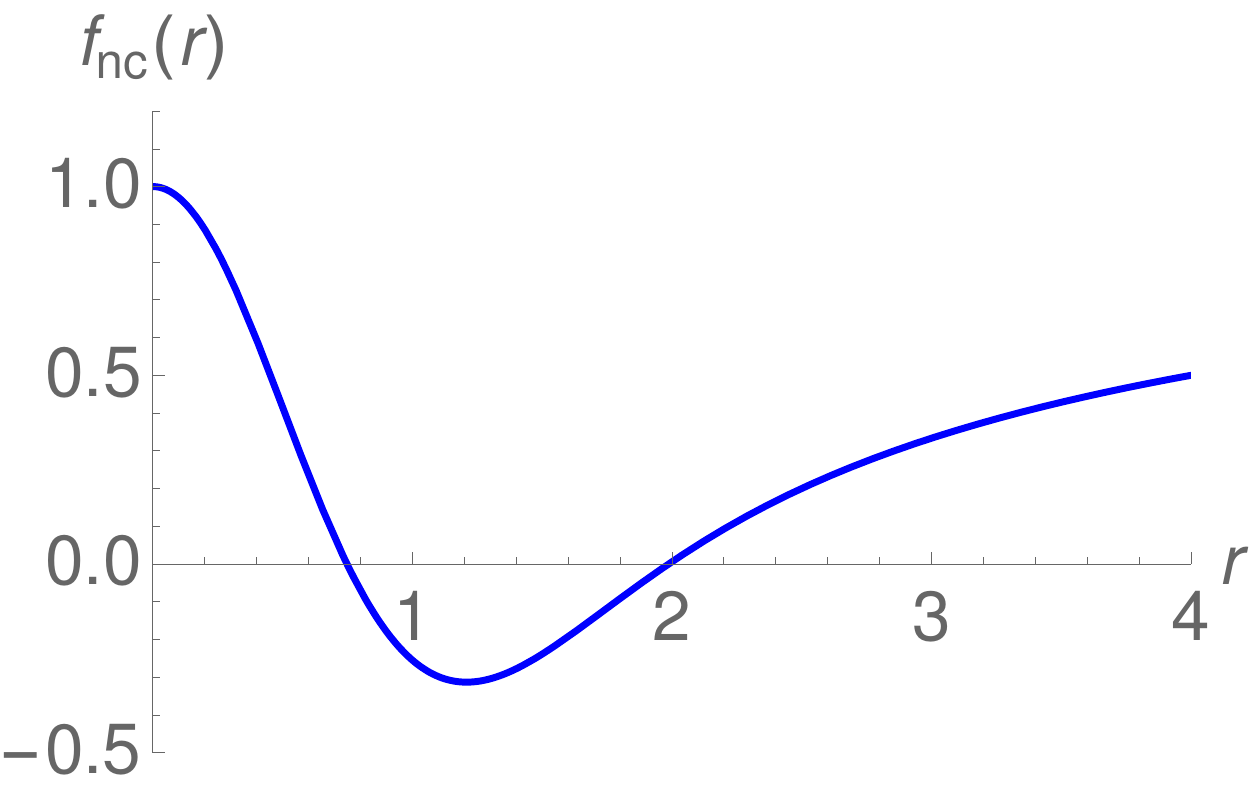}
		\caption{$\theta_{nc} = 0.16$}
	\end{subfigure}
	\begin{subfigure}[b]{0.49\linewidth}
		\includegraphics[width=\linewidth]{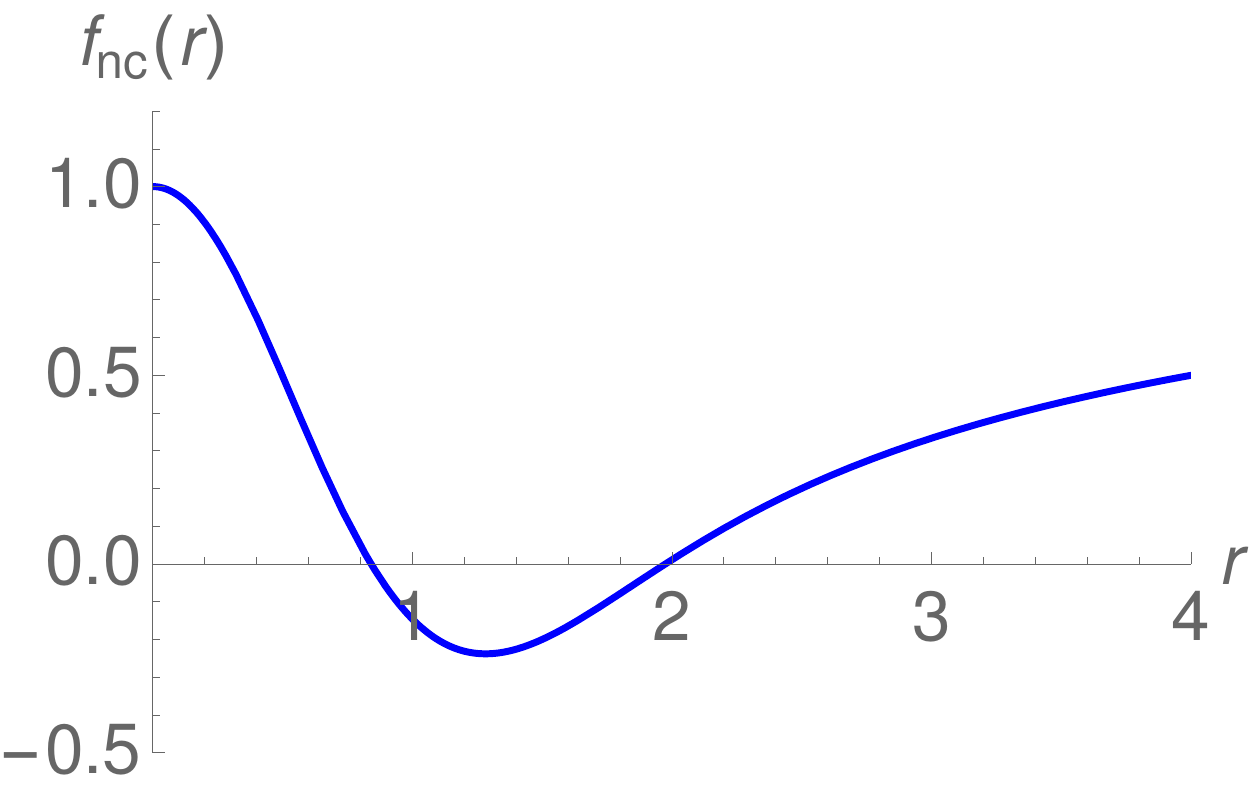}
		\caption{$\theta_{nc} = 0.18$}
	\end{subfigure}
	\begin{subfigure}[b]{0.49\linewidth}
		\includegraphics[width=\linewidth]{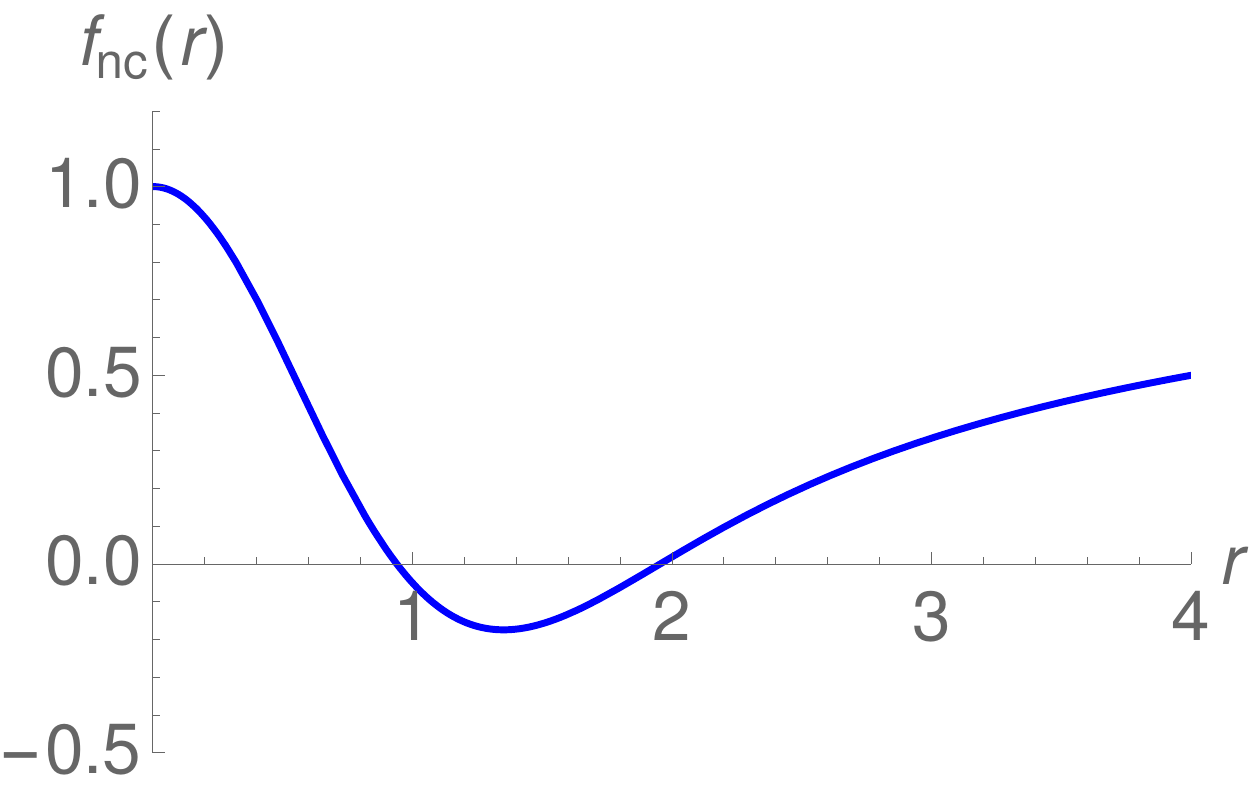}
		\caption{$\theta_{nc} = 0.20$}
	\end{subfigure}
	\caption{The figures show the variation of $f_{nc}(r)$ with $r$ for different values of $\theta_{nc}$ where the exterior horizon is at $r_H\approx2$.\label{fig:1}}
\end{figure}

\noindent
To demonstrate the characteristics of the motion of the particle, we first solve the equations of motion, namely equations (\ref{rdot})-(\ref{ptheta_dot}), numerically. Next, we present the Poincar\'e sections for our composite system in order to see the change in the dynamics of the particle motion with the variation of different parameters like the NC parameter $\theta_{nc}$ and the system energy $E$.  Although the exact form of the NC-deformed metric (\ref{Nicolini metric}) is valid for all $r$, the near-horizon condition is imposed in the numerical computation where, for our exterior black hole horizon at $r_{H}\approx 2$ (see Fig. \ref{fig:1}), the range of $r$ is fixed at $3.5>r>3.0$ (to make sure that the particle resides near the exterior horizon).

\subsubsection{Poincar\'e sections for NC-deformed metric}
\par\noindent
In the following figure, i.e., Fig. \ref{fig:2}, we show the Poincar\'e section of the particle trajectory projected over the $(r,p_r)$ plane for different energies but for a constant value of $\theta_{nc}$. These sections are plotted with the conditions $p_{\theta}>0$ and $\theta=0$. For the present case, we have also considered $M=1.0$, $K_{r}=100$, $K_{\theta}=25$, $r_{c}=3.2$ and $y_{c}=0$ solving the dynamical equations of motion of the particle, i.e., equations (\ref{rdot})-(\ref{ptheta_dot}). Following our earlier discussion, see  below  (\ref{energy density}), we adhere to the upper bound of the NC parameter $\theta_{nc}<\theta_{{nc}_0}$. In the present case, for the black hole mass $M=1.0$ this amounts to $\theta_{{nc}_0}\sim 0.27$.
\par\noindent
In these plots,  we have considered the energies $E=50, 55, 60$, and $65$. Now, looking  at these plots, it can be seen that for the lower energy value $E=50$, the Poincar\'e section exhibits the regular KAM torus, which suggests that our system is still periodic as only a single frequency is present in the system. However, as the total energy of the system is increased  $E=55~\text{and}~60$, the trajectory approaches  the horizon, and as a consequence of that, this torus starts getting distorted and finally breaks down, which indicates the appearance of chaos into the system. Furthermore, a further increase in the energy to $E=65$ results in the complete breaking of the regular torus and the appearance of scattered points in the plane. The emergence of these scattered points suggests that our system has reached a completely chaotic situation.
\begin{figure}[hbt!]
	\centering
	\begin{subfigure}[b]{0.4\linewidth}
		\includegraphics[width=\linewidth]{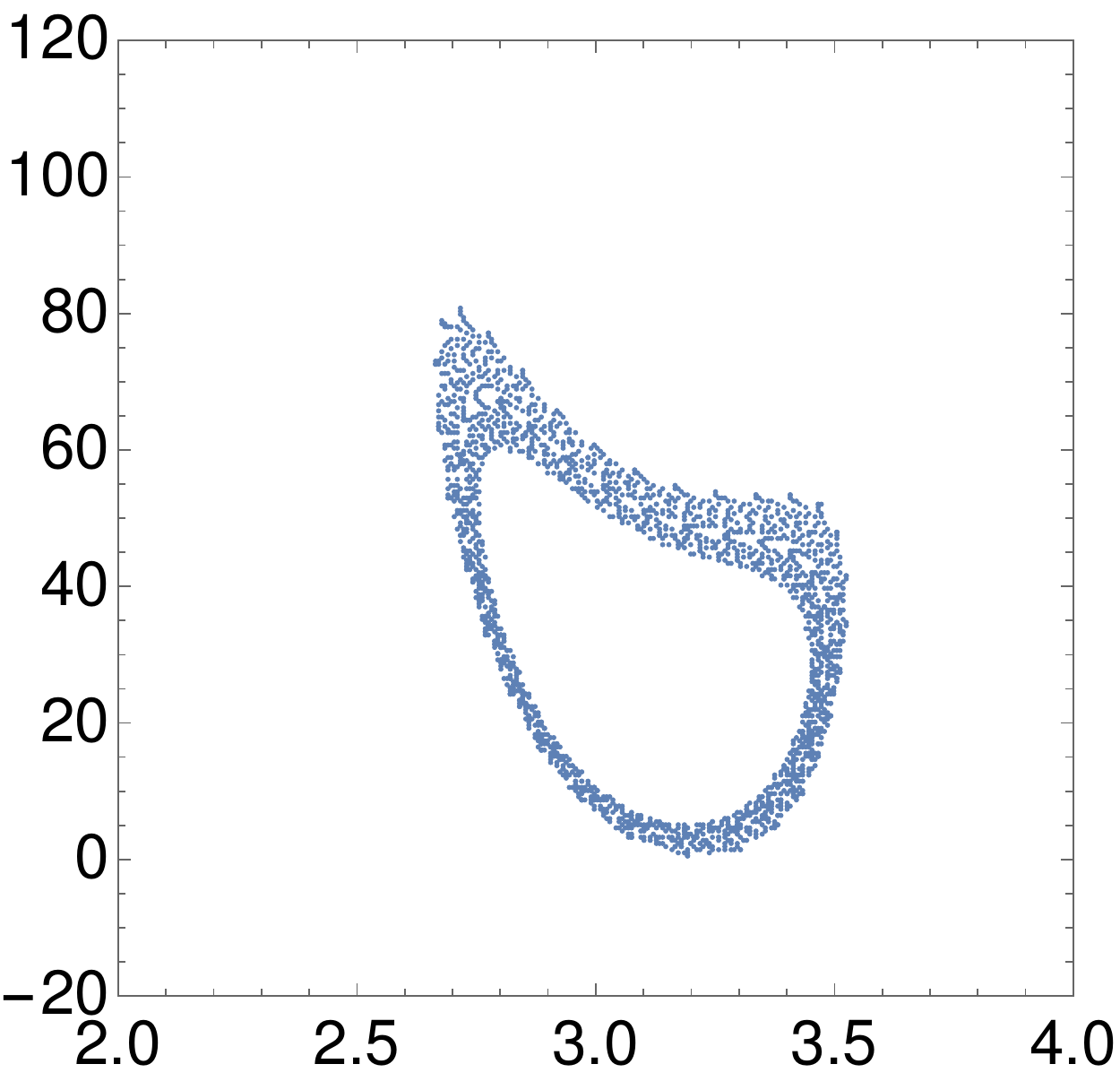}
		\caption{E=50}
	\end{subfigure}
	\begin{subfigure}[b]{0.4\linewidth}
		\includegraphics[width=\linewidth]{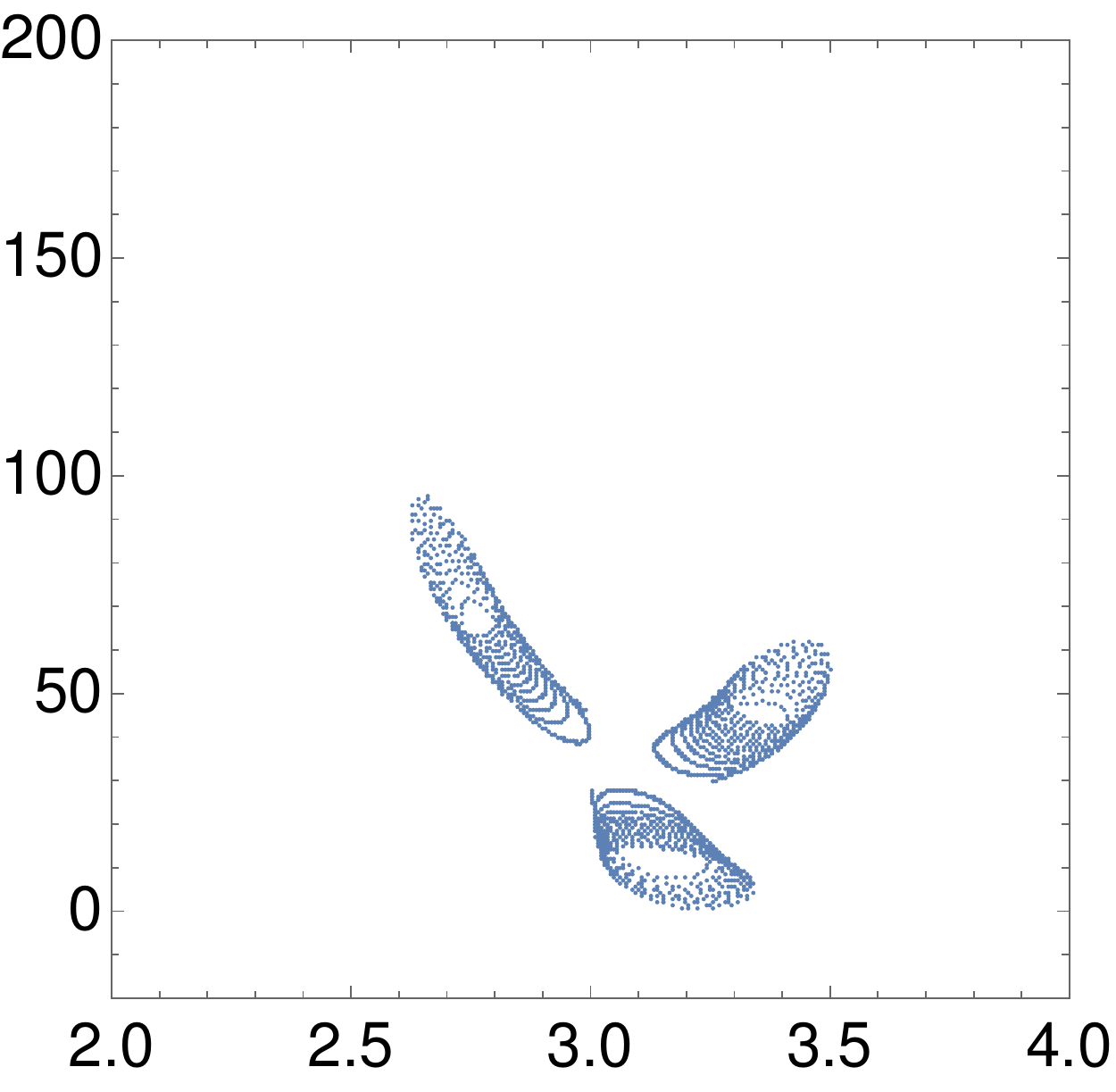}
		\caption{E=55}
	\end{subfigure}
	\begin{subfigure}[b]{0.4\linewidth}
		\includegraphics[width=\linewidth]{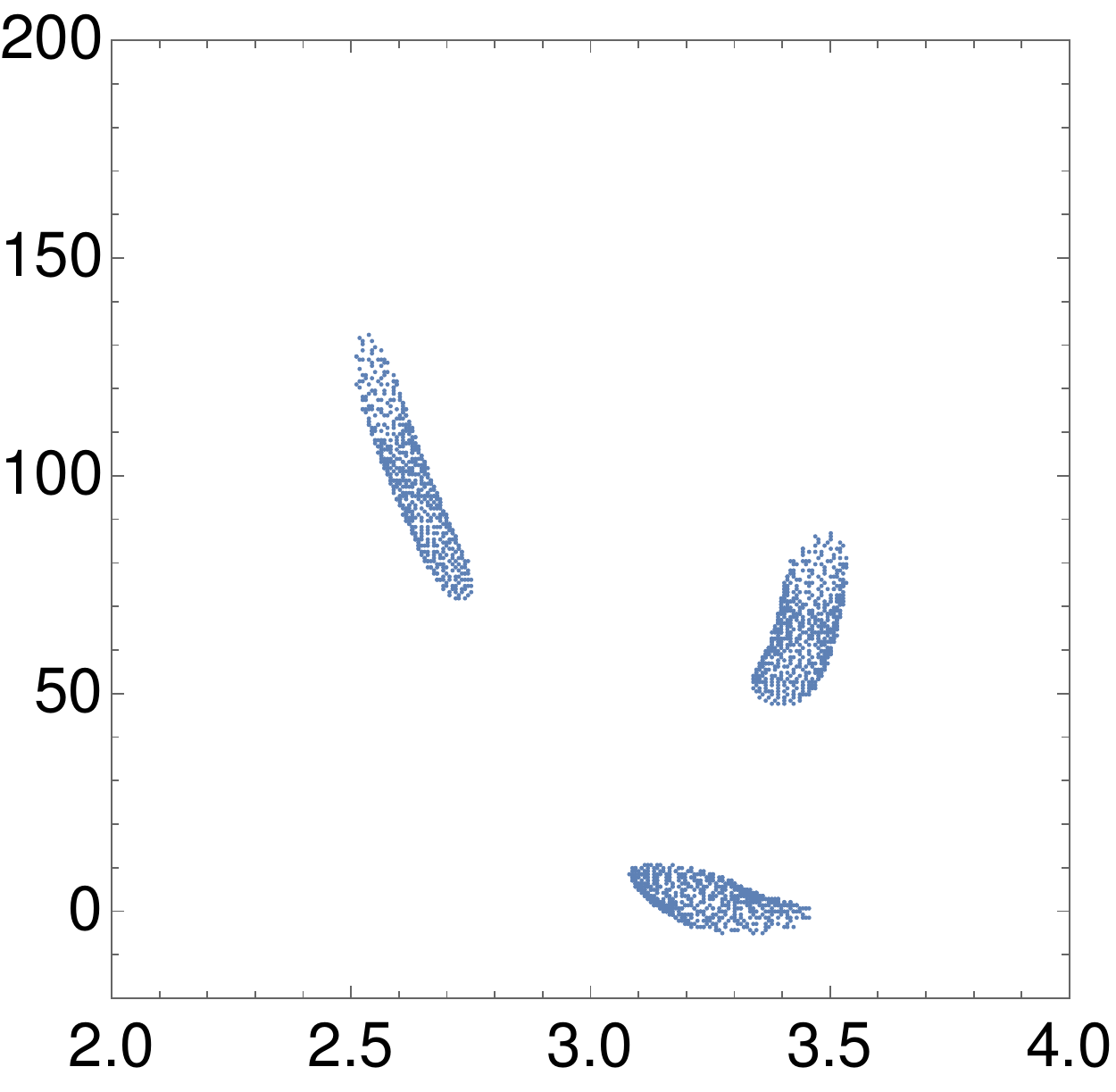}
		\caption{E=60}
	\end{subfigure}
	\begin{subfigure}[b]{0.4\linewidth}
		\includegraphics[width=\linewidth]{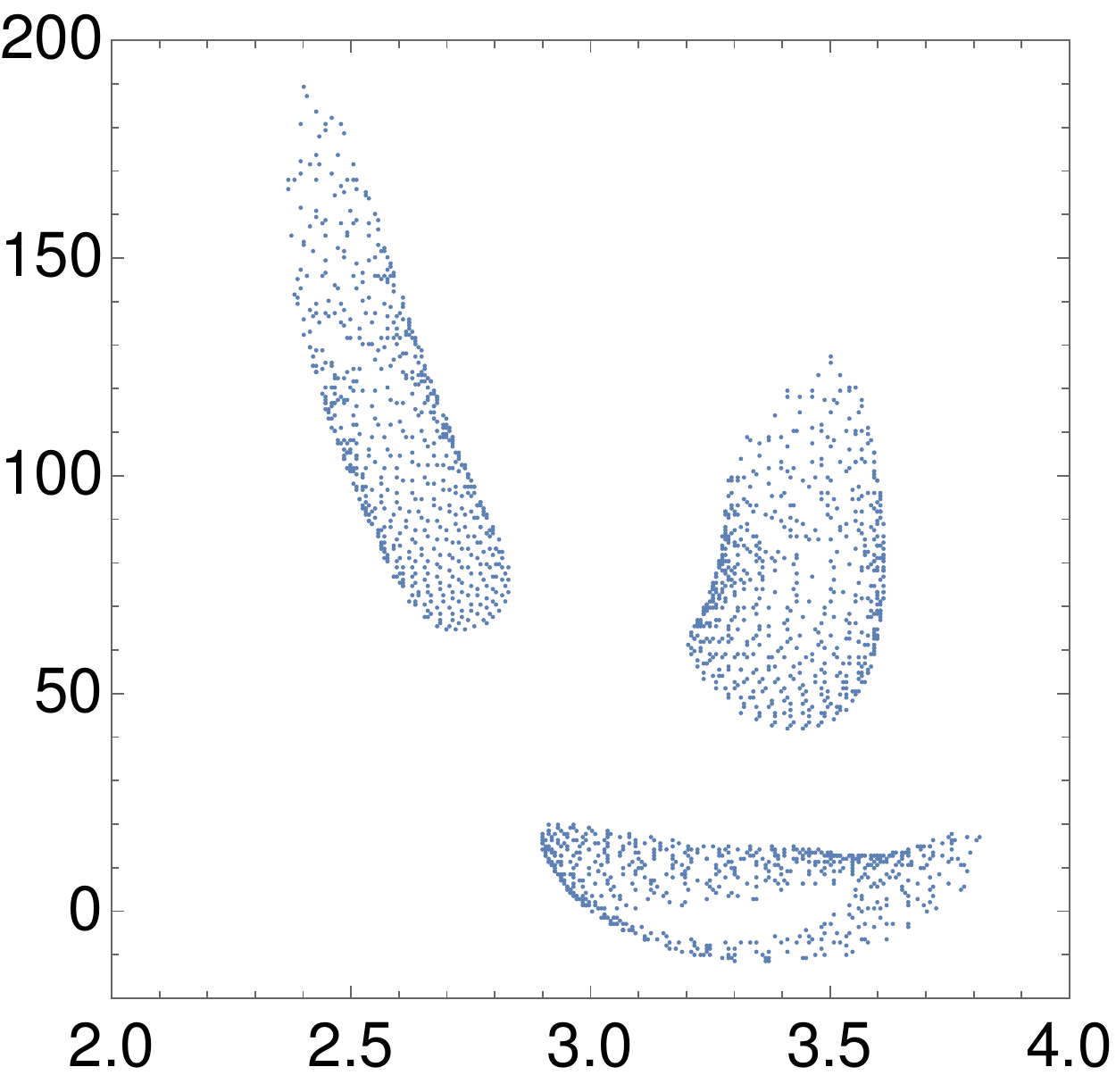}
		\caption{E=65}
	\end{subfigure}
	\caption{The Poincar\'e sections for $\theta_{nc} = 0.14$ in the ($r,p_r$) plane with $\theta = 0$ and $p_\theta > 0$ at different energies for quantum-corrected Schwarzschild black hole. The horizontal and vertical axes in each of the graphs correspond to $r$ and $p_r$, respectively.}
	\label{fig:2}
\end{figure}
Next we plot the Poincar\'e sections for increasing values of $\theta_{nc}=0.16$ (Fig. \ref{fig:3}), 0.18 (Fig. \ref{fig:4}) and 0.20 (Fig. \ref{fig:5}) but for the same energy values, i.e.,  $E=50,55,60$, and $65$, and try to analyze them in the following figures. 
\begin{figure}[hbt!]
	\centering
	\begin{subfigure}[b]{0.4\linewidth}
		\includegraphics[width=\linewidth]{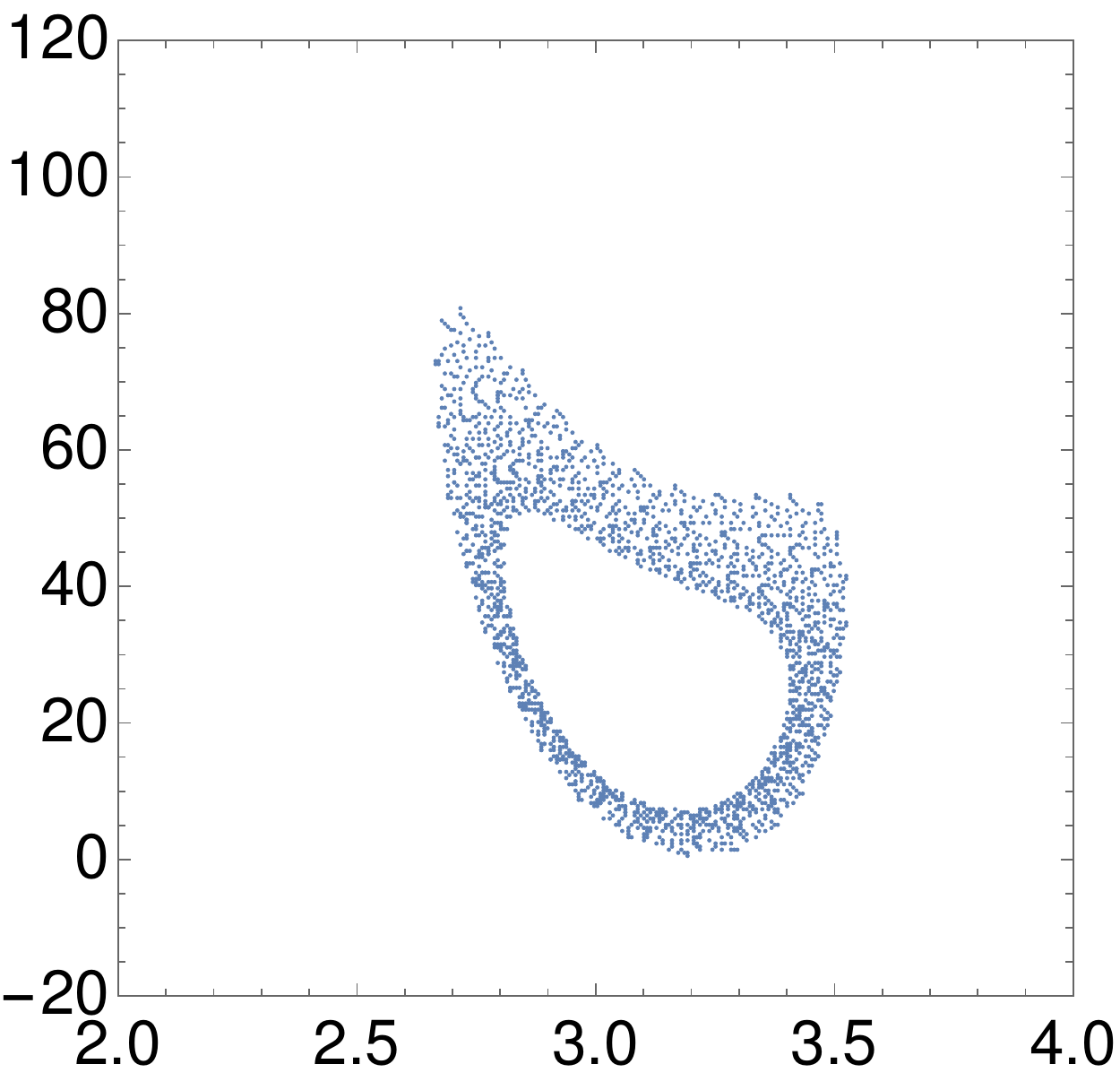}
		\caption{E=50}
	\end{subfigure}
	\begin{subfigure}[b]{0.4\linewidth}
		\includegraphics[width=\linewidth]{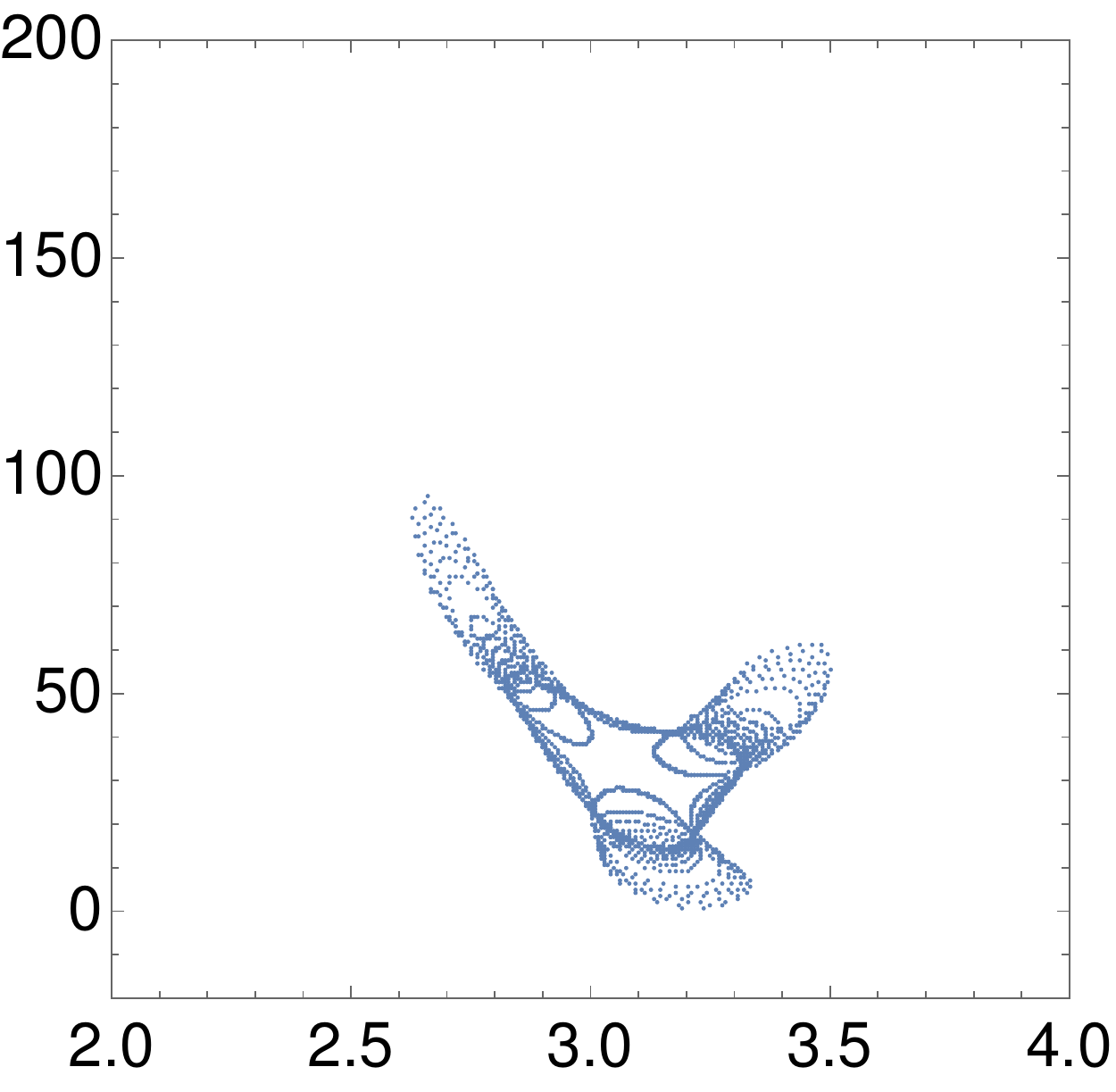}
		\caption{E=55}
	\end{subfigure}
	\begin{subfigure}[b]{0.4\linewidth}
		\includegraphics[width=\linewidth]{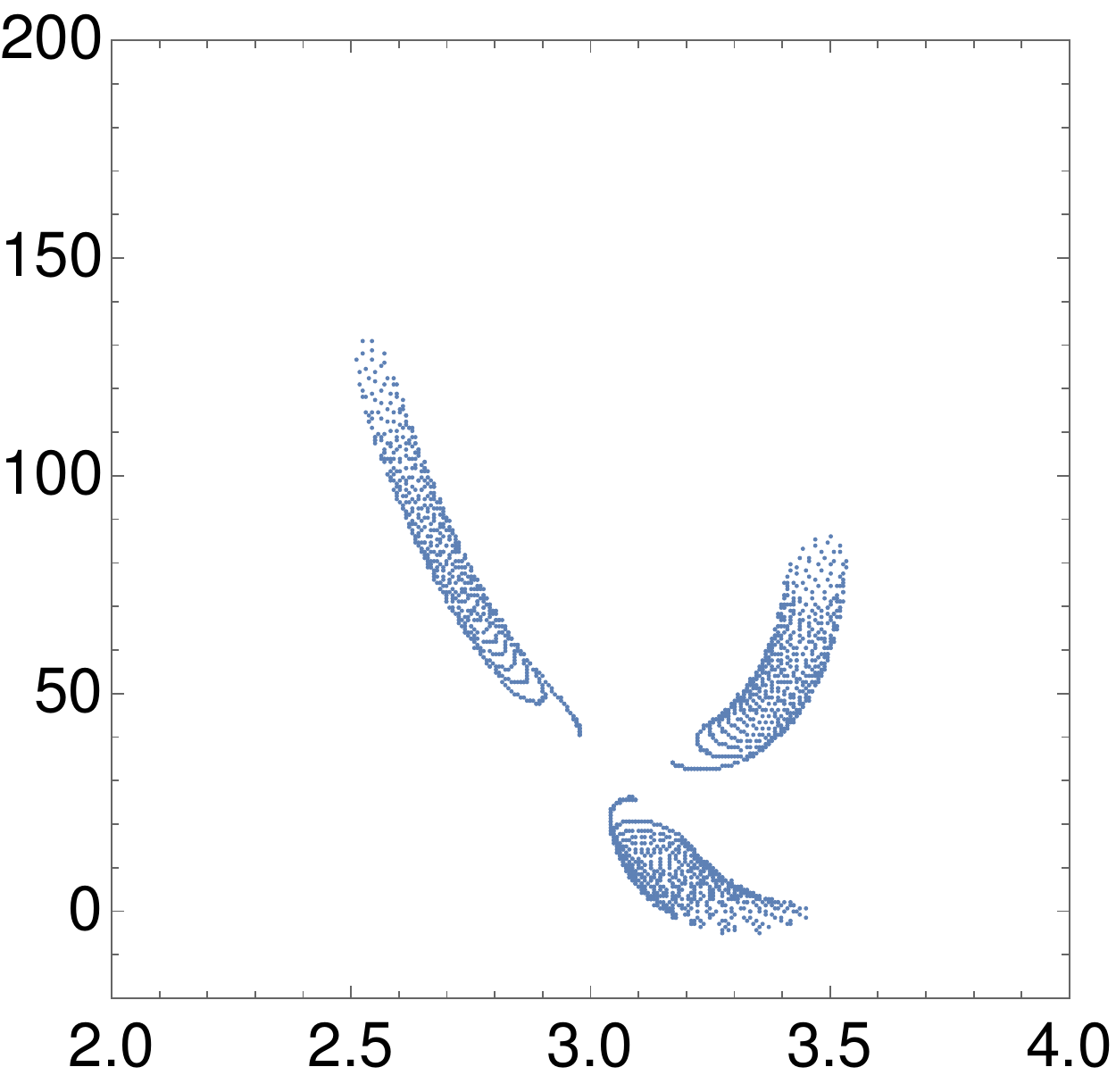}
		\caption{E=60}
	\end{subfigure}
	\begin{subfigure}[b]{0.4\linewidth}
		\includegraphics[width=\linewidth]{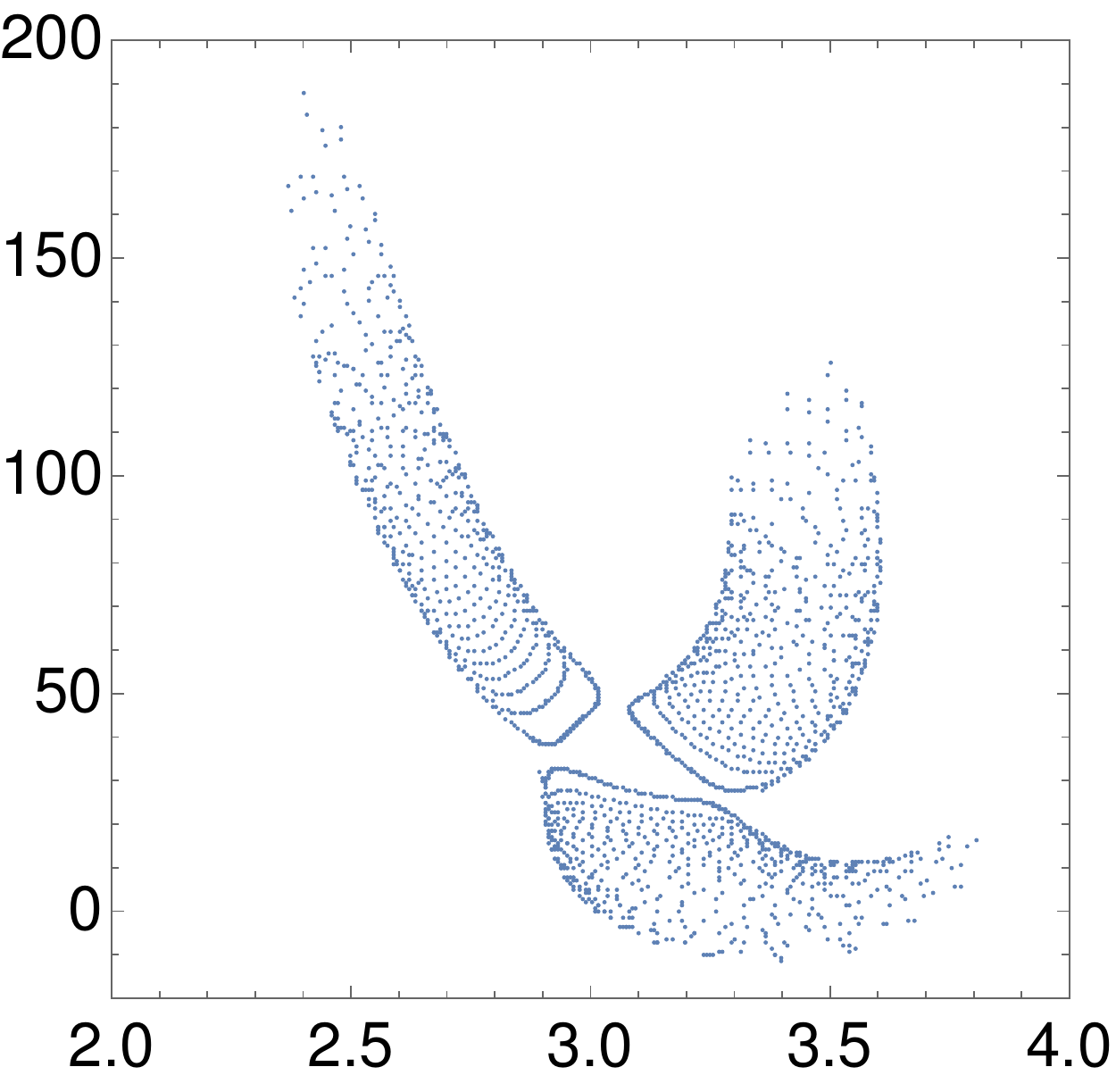}
		\caption{E=65}
	\end{subfigure}
	\caption{The Poincar\'e sections for $\theta_{nc} = 0.16$ in the ($r,p_r$) plane with $\theta = 0$ and $p_\theta > 0$ at different energies for the quantum-corrected Schwarzschild black hole. The horizontal and vertical axes in each of the graphs correspond to $r$ and $p_r$, respectively.}
	\label{fig:3}
\end{figure}
%
%
\begin{figure}[hbt!]
	\centering
	\begin{subfigure}[b]{0.4\linewidth}
		\includegraphics[width=\linewidth]{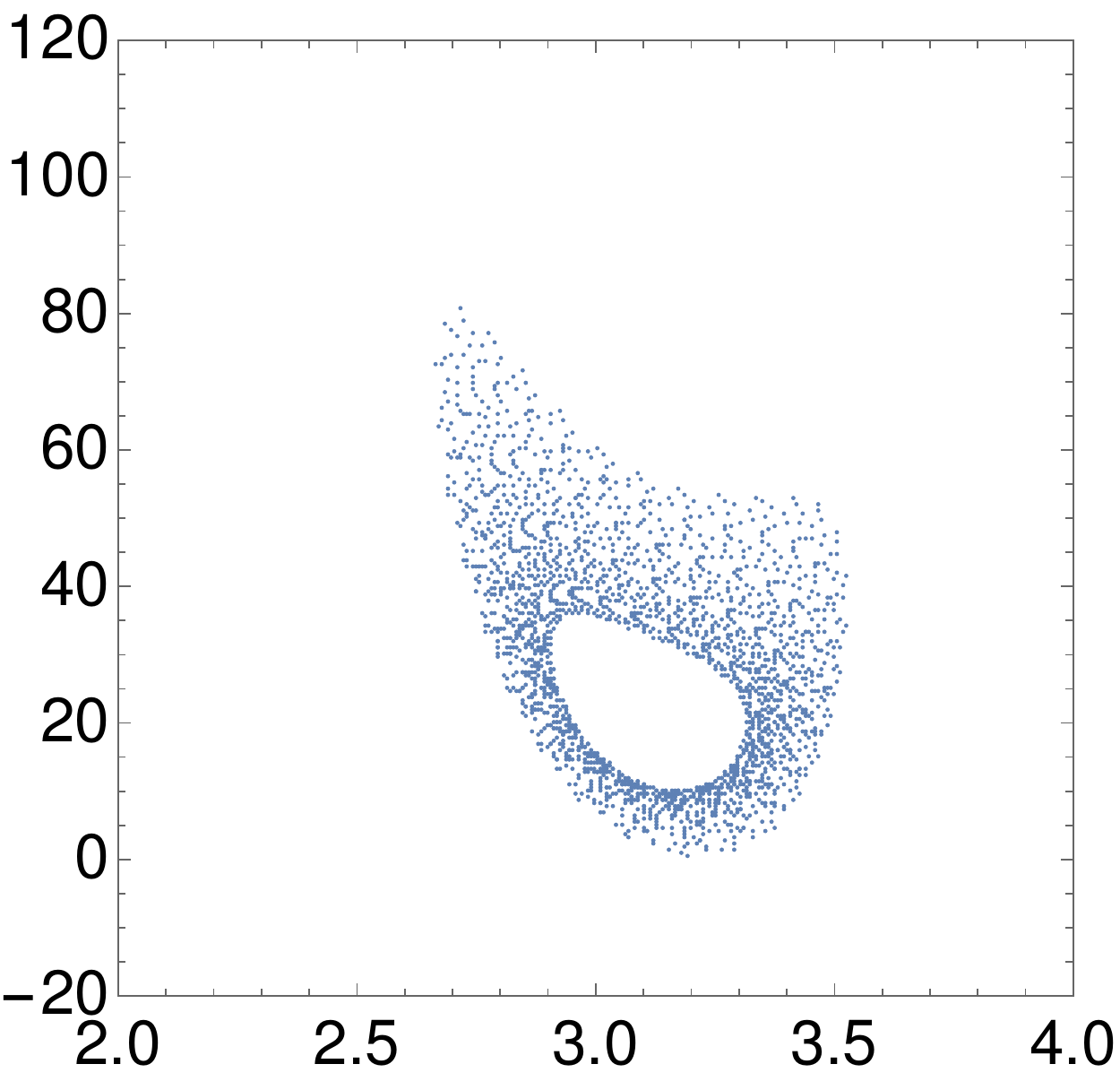}
		\caption{E=50}
	\end{subfigure}
	\begin{subfigure}[b]{0.4\linewidth}
		\includegraphics[width=\linewidth]{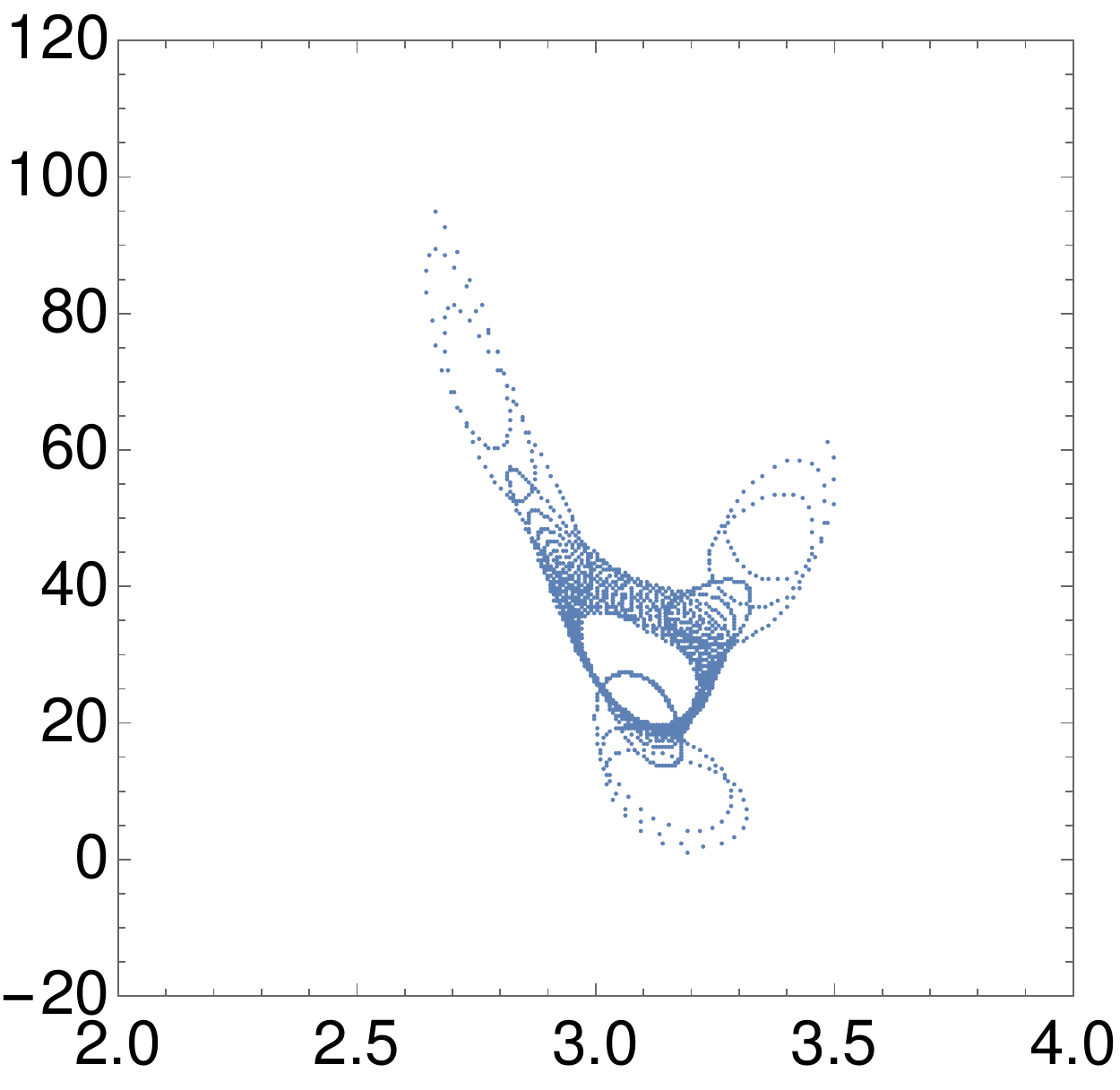}
		\caption{E=55}
	\end{subfigure}
	\begin{subfigure}[b]{0.4\linewidth}
		\includegraphics[width=\linewidth]{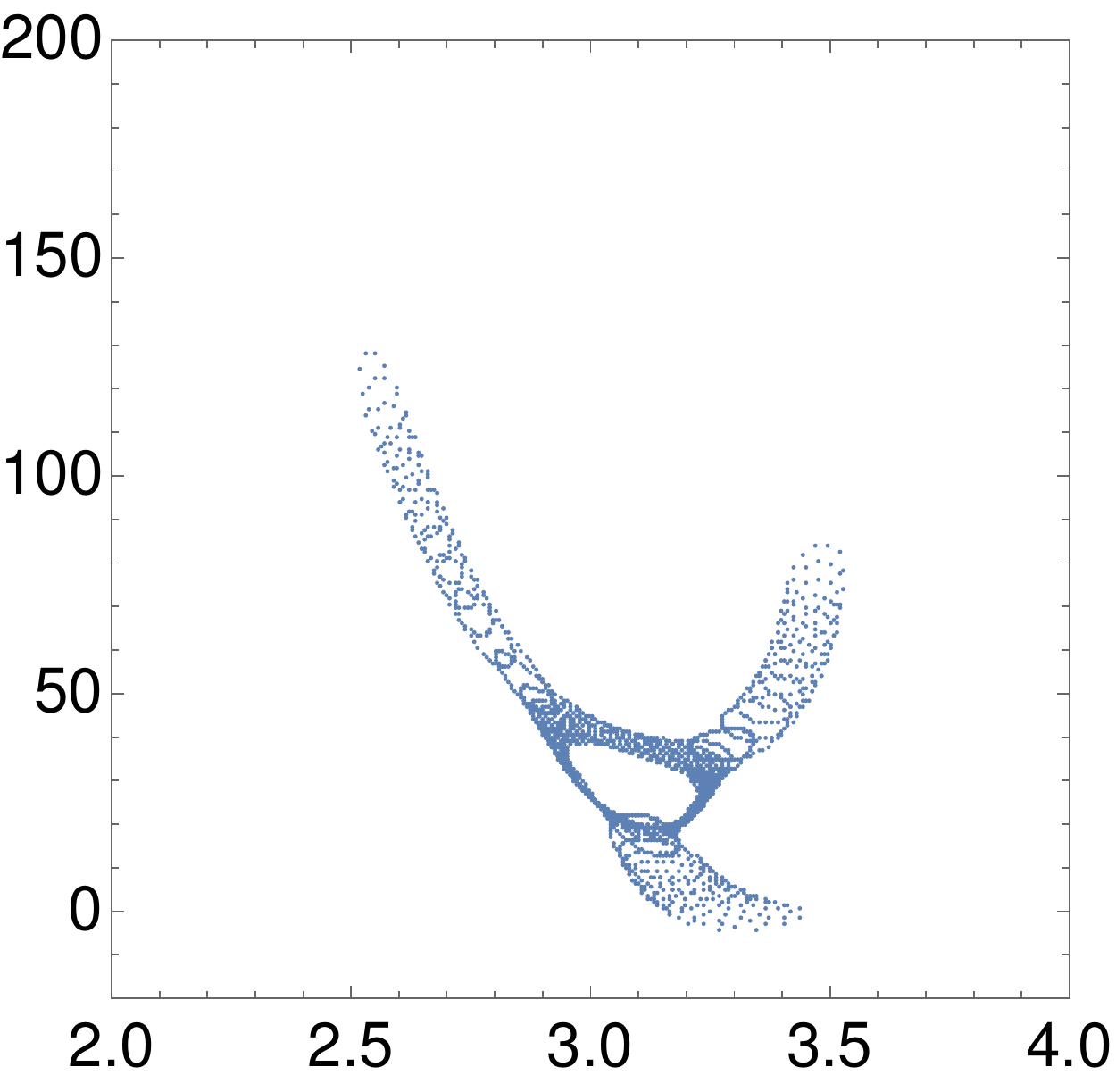}
		\caption{E=60}
	\end{subfigure}
	\begin{subfigure}[b]{0.4\linewidth}
		\includegraphics[width=\linewidth]{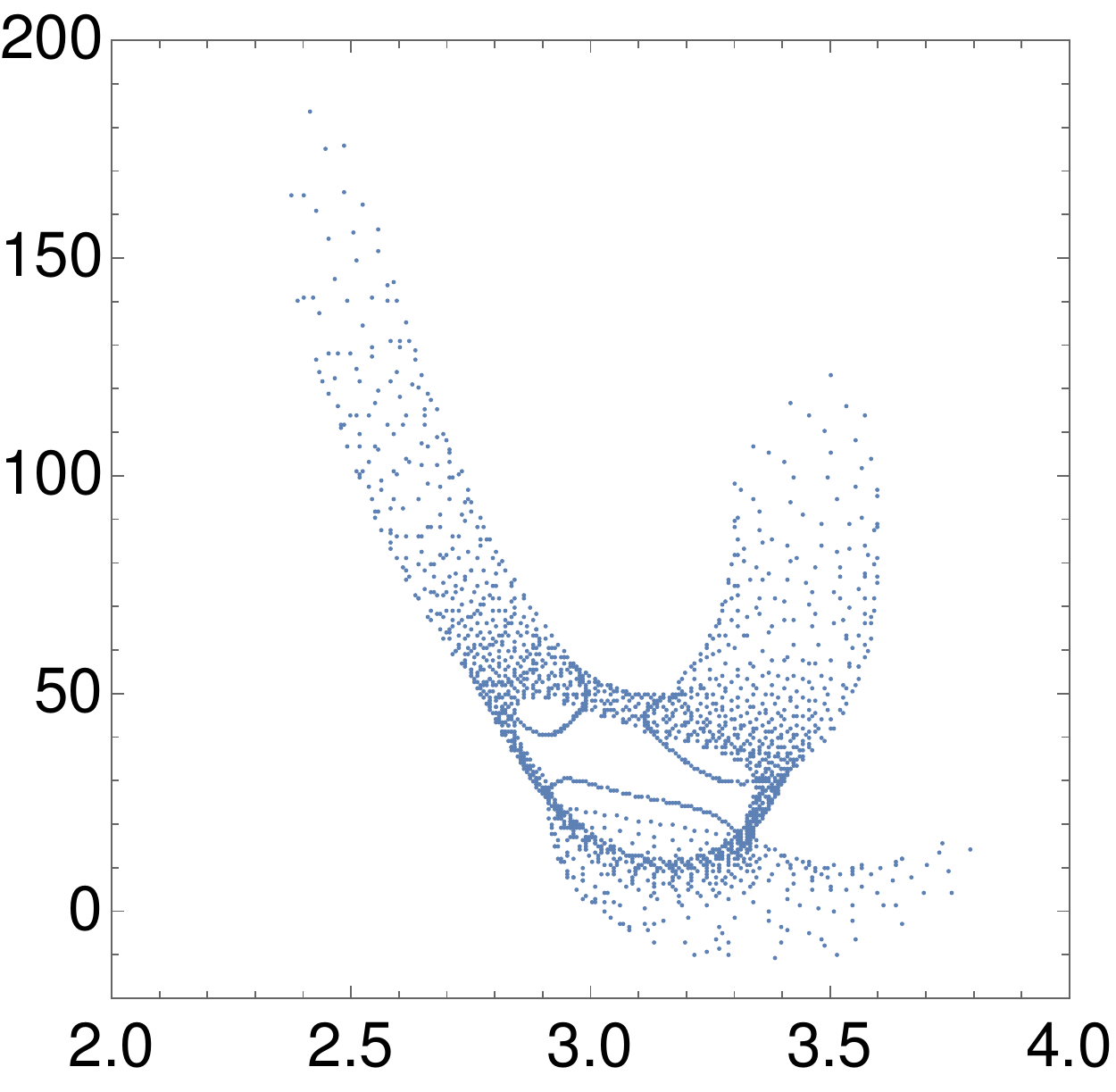}
		\caption{E=65}
	\end{subfigure}
	\caption{The Poincar\'e sections for $\theta_{nc} = 0.18$ in the ($r,p_r$) plane with $\theta = 0$ and $p_\theta > 0$ at different energies for the quantum-corrected Schwarzschild black hole. The horizontal and vertical axes in each of the graphs correspond to $r$ and $p_r$, respectively.}
	\label{fig:4}
\end{figure}
\begin{figure}[hbt!]
	\centering
	\begin{subfigure}[b]{0.4\linewidth}
		\includegraphics[width=\linewidth]{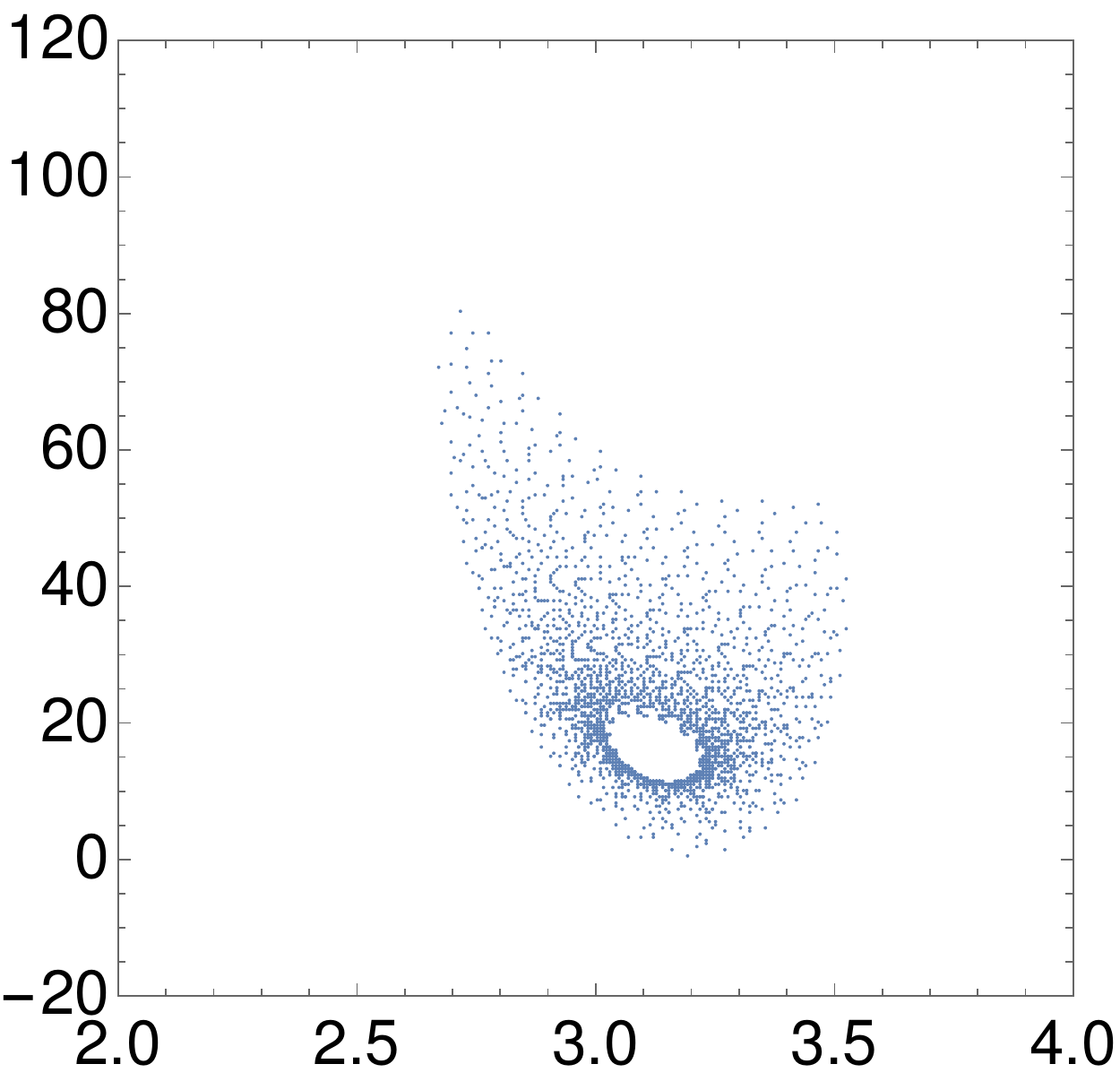}
		\caption{E=50}
	\end{subfigure}
	\begin{subfigure}[b]{0.4\linewidth}
		\includegraphics[width=\linewidth]{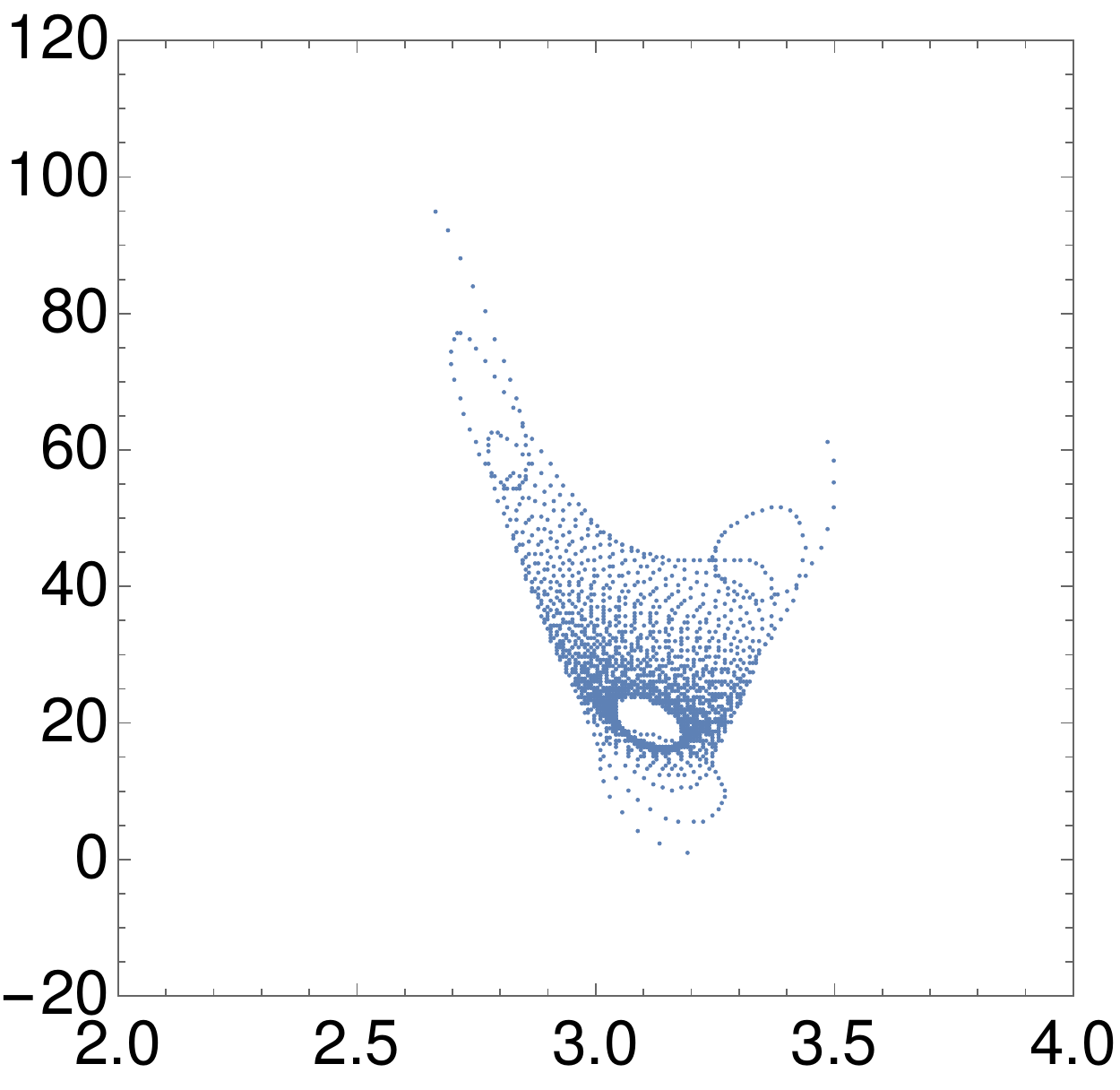}
		\caption{E=55}
	\end{subfigure}
	\begin{subfigure}[b]{0.4\linewidth}
		\includegraphics[width=\linewidth]{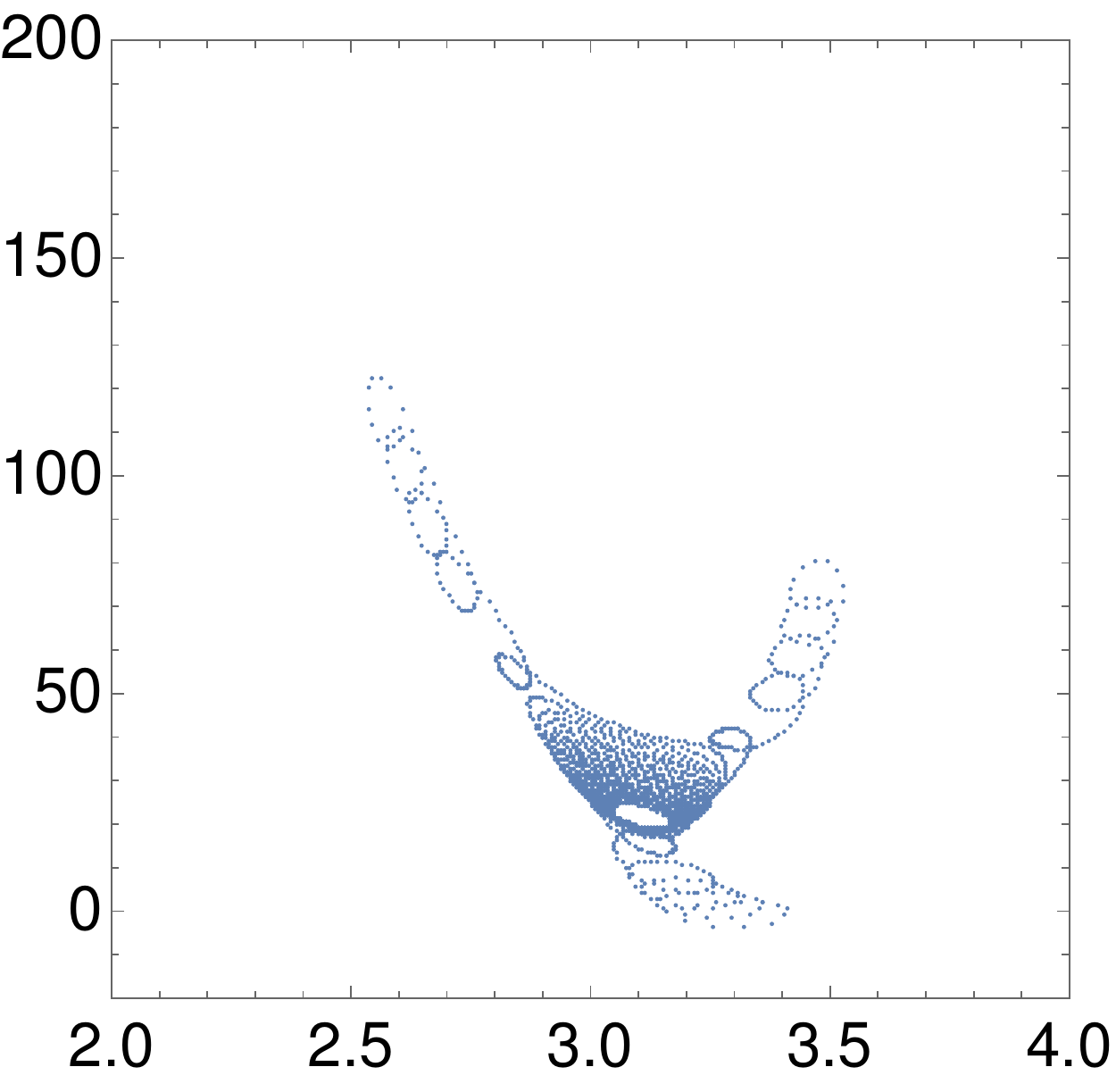}
		\caption{E=60}
	\end{subfigure}
	\begin{subfigure}[b]{0.4\linewidth}
		\includegraphics[width=\linewidth]{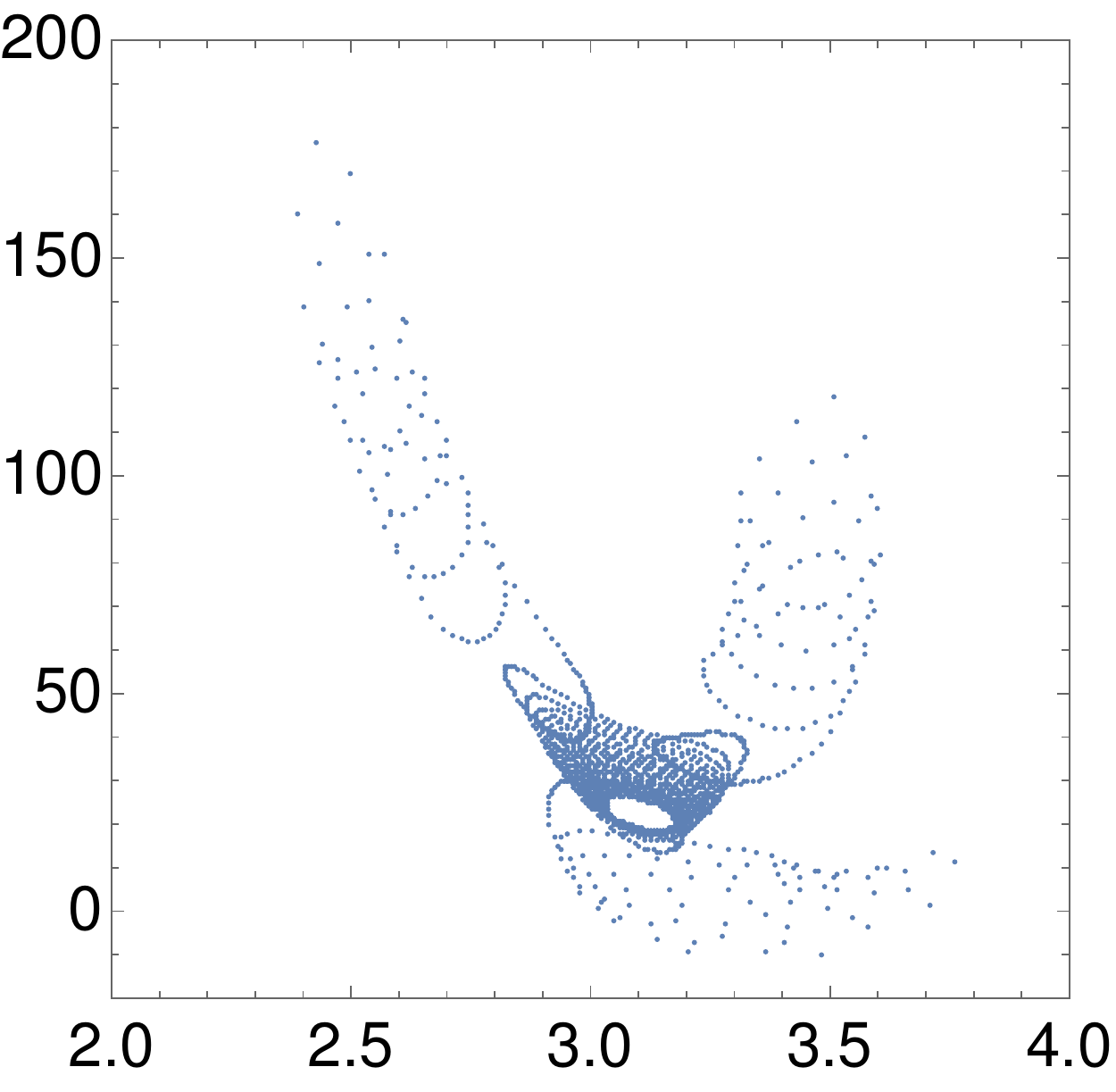}
		\caption{E=65}
	\end{subfigure}
	\caption{The Poincar\'e sections for $\theta_{nc} = 0.20$ in the ($r,p_r$) plane with $\theta = 0$ and $p_\theta > 0$ at different energies for quantum-corrected Schwarzschild black hole. The horizontal and vertical axes in each of the graphs correspond to $r$ and $p_r$, respectively.}
	\label{fig:5}
\end{figure}
\par\noindent
In order to see the effect of $\theta_{nc}$ on the particle dynamics, we have gradually increased the value of $\theta_{nc}$ in the above figures, namely Fig. \ref{fig:3}, Fig. \ref{fig:4}, and Fig. \ref{fig:5}. Interestingly, we notice that with the increase of the value of $\theta_{nc}=0.16,~0.18,~0.20$, the chaotic fluctuations start to appear in lower energy values. This means that the increment in the value of $\theta_{nc}$ induces more chaos into the system, which is evident from the plots. The torus gets more deformed for higher values of $\theta_{nc}$ in  lower energy values. \\
%
%
%
%
%
At this point a couple of comments are in order.
First, it should be noted an important aspect regarding the changing values of the spring constants, i.e., $K_{r}$ and $K_{\theta}$, of the harmonic potential. The strength of the harmonic potential is determined by its spring constants. If we decrease the value of the spring constants, then the contribution of the horizon in the total energy value starts to dominate, and as a result of that, the system starts showing the chaotic dynamics in lower energy values. On the contrary, if we increase the values of the spring constants, then the  system starts showing chaotic fluctuations in higher energy values. For a detailed discussion, we refer to Appendix \ref{App3} and Appendix \ref{App4}.\\
%
%
%
%
%
	Second, we would like to mention the connection between redshift and chaos in a near-horizon region. In \cite{Hashimoto:2016dfz}, it has been argued that the motion of a particle near the horizon becomes extremely slow, and that is the physical reason for the emergence of chaos in the near-horizon region. Specifically, in the vicinity of the horizon, the particle motion gets infinitely redshifted, and this exponentially slow-motion enlarges the difference in phase space motion in a later time. The authors of \cite{Hashimoto:2016dfz} have numerically shown that the KAM torus starts to break as the particle trajectories approach the horizon with the increase of particle's energy and pointed towards a possibility that the redshift is causing the chaos.
	In our case, we have also observed a similar phenomenon with the increase in the system energy $E$. As we increase the value of the system energy for a constant value of the NC parameter $\theta_{nc}$, the massless particle trapped in the harmonic potential approaches towards the horizon and the regular torus of the Poncar\'e sections starts getting distorted and separated into three distinct islands (see Fig. \ref{fig:2}). This means that with the increase in the system energy $E$ and keeping all the other parameters fixed, namely a constant value of $\theta_{nc}$, $K_r$, and $K_{\theta}$,  the effect of the horizon in the particle dynamics is getting increased. Now, as the particle approaches towards the horizon with the increment of $E$, its motion gets redshifted with respect to  an asymptotic observer because, with a similar argument of  \cite{Hashimoto:2016dfz}, in our case also the metric component $g_{rr}$ diverges near the horizon. So, the particle trapped in the harmonic potential near the black hole horizon exhibits chaos as well as it gets redshifted with respect to  a distant observer. This suggests that these two phenomena are closely related to each other, and there might be a very intricate relationship between chaos and redshift in the context of the horizon, which needs further investigation. Furthermore, it would be interesting from observational point of view if the redshift can have a handle on the NC parameter $\theta_{nc}$. 
\subsection{Approximate form of NC-deformed metric}
%
%
%
%
%
\par\noindent
Besides Poincar\'e section, another well known parameter utilised for the study of chaos is the Lyapunov exponent which is very useful in quantifying chaos. However, as we have experienced, calculation of the Lyapunov exponent using  the exact form of the Nicolini metric (\ref{Nicolini metric}) is  much harder in a numerical framework (due to the incomplete Gamma function).  On the other hand, an approximate analytical form of the Nicolini metric has also been provided in \cite{Nicolini:2008aj} and can be exploited in the computation of the Lyapunov exponent. However, caution is necessary since we are dealing with highly non-linear expressions in a numerical framework and seemingly mild changes might generate spurious effects. We must ensure that the approximate form of the incomplete $\gamma$ function  at large distances $\frac{r^{2}}{4\theta_{nc}}>>1$ \cite{Nicolini:2008aj} 
\begin{eqnarray}\label{app}
\gamma\left(\frac{3}{2},\frac{r^{2}}{4\theta_{nc}}\right)\thickapprox \frac{\sqrt{\pi}}{2}+\frac{1}{2}\frac{r}{\sqrt{\theta_{nc}}}e^{-\frac{r^{2}}{4\theta_{nc}}}
\end{eqnarray}
yields the correct behaviour (to the level of accuracy we are interested in) within the parameter window we are concerned with. This will be performed in two steps: (i) comparing  positions of the (outer) horizon using exact and approximate forms of metric and (ii) comparing Poincar\'e sections using the above metrics.

\begin{figure}[htb!]
	\centering
	\begin{subfigure}[b]{0.49\linewidth}
		\includegraphics[width=\linewidth]{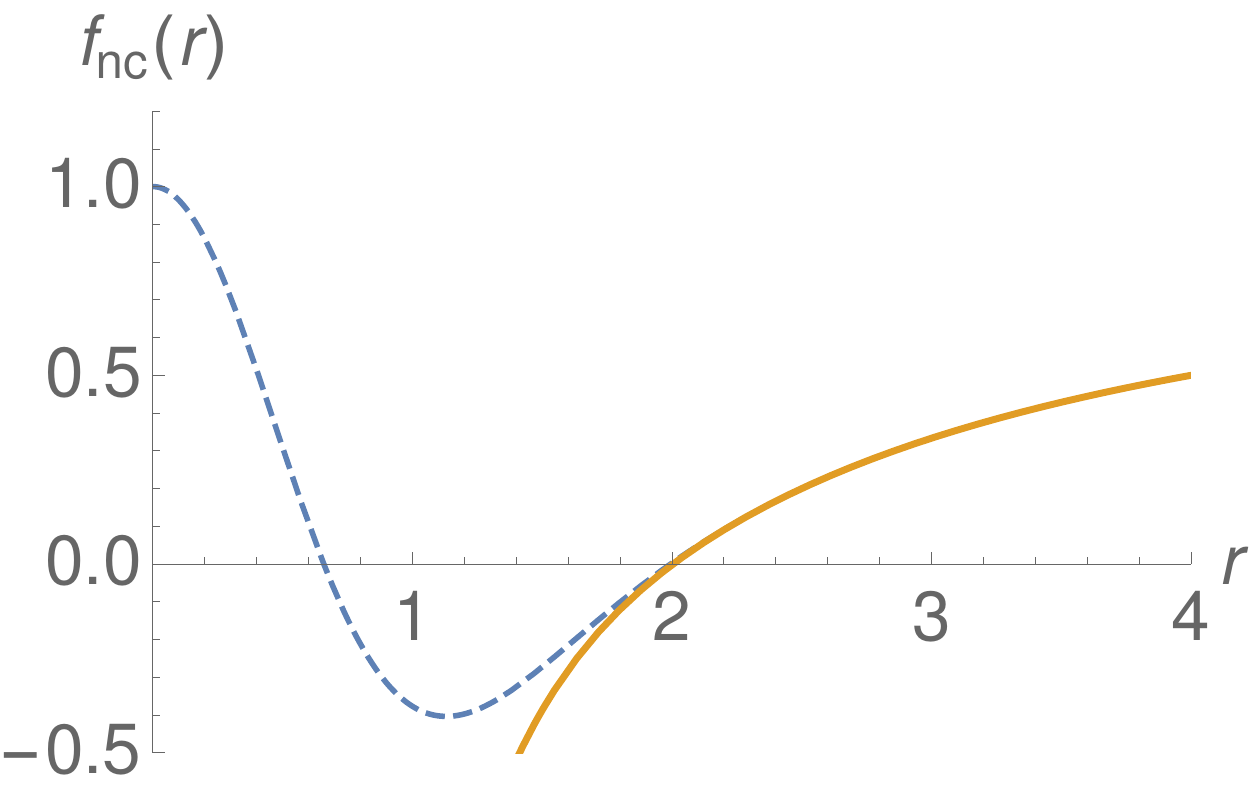}
		\caption{$\theta_{nc} = 0.14$}
	\end{subfigure}
	\begin{subfigure}[b]{0.49\linewidth}
		\includegraphics[width=\linewidth]{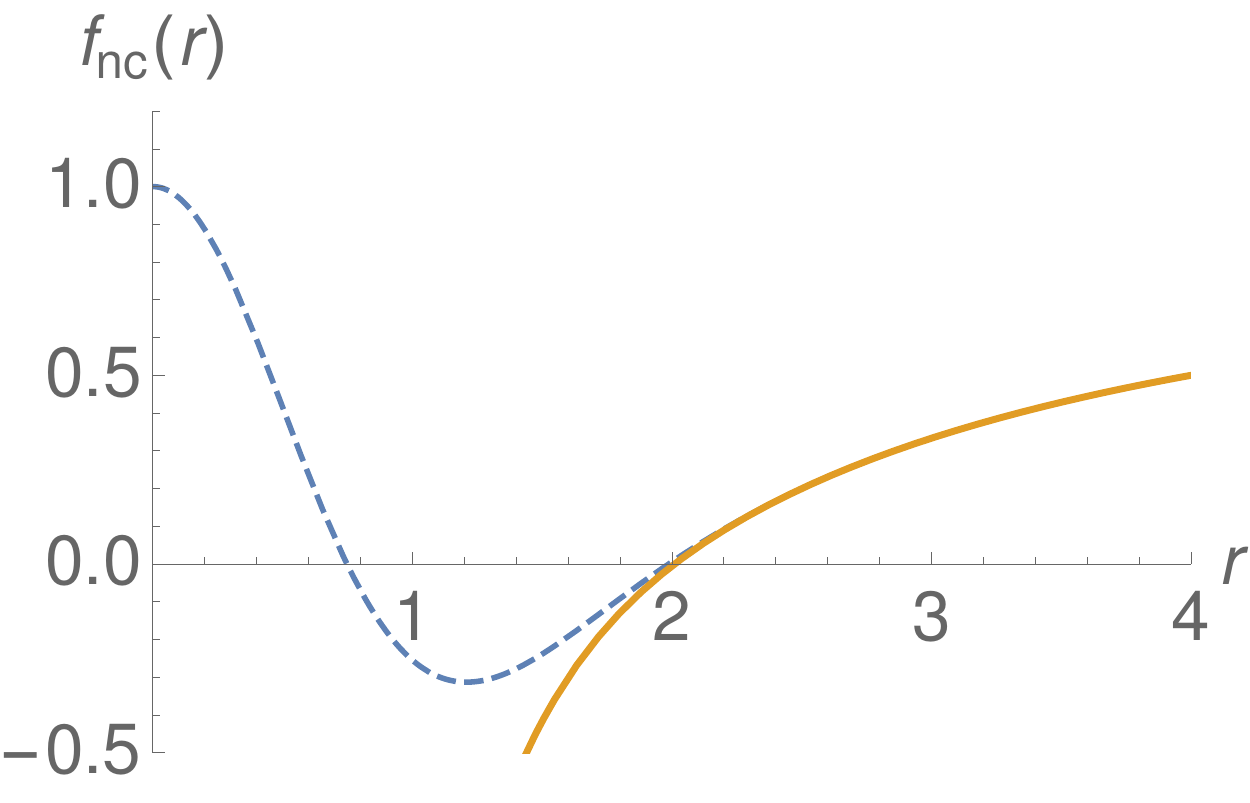}
		\caption{$\theta_{nc} = 0.16$}
	\end{subfigure}
	\begin{subfigure}[b]{0.49\linewidth}
		\includegraphics[width=\linewidth]{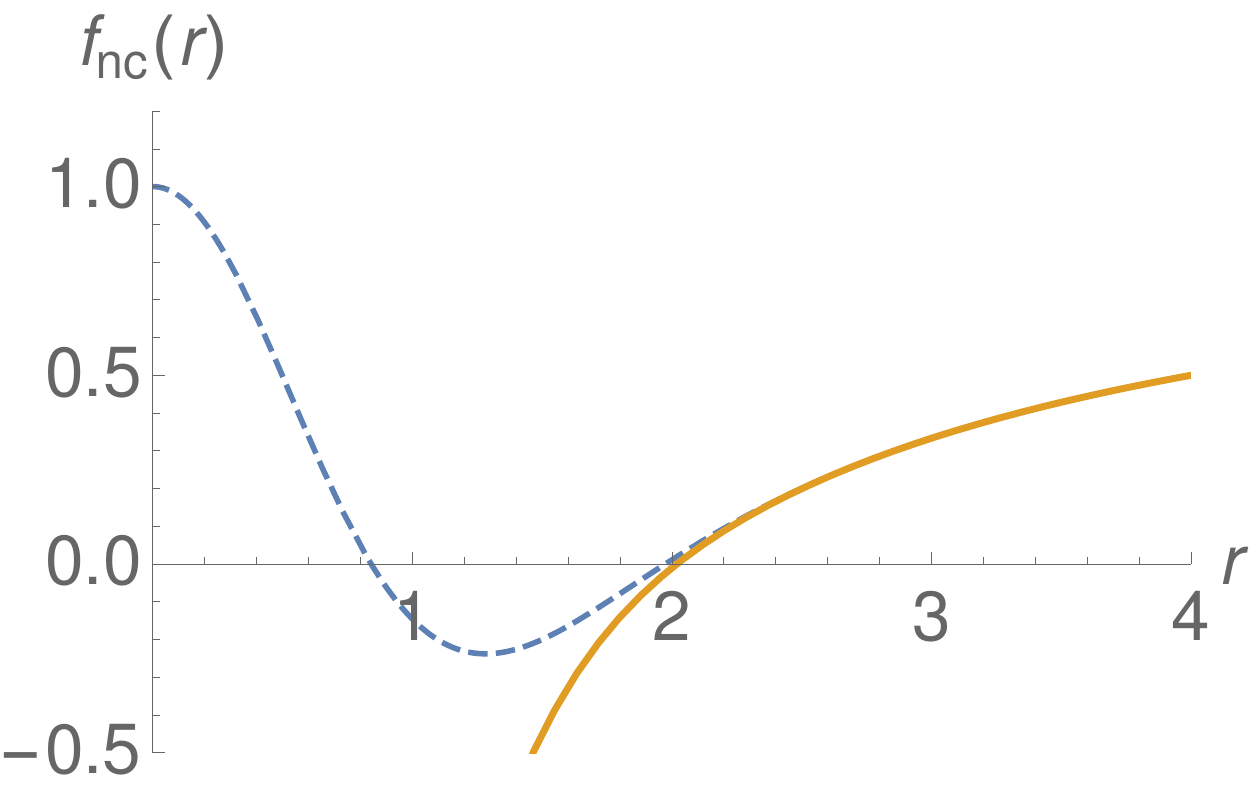}
		\caption{$\theta_{nc} = 0.18$}
	\end{subfigure}
	\begin{subfigure}[b]{0.49\linewidth}
	\includegraphics[width=\linewidth]{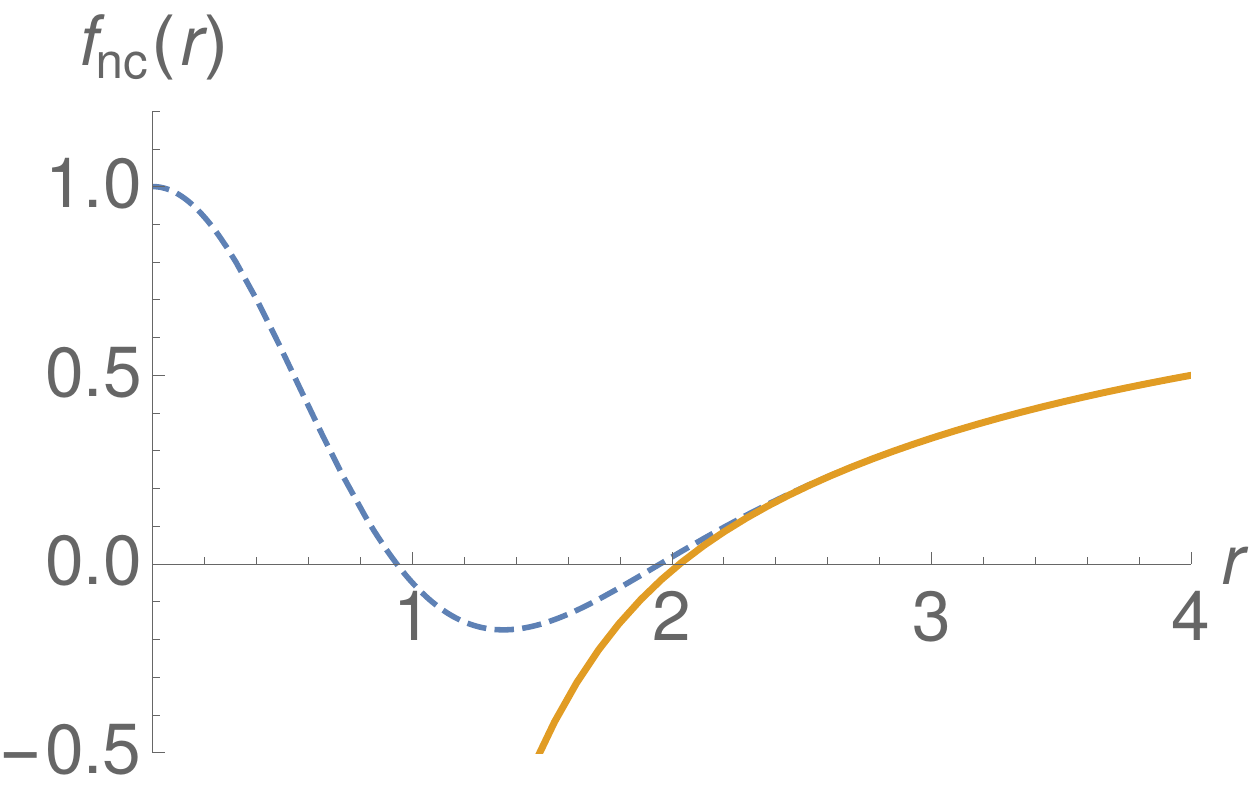}
	\caption{$\theta_{nc} = 0.20$}
\end{subfigure}
\caption{The figures show the variation of $f_{nc}(r)$ with $r$. The dashed and the thick curves correspond to the exact metric and approximated metric, respectively. For the approximated metric the horizon is at $r_H\approx2$.}
\label{fig:6}
\end{figure}

Incorporating (\ref{app}) into the exact metric (\ref{Nicolini metric}) we obtain the analytic form  \cite{Nicolini:2008aj}
\begin{widetext}
	\begin{eqnarray}
	ds^{2}=-\left(1-\frac{2M}{r}-\frac{2M}{\sqrt{\pi\theta_{nc}}}e^{-\frac{r^{2}}{4\theta_{nc}}}\right)dt^{2}+\left(1-\frac{2M}{r}-\frac{2M}{\sqrt{\pi\theta_{nc}}}e^{-\frac{r^{2}}{4\theta_{nc}}}\right)^{-1}dr^{2}+r^{2}d\Omega^{2}\label{approx nicolini metric}~.
	\end{eqnarray}
\end{widetext}
{\it{Position of horizon}}: In Fig. \ref{fig:6}, we show that if we consider the approximated form of the metric, i.e., equation (\ref{approx nicolini metric}), the position of the outer horizon does not change with the changing value of $\theta_{nc}$ for some constant value of $M$ ($M=1.0$ in this case). However, as we mentioned before, there is a limiting value of $M=M_0=1.90\sqrt{\theta_{nc_0}}$, therefore in this case also we cannot increase the value of $\theta_{nc}$ beyond the allowed range. The limiting value of $\theta_{nc_0}$ is $\sim 0.27$ as we mentioned earlier due to our consideration of the value of the mass $M=1.0$ in this case.  
%
%
%
%
%
\par
\begin{figure}[H]
	\centering
	\begin{subfigure}[b]{0.4\linewidth}
		\includegraphics[width=\linewidth]{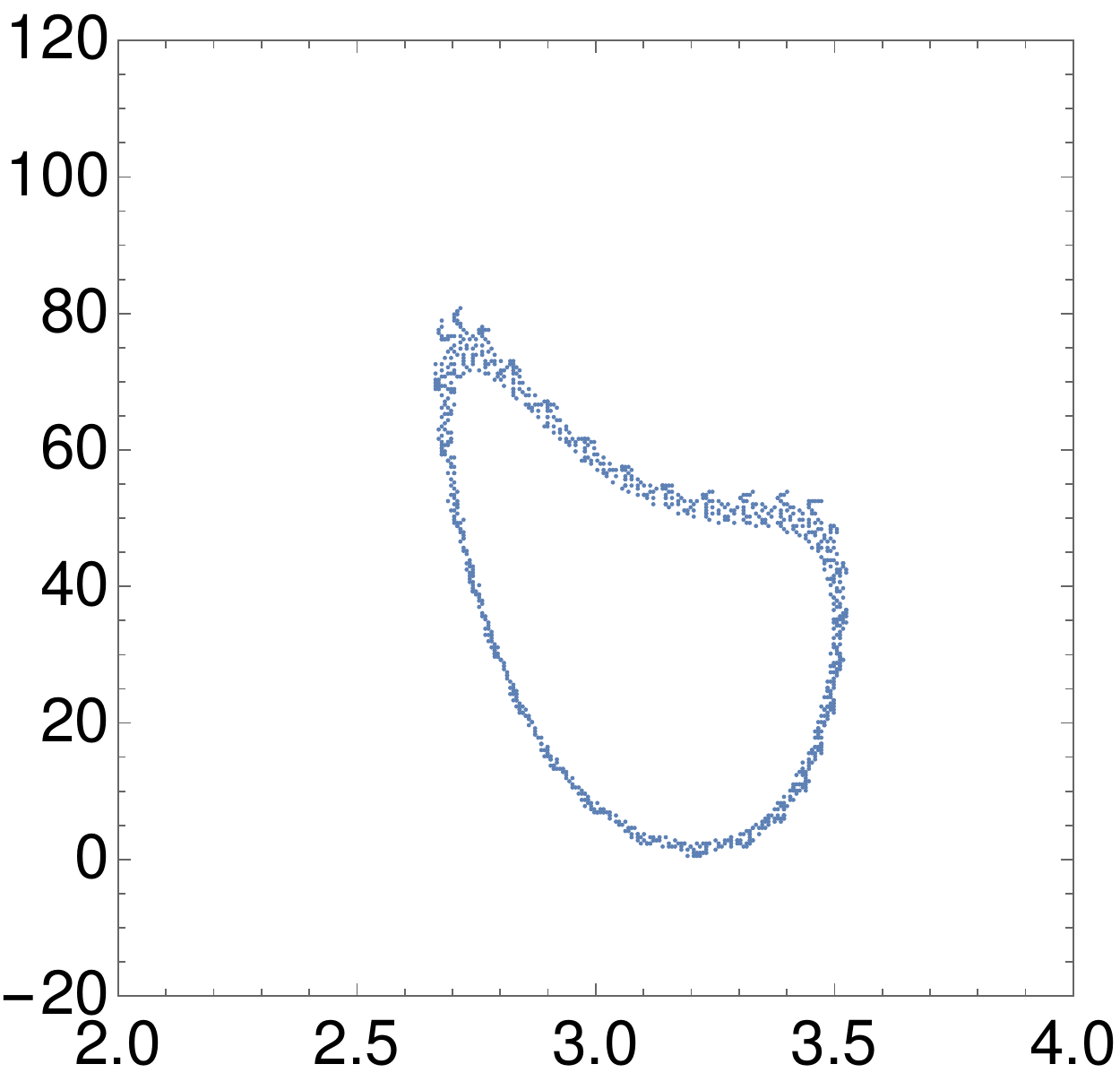}
		\caption{E=50}
	\end{subfigure}
	\begin{subfigure}[b]{0.4\linewidth}
		\includegraphics[width=\linewidth]{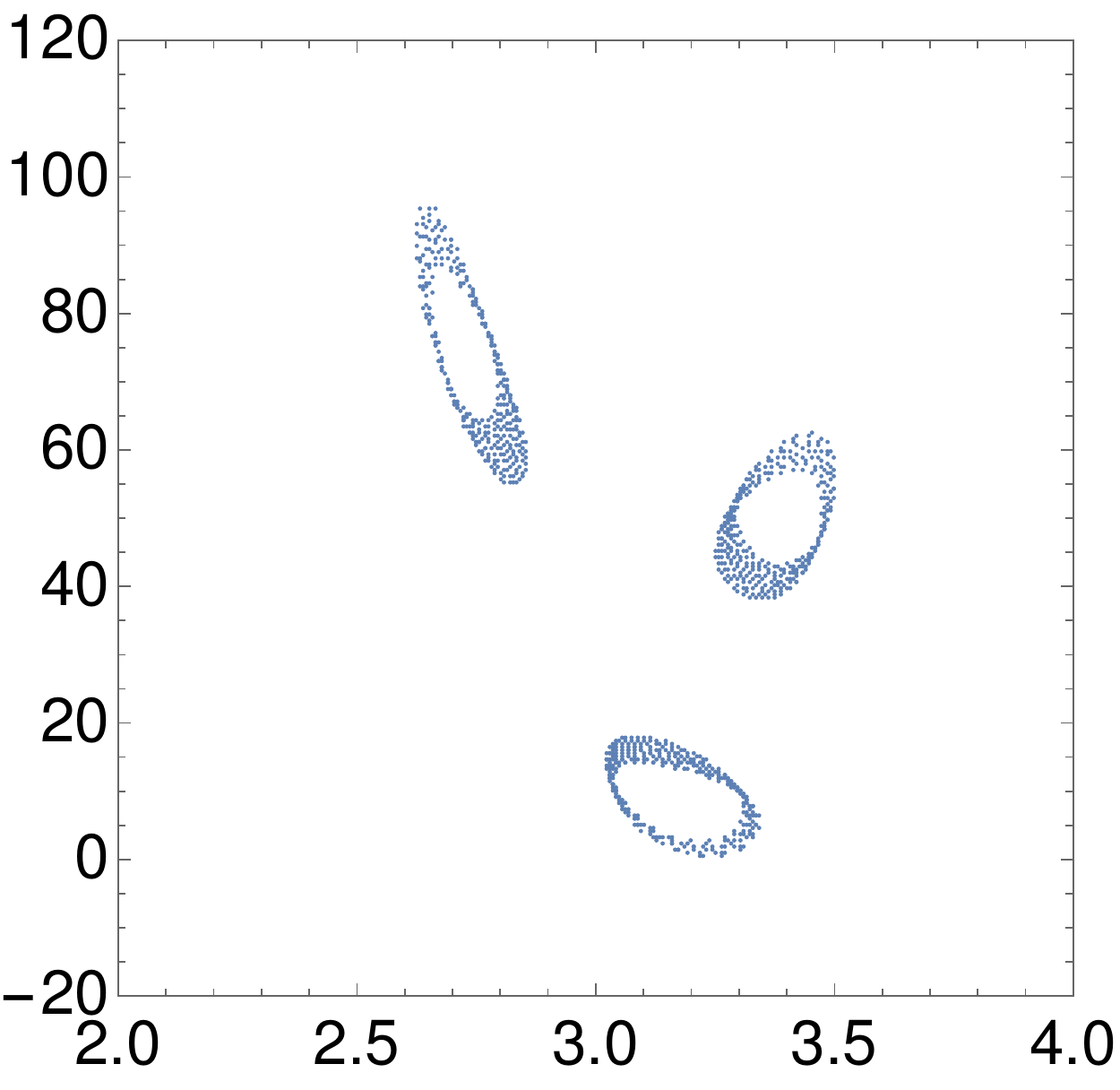}
		\caption{E=55}
	\end{subfigure}
	\begin{subfigure}[b]{0.4\linewidth}
		\includegraphics[width=\linewidth]{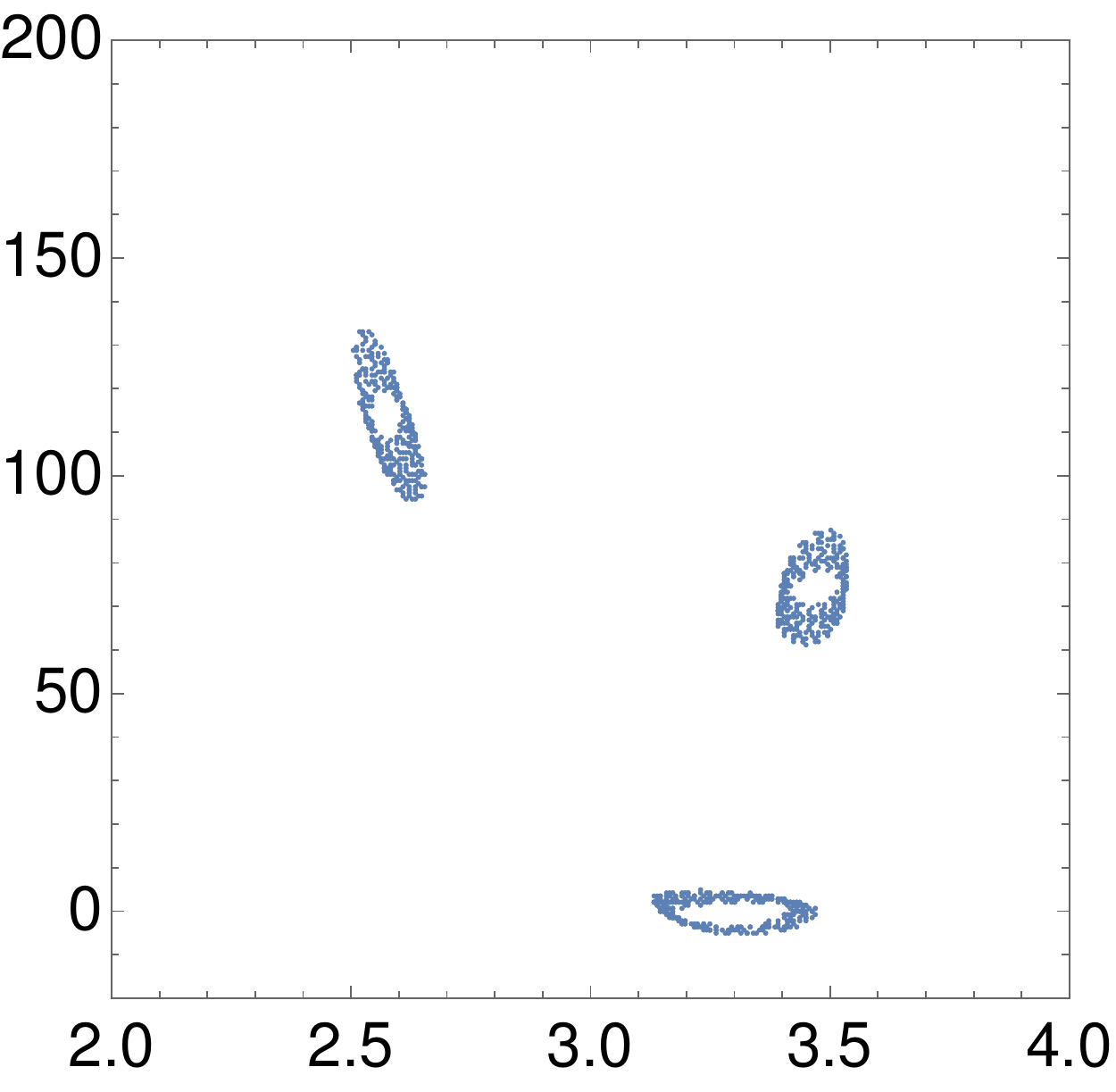}
		\caption{E=60}
	\end{subfigure}
	\begin{subfigure}[b]{0.4\linewidth}
		\includegraphics[width=\linewidth]{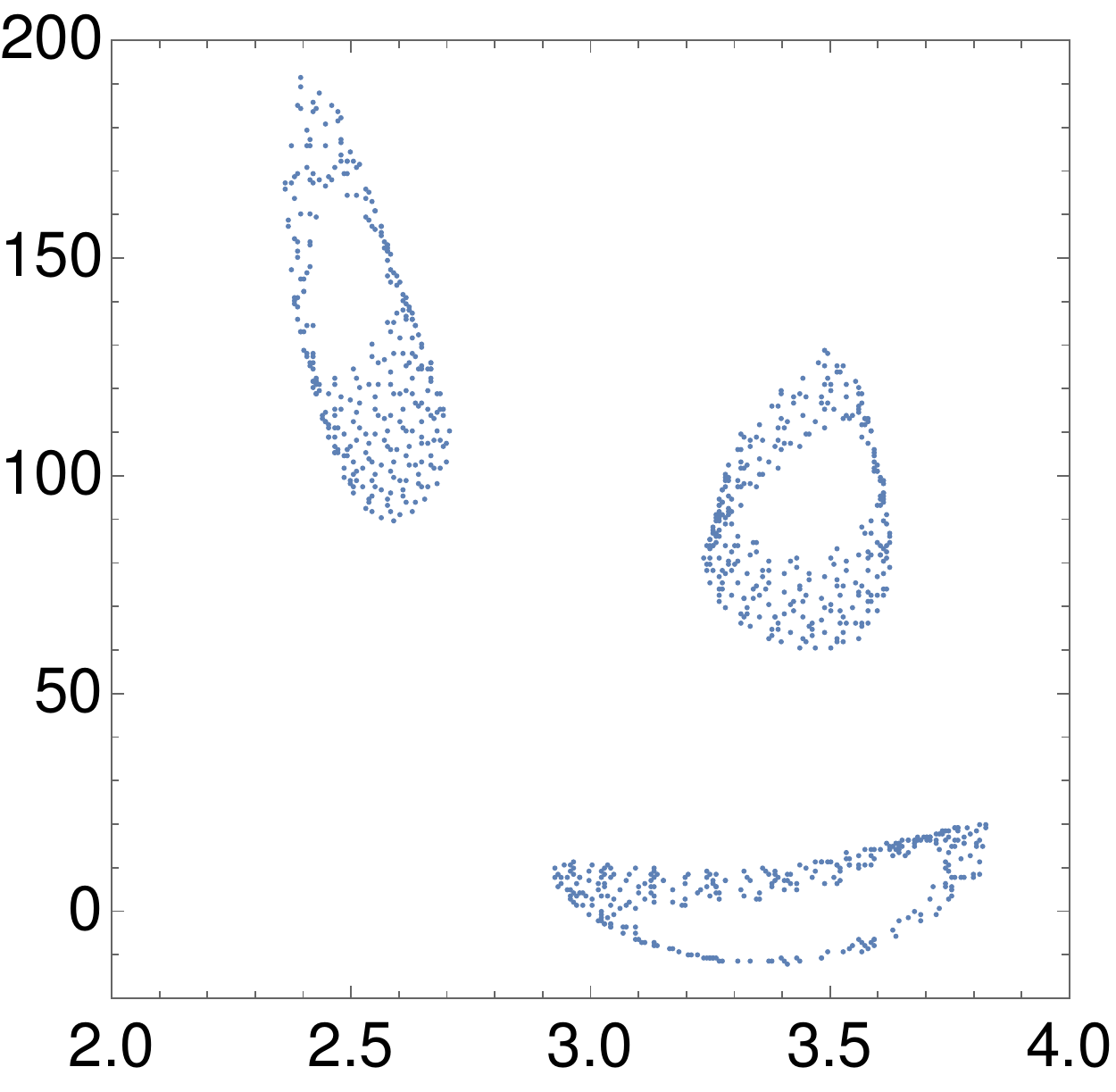}
		\caption{E=65}
	\end{subfigure}
	\caption{The Poincar\'e sections for $\theta_{nc} = 0.16$ in the ($r,p_r$) plane with $\theta = 0$ and $p_\theta > 0$ at different energies for the quantum-corrected Schwarzschild black hole. The horizontal and vertical axes in each of the graphs correspond to $r$ and $p_r$, respectively.}
	\label{fig:7}
\end{figure}

\begin{figure}[H]
	\centering
	\begin{subfigure}[b]{0.4\linewidth}
		\includegraphics[width=\linewidth]{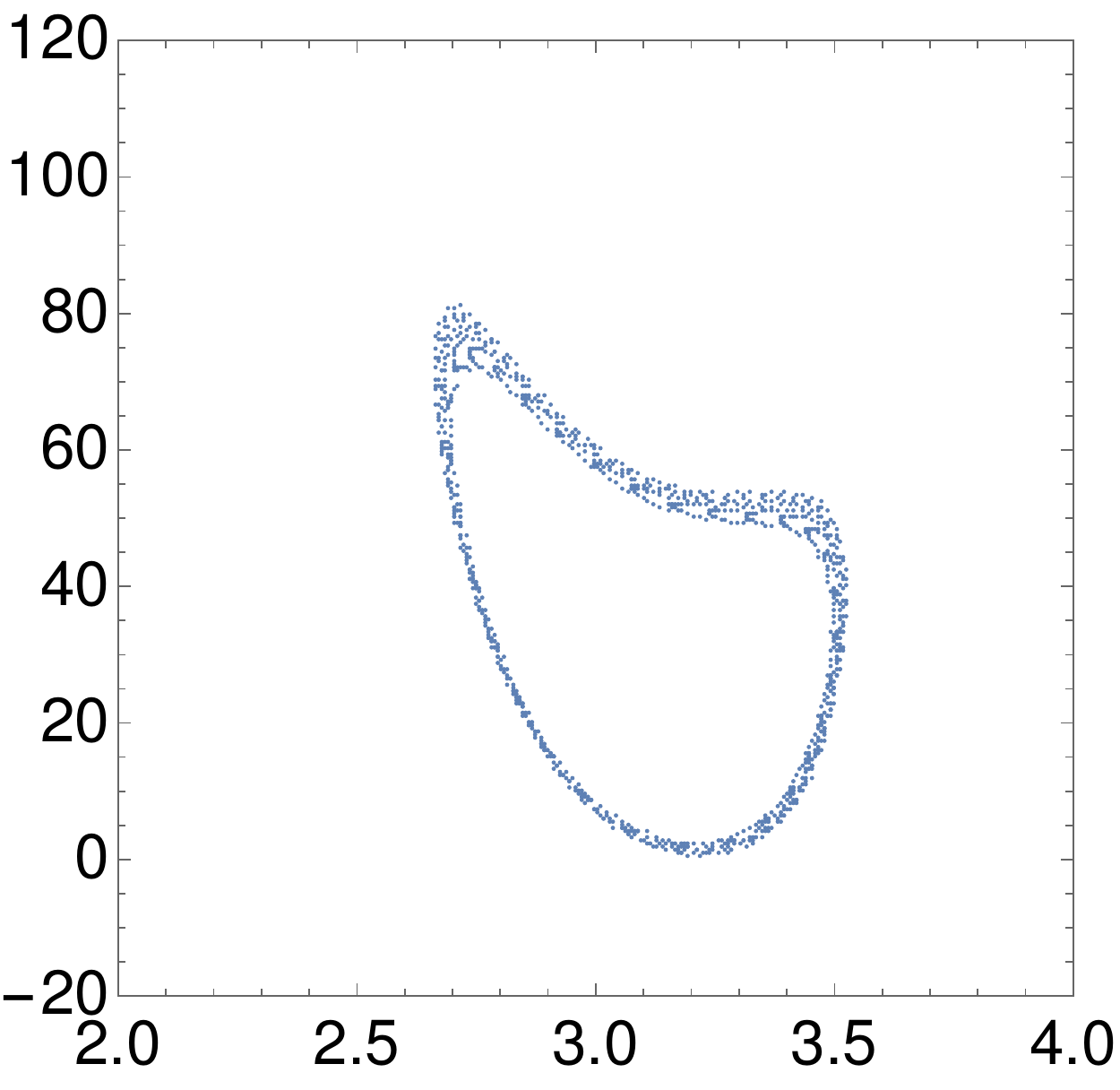}
		\caption{E=50}
	\end{subfigure}
	\begin{subfigure}[b]{0.4\linewidth}
		\includegraphics[width=\linewidth]{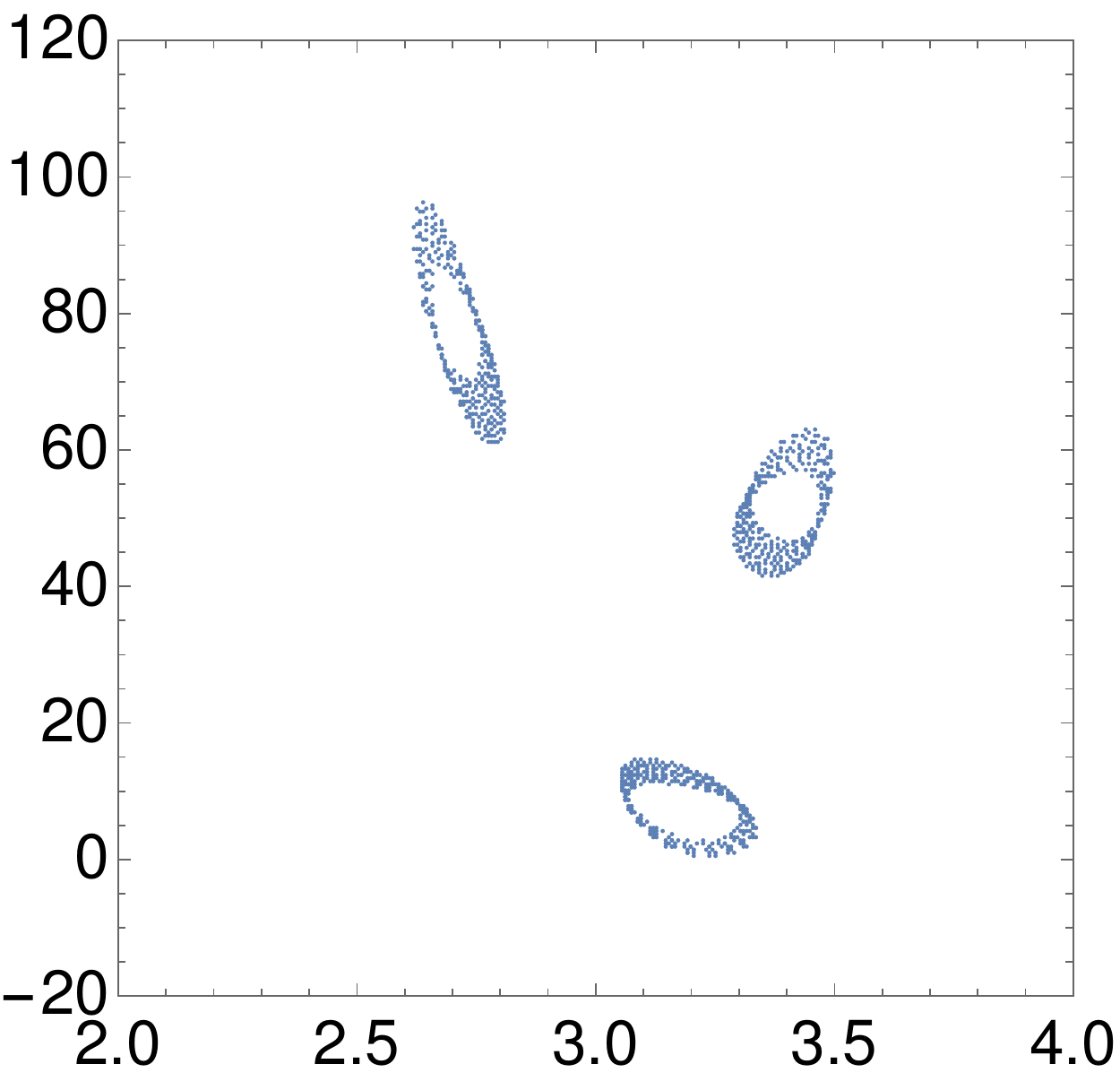}
		\caption{E=55}
	\end{subfigure}
	\begin{subfigure}[b]{0.4\linewidth}
		\includegraphics[width=\linewidth]{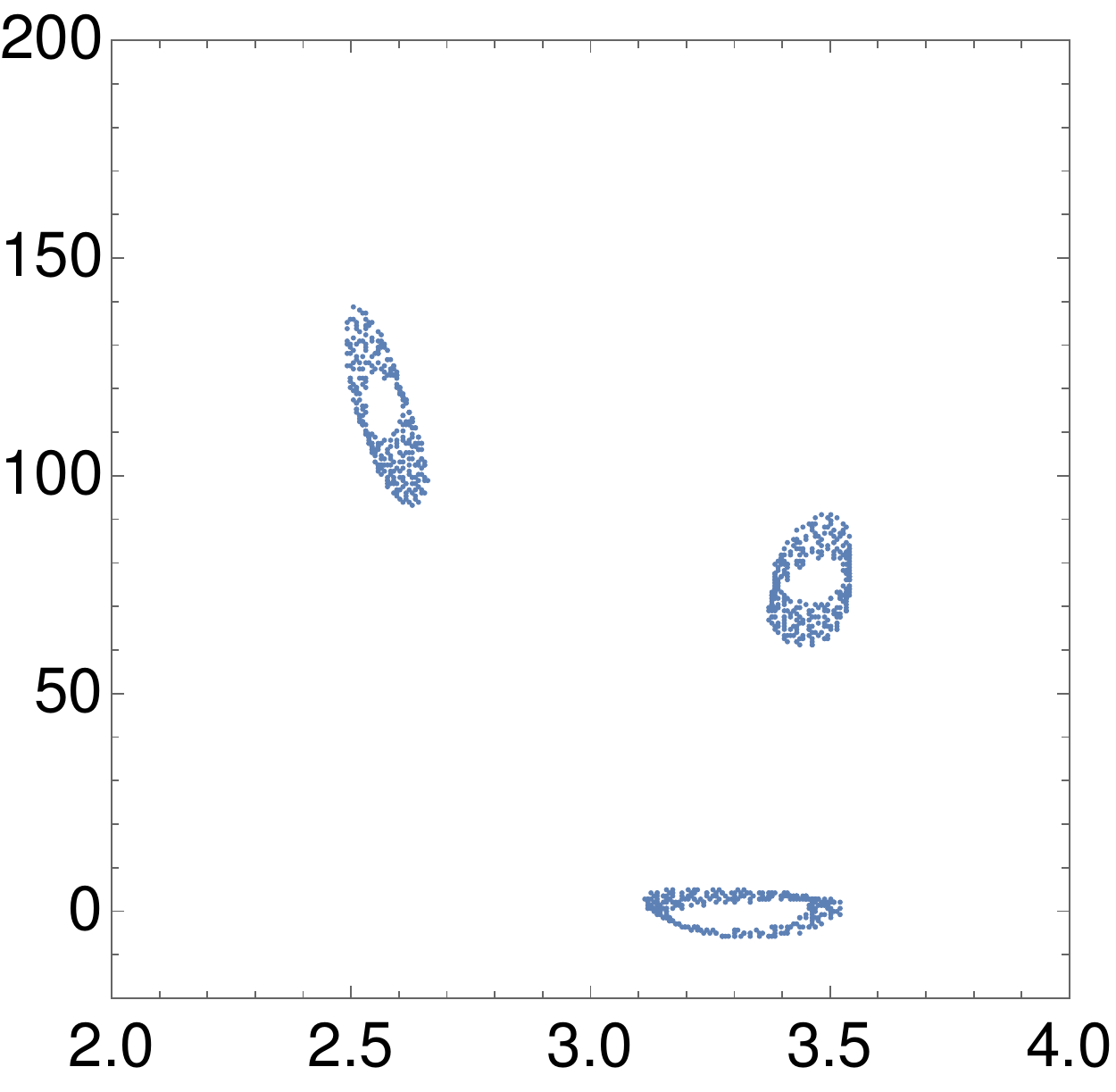}
		\caption{E=60}
	\end{subfigure}
	\begin{subfigure}[b]{0.4\linewidth}
		\includegraphics[width=\linewidth]{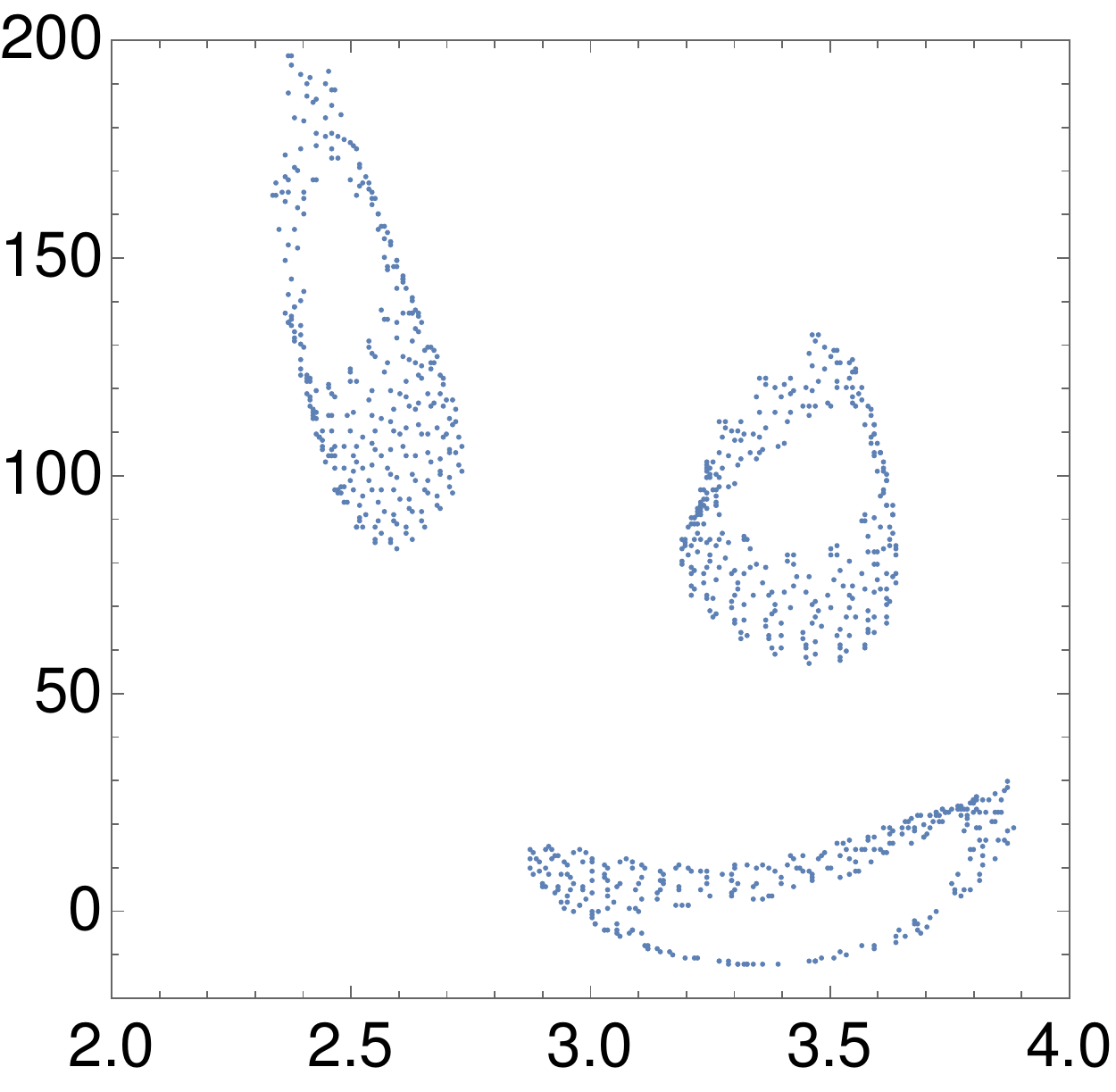}
		\caption{E=65}
	\end{subfigure}
	\caption{The Poincar\'e sections for $\theta_{nc} = 0.20$ in the ($r,p_r$) plane with $\theta = 0$ and $p_\theta > 0$ at different energies for the quantum-corrected Schwarzschild black hole. The horizontal and vertical axes in each of the graphs correspond to $r$ and $p_r$, respectively.}
	\label{fig:8}
\end{figure}

\noindent
{\it{Poincar\'e sections for approximate NC-deformed metric}}:    In order to investigate the near-horizon particle dynamics, we plot  the Poincar\'e sections following the same procedure as before utilising the approximated metric (\ref{approx nicolini metric}). We consider  different values of energy of our system consisting of the massless test particle. Fig. \ref{fig:7} and Fig. \ref{fig:8} are plotted for some constant values of $\theta_{nc}=0.16$ and $0.20$ and, in both cases, we obtain that with the increase in the value of the system energy, our system gradually turns out to be chaotic. Not only that, with the increase in the value of $\theta_{nc}$, chaos comes into the picture in lower energy values, just like the case for the exact metric. Therefore, it turns out that both the exact metric (\ref{Nicolini metric}) and the approximated one (\ref{approx nicolini metric}) generate similar  physical behaviour in our region of interest. Therefore, for these specific values of $\theta_{nc}$ and in this mass limit $(M_0=1.0)$, this approximation still holds.
%
%
\begin{figure}[hbt!]
	\centering
	\begin{subfigure}[b]{0.6\linewidth}
		\includegraphics[width=\linewidth]{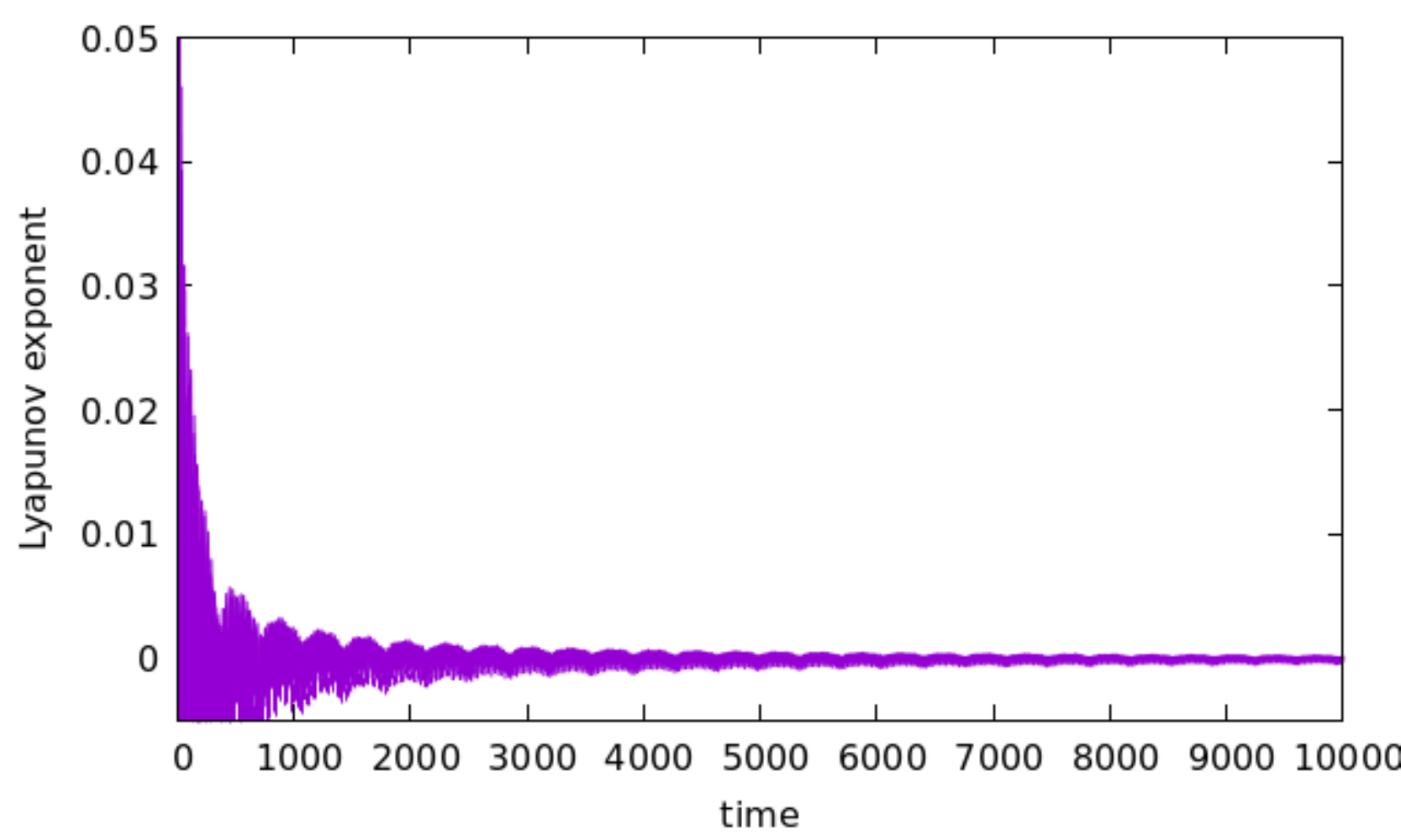} \caption{$E=50$}
	\end{subfigure}
	\begin{subfigure}[b]{0.6\linewidth}
		\includegraphics[width=\linewidth]{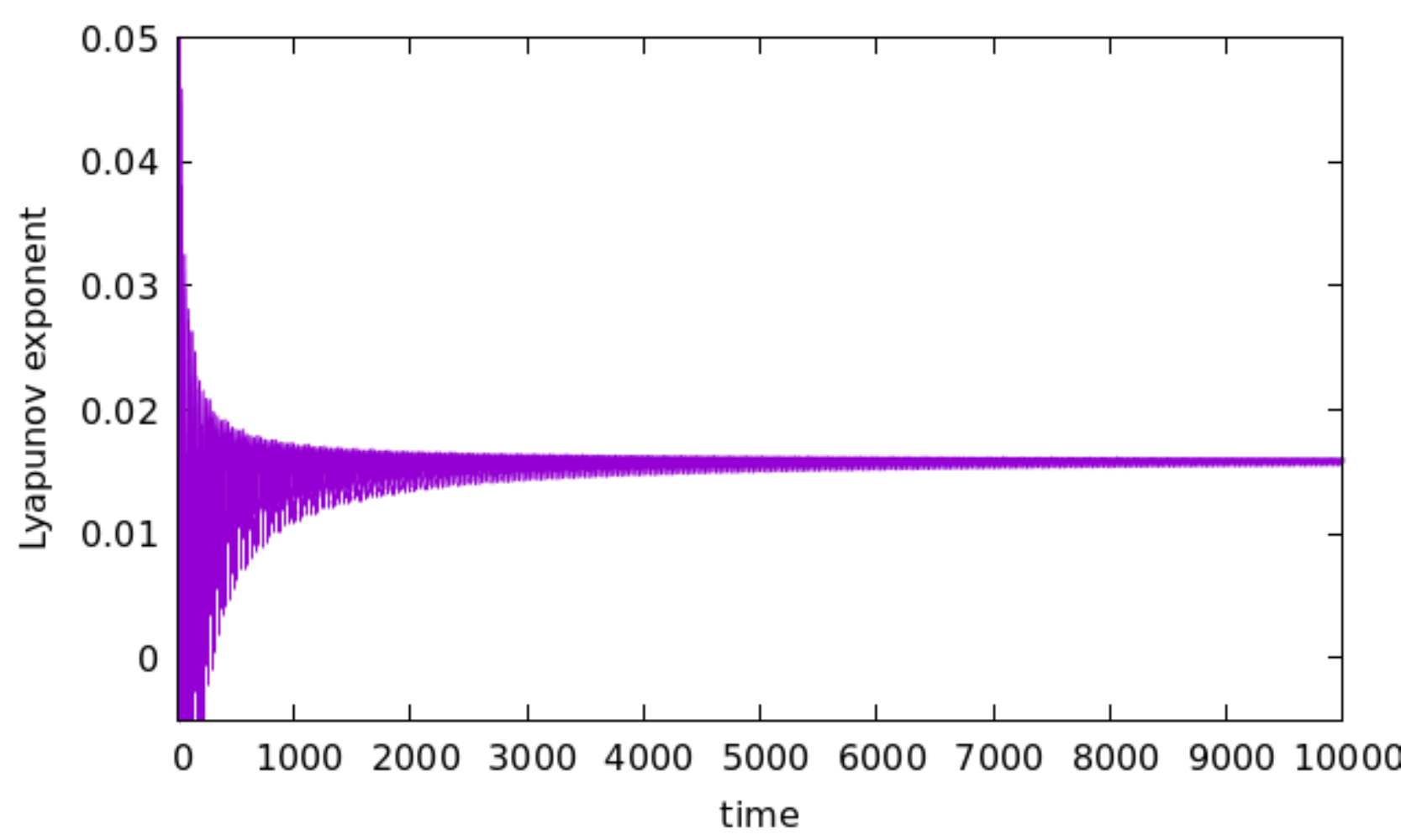} \caption{$E=55$}
	\end{subfigure}
	\begin{subfigure}[b]{0.6\linewidth}
		\includegraphics[width=\linewidth]{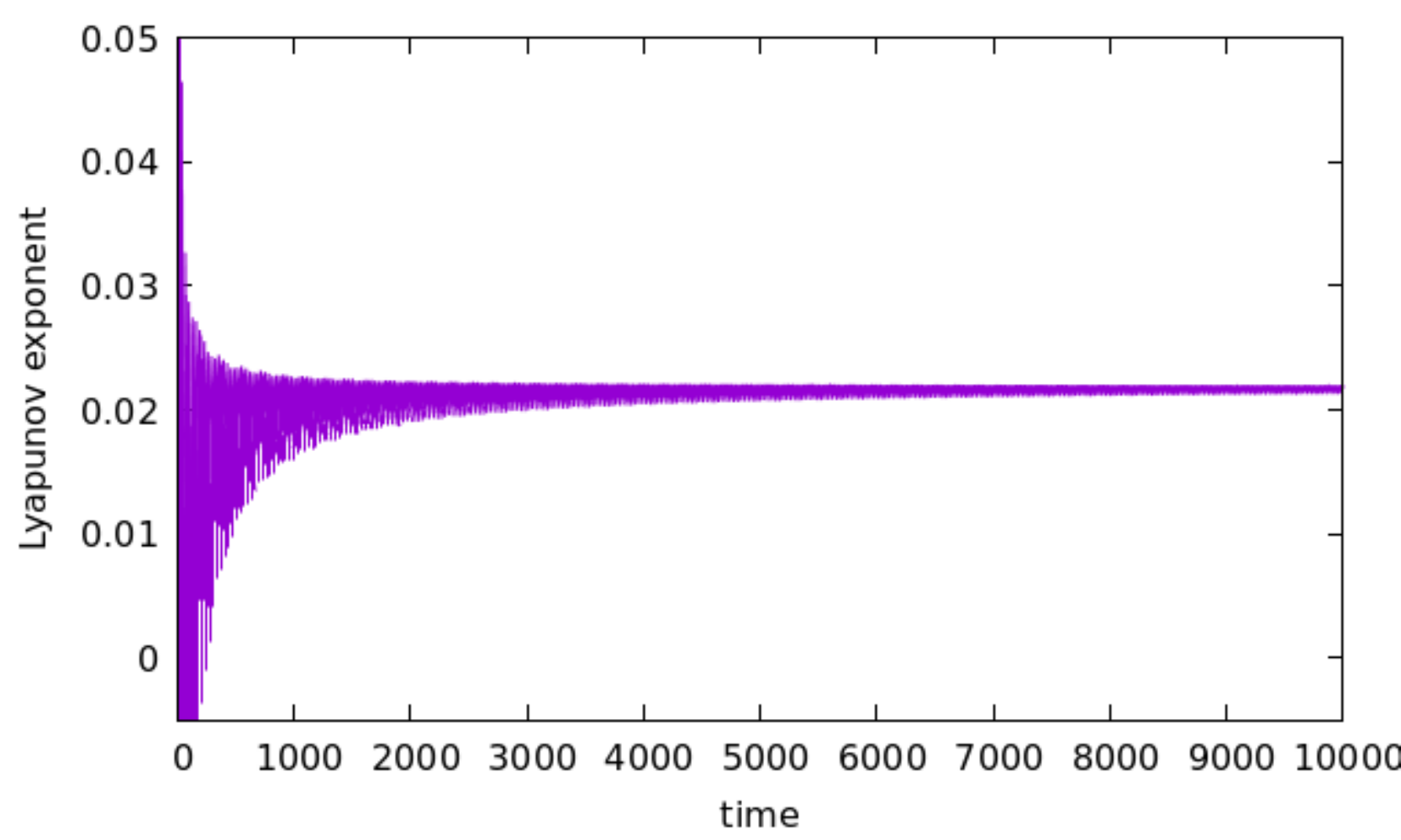} \caption{$E=60$}
	\end{subfigure}
	\begin{subfigure}[b]{0.6\linewidth}
		\includegraphics[width=\linewidth]{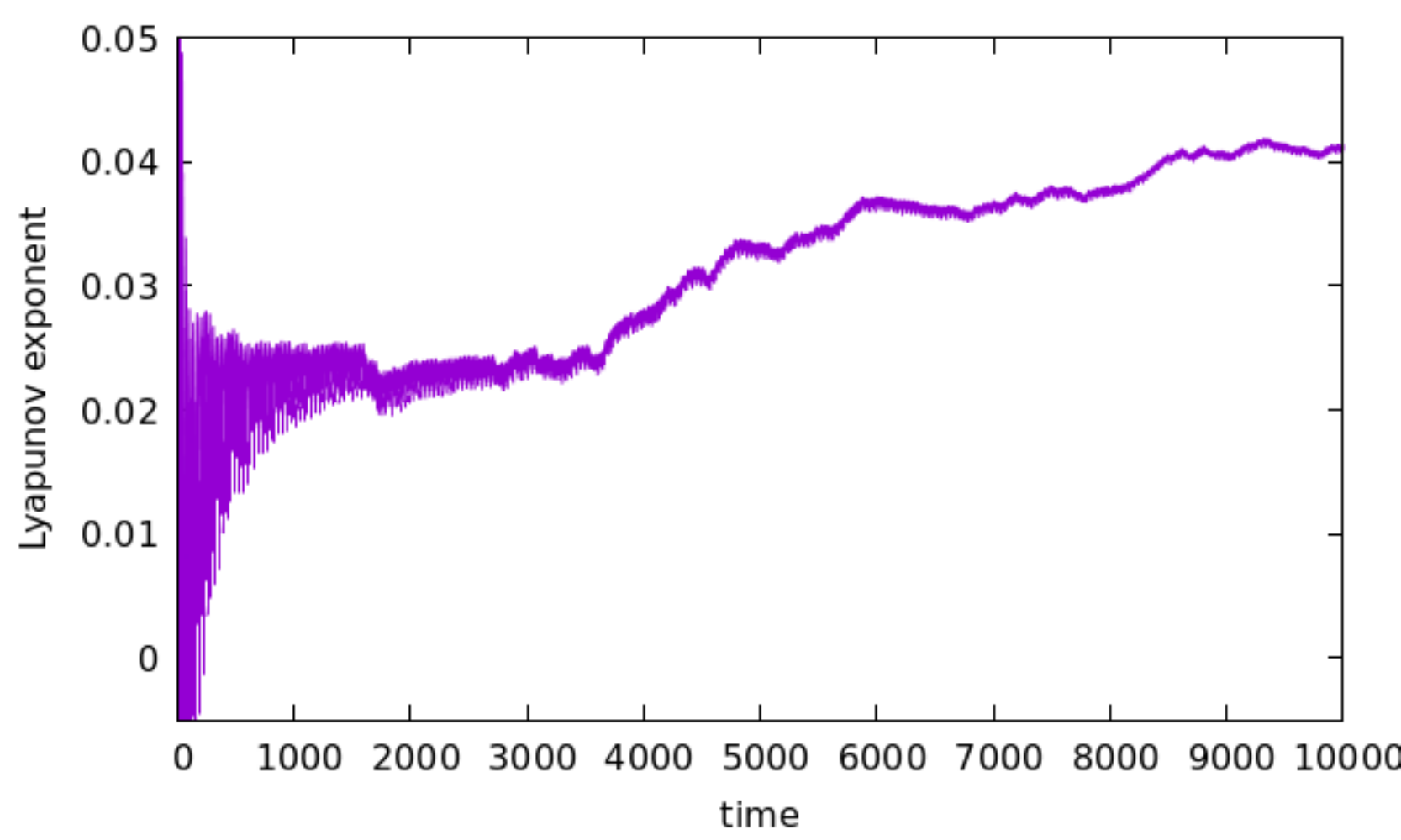} \caption{$E=65$}
	\end{subfigure}
	\caption{Largest Lyapunov exponents for different values of energy of the system $E$ but for a particular value of $\theta_{nc}=0.20$.}
	\label{fig:9}
\end{figure} 	 
\begin{figure}[hbt!]
	\centering
	\begin{subfigure}[b]{0.6\linewidth}
		\includegraphics[width=\linewidth]{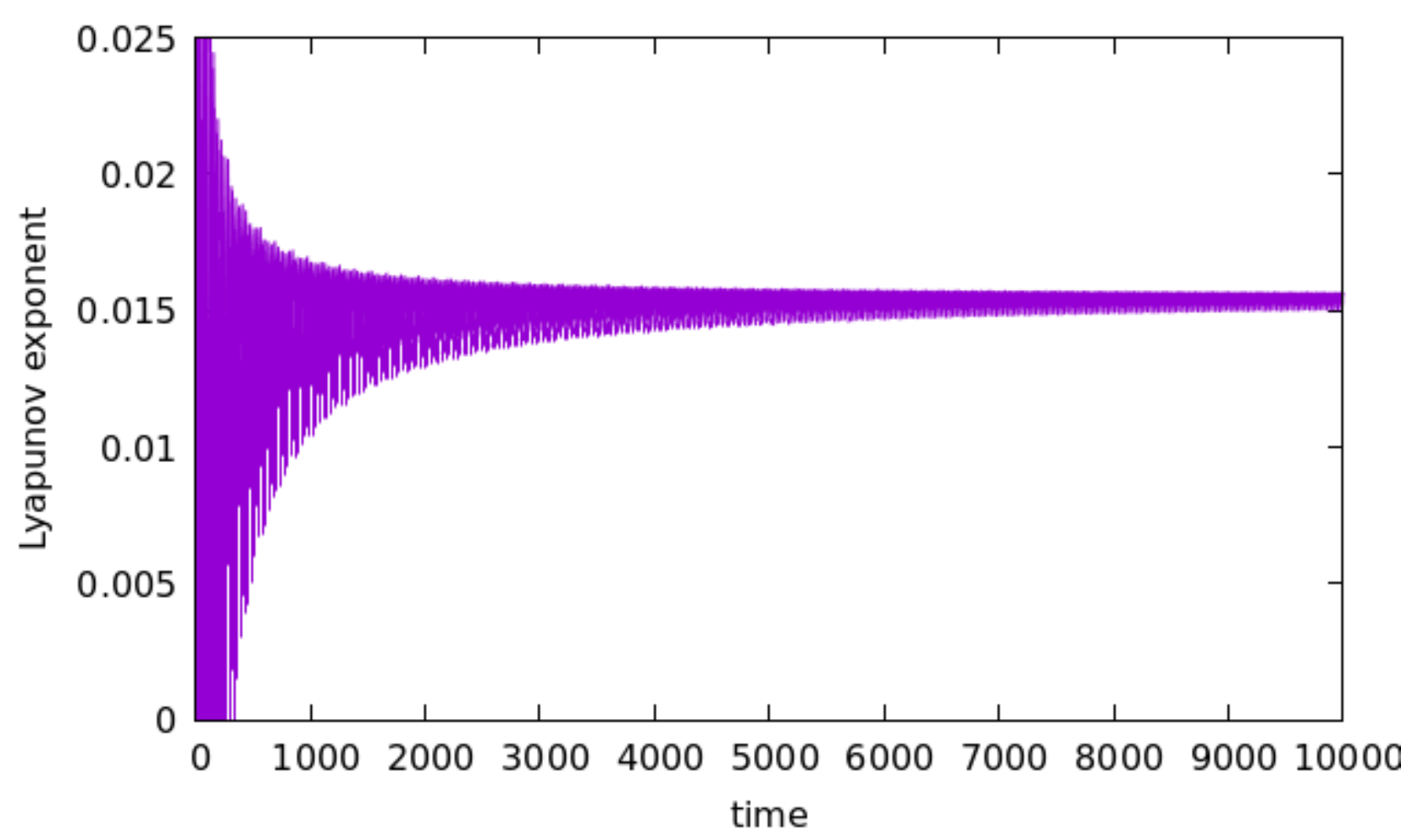} \caption{$\theta_{nc}=0.19$}
	\end{subfigure}
	\begin{subfigure}[b]{0.6\linewidth}
		\includegraphics[width=\linewidth]{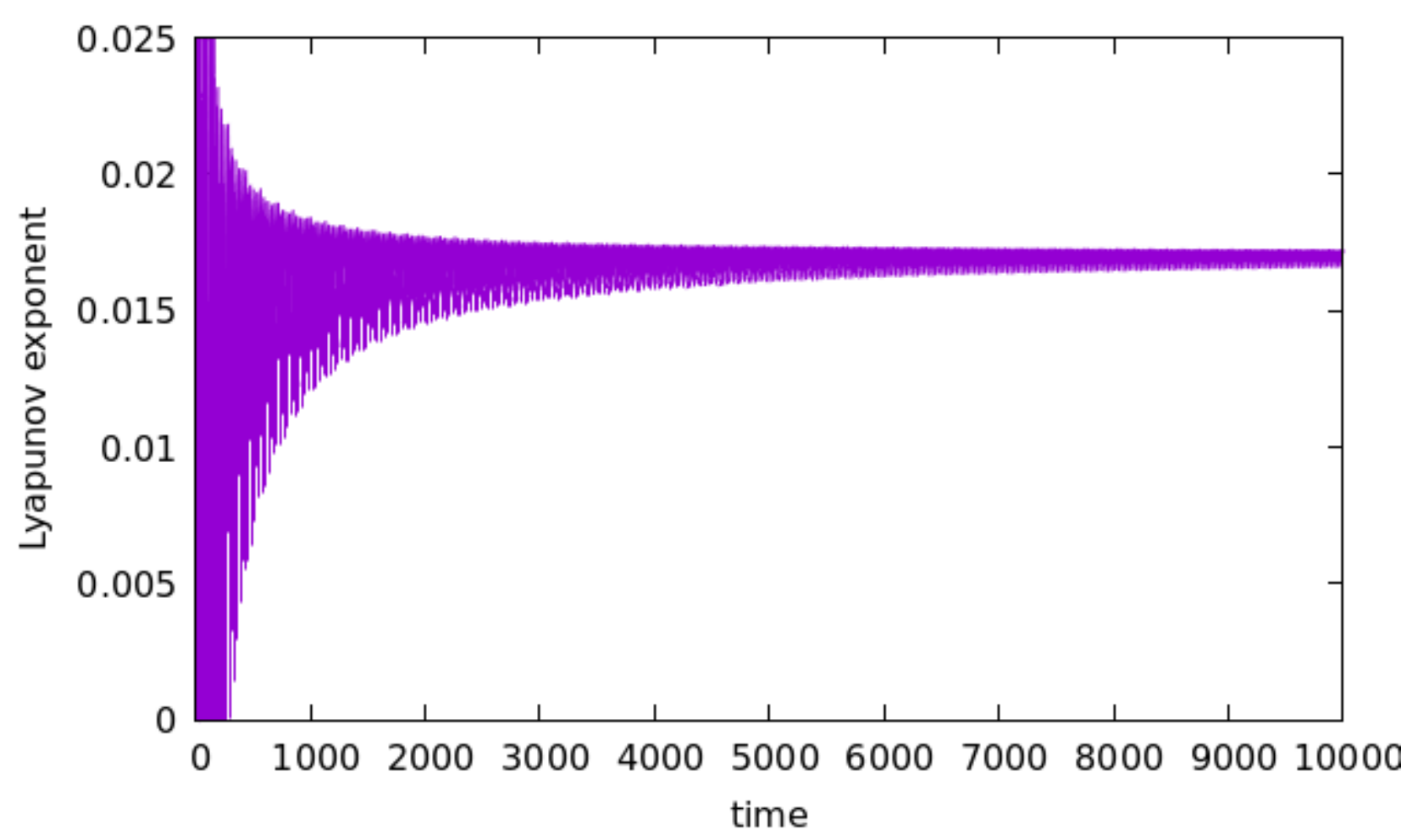} \caption{$\theta_{nc}=0.22$}
	\end{subfigure}
	\begin{subfigure}[b]{0.6\linewidth}
		\includegraphics[width=\linewidth]{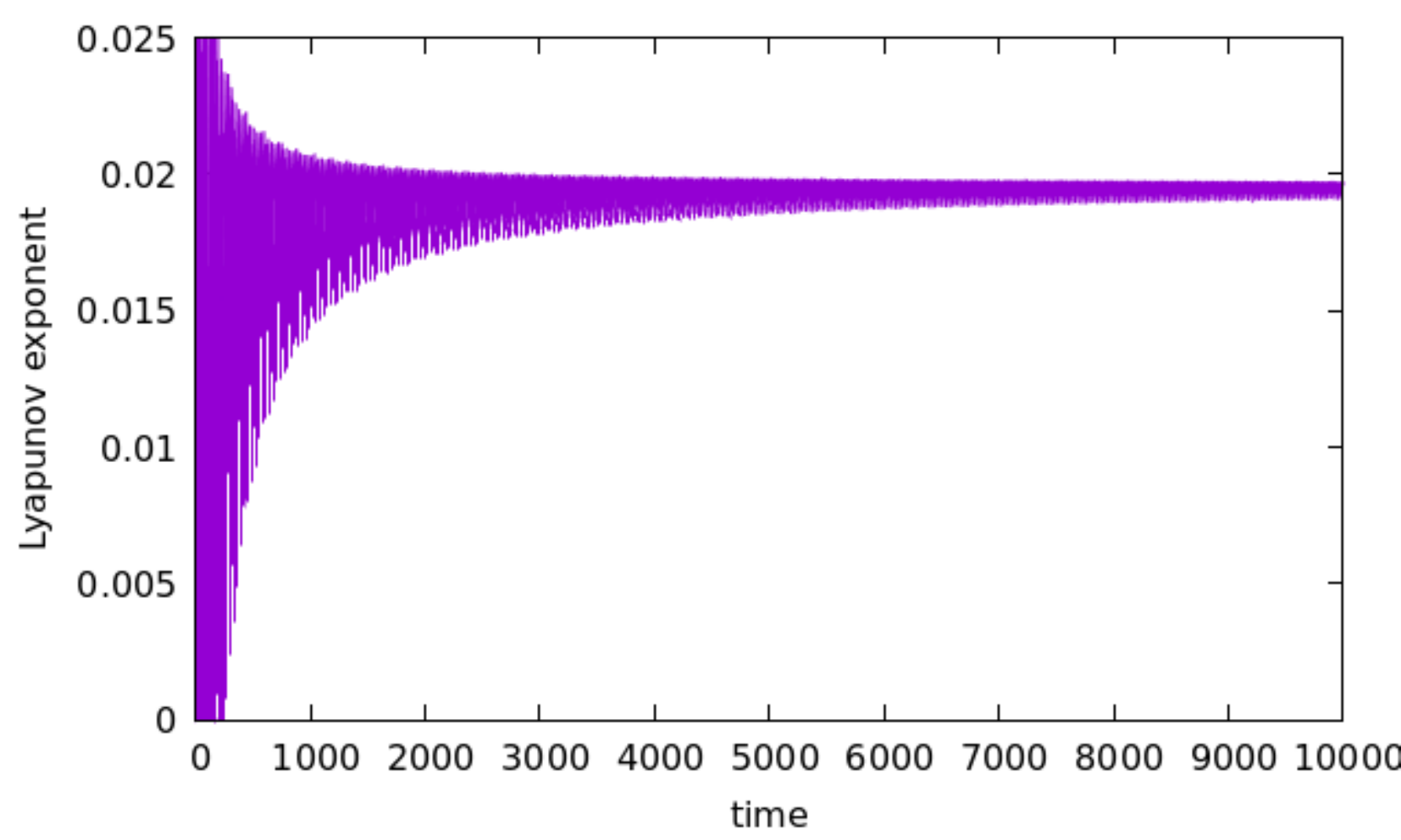} \caption{$\theta_{nc}=0.25$}
	\end{subfigure}
	\begin{subfigure}[b]{0.6\linewidth}
		\includegraphics[width=\linewidth]{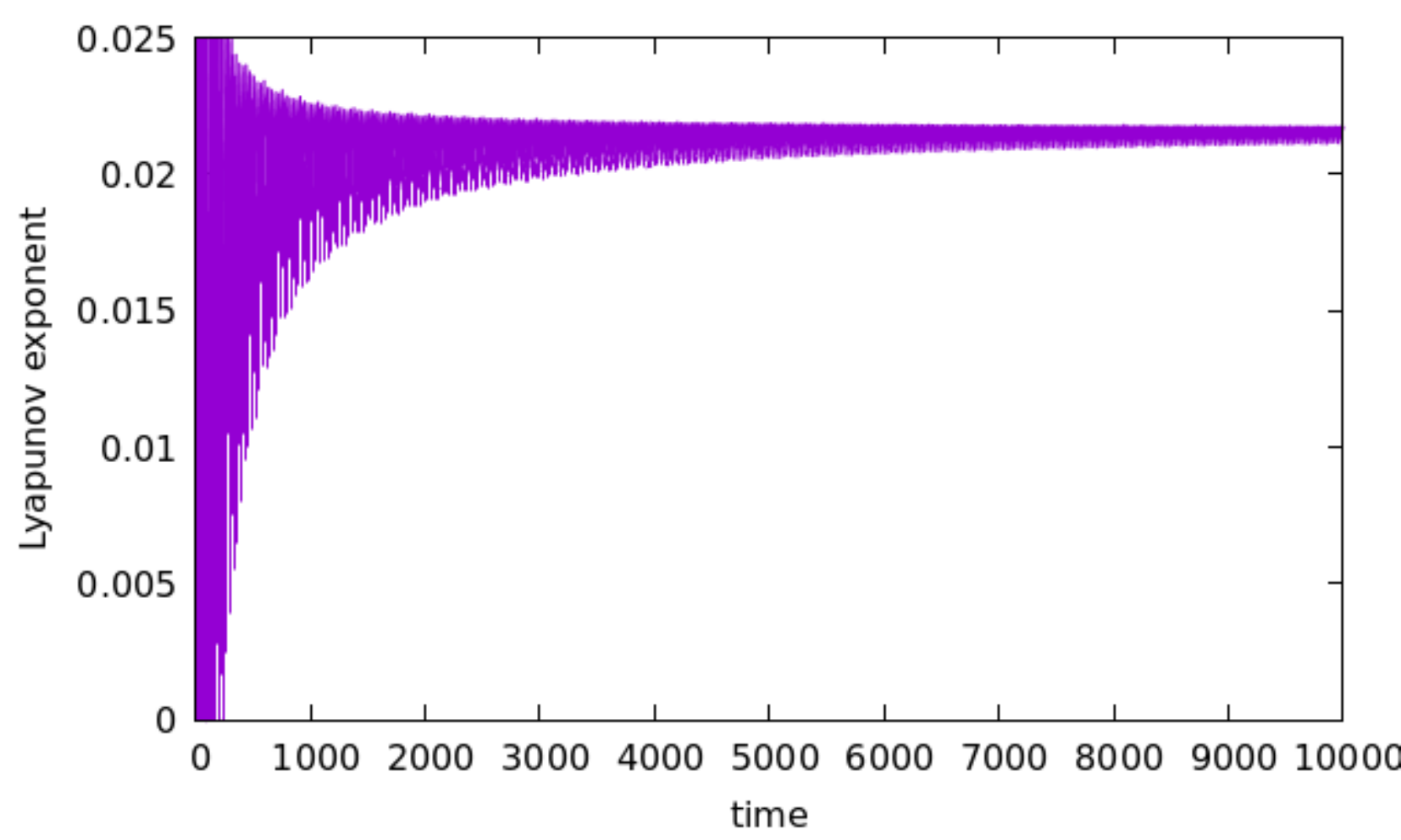} \caption{$\theta_{nc}=0.27$}
	\end{subfigure}
	\caption{Largest Lyapunov exponents for different values of $\theta_{nc}$ but for a particular value of the system energy $E=55$.}
	\label{fig:10}
\end{figure} 	 
\par\noindent
It is clear to see  that for the approximate case, chaos emerges into the system for the same energy $E$ and NC parameter $\theta_{nc}$ values as that we obtained in the case for the exact one. Therefore, we are reassured that the approximate form is equally valid from the point of view of physics, and we now use this approximated metric for the investigation of  Lyapunov exponents.
%
%
%
\subsubsection{Lyapunov exponent for approximated  NC-deformed metric}
%
%
%
%
\par\noindent
The appearance of chaos is evident whether we increase the value of the energy of the system $E$ or the value of $\theta_{nc}$  and that is what we have learned from the study of the Poincar\'e sections in the previous section. Now, this gradual emergence of chaotic behavior in the system can be shown more unambiguously in a quantifying manner with the help of studying the largest Lyapunov exponent of the system. We adopted the standard algorithm to compute the largest Lyapunov exponent which is related to the rate of separation of the trajectories for two nearby points \cite{sandri:96}. If we consider two trajectories with initial separation $\delta x_{0}$, the rate of divergence within the linearized approximation is given by 
\begin{eqnarray}
\vert\delta x(t)\vert\thickapprox e^{\lambda_{L}t}\vert\delta x_{0}\vert
\end{eqnarray}
where $\lambda_{L}$ is the Lyapunov exponent. Here we have reproduced two  graphs of the Lyapunov exponents (Fig. \ref{fig:9} and Fig. \ref{fig:10}). In order to plot these graphs, first we numerically solved the equations of motion of the particle, namely equations (\ref{rdot}) -  (\ref{ptheta_dot}), for $r$ with two initial conditions which are initially separated infinitesimally, i.e., $\delta r_{0}$. Then, the separation of these two trajectories has been studied for a long period of time which essentially gives us the saturated value of the maximum Lyapunov exponent. Here, we have studied the characteristics of the Lyapunov exponent for two cases. In Fig. \ref{fig:9}, we have plotted for different values of the system energies but for a constant value of $\theta_{nc}$ (for $\theta_{nc}=0.20$). Fig. \ref{fig:10} is plotted for different values of $\theta_{nc}$ but for a particular value of the system energy $E$ (for $E=55$).

From Fig. \ref{fig:9}, we find that with the increase in the value of  energy $E$ for a particular value of $\theta_{nc}=0.20$ the value of the largest Lyapunov exponent also increases. At $E=50$, the value of the largest Lyapunov exponent saturates at a value around $0$ which suggests that our system is still periodic. Whereas with the increase in the energy value, i.e., for $E=55,~60$, and $65$, the largest Lyapunov exponent value saturates at the value $\sim 0.015,~\sim 0.02$, and $\sim 0.04$, respectively. The positive increment in the largest Lyapunov exponent values with the increase in the energy in the system indicates that our system becomes more chaotic with the increase in  the energy. One important point to be noted here is that the Lyapunov exponent has an upper bound \cite{Maldacena:2015waa} which in this case is $\lambda_{L_{max}}=\kappa=1/2f'(r_{H})$, where $\kappa$ is the surface gravity of the black hole. Finally, it should be noted that the obtained values of the Lyapunov exponent for different values of the system energy $E$ are much lower than the corresponding upper bound (in this case the upper bound is $\lambda_{L_{max}}\approx 0.21$).   

Similarly, from Fig. \ref{fig:10}, we can see that for a particular value of the energy of the system, i.e., $E=55$, with the changing value in $\theta_{nc}$ some changes occur into the system. With the increment in the value of $\theta_{nc}=0.19,~0.22,~0.25$, and $0.27$, our system turns into a more chaotic one which suggests that the increased value of the parameter $\theta_{nc}$ induces more chaos into the system. In addition, in this case we also found out that the acquired value of the Lyapunov exponents for different values of $\theta_{nc}$ are lower than the upper bounds discovered in \cite{Maldacena:2015waa} ($\lambda_{L_{max}}\approx 0.22,~0.20,~0.18$ and $0.16$ for $\theta_{nc}=0.19,~0.22,~0.25$, and $0.27$, respectively). Notice that the upper bound depends on $\theta_{nc}$.
%
%
%
%
\section{Chaos for quantum-corrected Schwarzschild metric}
%
%
%
%
\par\noindent
The metric for quantum-corrected (QC) Schwarzschild black hole has appeared in \cite{don}
\begin{widetext}
	\begin{eqnarray}
	ds^2 =- \left(1-\frac{2GM}{c^2r}+\epsilon(r)\right)dt^2 + \left(1-\frac{2GM}{c^2r}+\epsilon(r)\right)^{-1} dr^2 + r^2 d\Omega^2
	\end{eqnarray} 
\end{widetext}
where $\epsilon(r)$ is the quantum correction term  given by $\epsilon(r)= -6\left(G^2M^2/c^4r^2\right)(1+(m/M))-(41/5\pi)(GM/c^2)(l_p^2/r^3)$, where $l_p$ is the Planck length and $d\Omega^2 = d\theta^2 + \sin^2\theta d\phi^2$. The black hole mass is  denoted as $M$  while $m$ stands for the particle mass. However, in our analysis our main focus is on the dynamics of the massless particle, so $m=0$ in our case.\\

Due to smallness, the last term has not been taken into account in our analysis and for a massless particle the quantum-corrected metric takes the form
\begin{widetext}
	\begin{eqnarray}
	ds^2 = -\left(1-2\frac{k}{r}-6\frac{k^2}{r^2}\right)dt^2 + \left(1-2\frac{k}{r}-6\frac{k^2}{r^2}\right)^{-1} dr^2 + r^2 d\Omega^2\label{QC metric}
	\end{eqnarray}
\end{widetext}
%
%
%
where $k=GM/c^2$ and $k$ is taken to be unity for further analysis according to the unit convention $(G=c=\hbar=1)$ that we have been following so far and the value of the black hole mass, i.e., $M=1.0$. Next, we shall study the dynamics of a massless particle near the horizon of the quantum-corrected Schwarzschild metric (\ref{QC metric}). Considering all these facts, we obtain the position of the horizon at $r_H\approx 3.6$. In this case,  we shall also follow the same formalism as we performed in the previous case of NC-deformed Schwarzschild metric in Section II. With the introduction of the harmonic potentials, we obtain the total energy of the system in the background of the QC Schwarzschild metric as
\begin{eqnarray}
E = &&-\sqrt{(1-f_{QC}(r))}p_r +  \sqrt{p_r^2 + \frac{ p_\theta^2}{r^2}} + \frac{1}{2}K_r (r-r_c)^2 \nonumber\\
&&+ \frac{1}{2}K_\theta (y-y_c)^2\label{energy qc}
\end{eqnarray}  
where $f_{QC}(r)=1-2k/r-6k^2/r^2$ and the other parameters are defined the same as in the case of the NC-deformed metric. Using  the Hamilton's equations of motion, we obtain 
\begin{eqnarray}
\dot{r}&=&\frac{\partial E}{\partial p_{r}}=-\sqrt{1-f_{QC}(r)}+\frac{p_{r}}{\sqrt{p_{r}^{2}+\frac{p_{\theta}^{2}}{r^{2}}}}\label{rdot qc}\\
\dot{p_r}&=&- \frac{\partial E}{\partial r} = -\frac{f_{QC}'(r)}{2 \sqrt{(1-f_{QC}(r))}}p_r + \frac{p_\theta ^2 /r^3}{\sqrt{p_r^2 + \frac{ p_\theta ^2}{r^2}}}\nonumber\\
&&- K_r(r-r_c)\label{prdot qc}\\
\dot{\theta}&=&\frac{\partial E}{\partial p_\theta} = \frac{p_\theta /r^2}{\sqrt{p_r^2 + \frac{ p_\theta ^2}{r^2}}}\\\label{theta_dot qc}
\dot{p_\theta}&=& - \frac{\partial E}{\partial \theta} = -K_\theta r_H(y-y_c)\label{ptheta_dot qc}
\end{eqnarray} 
where the derivative is taken with respect to some affine parameter.

Now, solving numerically the equations of motion of the particle, we shall plot the Poincar\'e sections and Lyapunov exponents for the particle motion in the near-horizon region and then we shall try to analyze both of them.

\subsection{Poincar\'e sections for quantum-corrected metric}
\par\noindent

\begin{figure}[htb!]
	\centering
	\begin{subfigure}[b]{0.39\linewidth}
		\includegraphics[width=\linewidth]{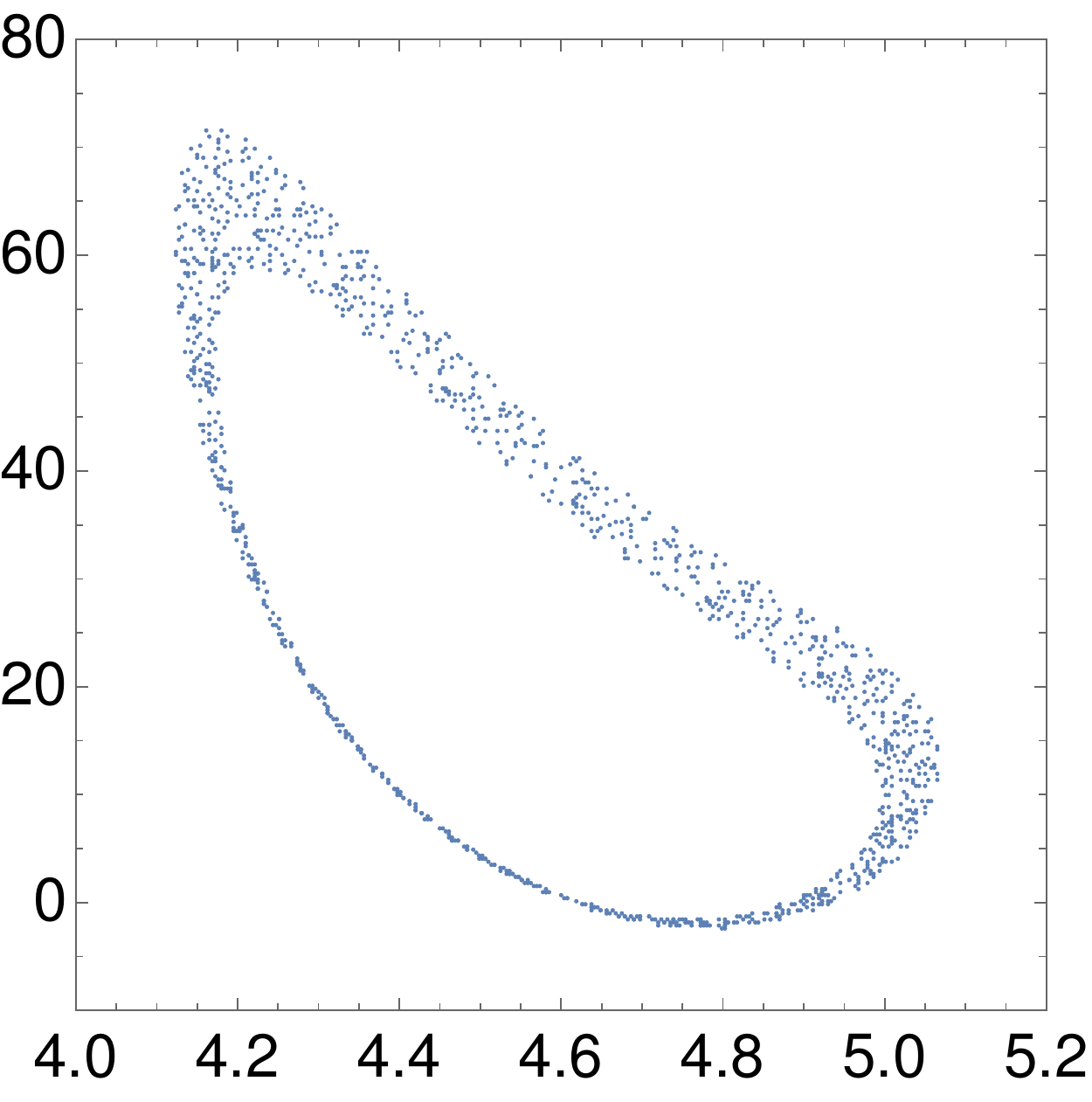}
		\caption{E=60}
	\end{subfigure}
	\begin{subfigure}[b]{0.4\linewidth}
		\includegraphics[width=\linewidth]{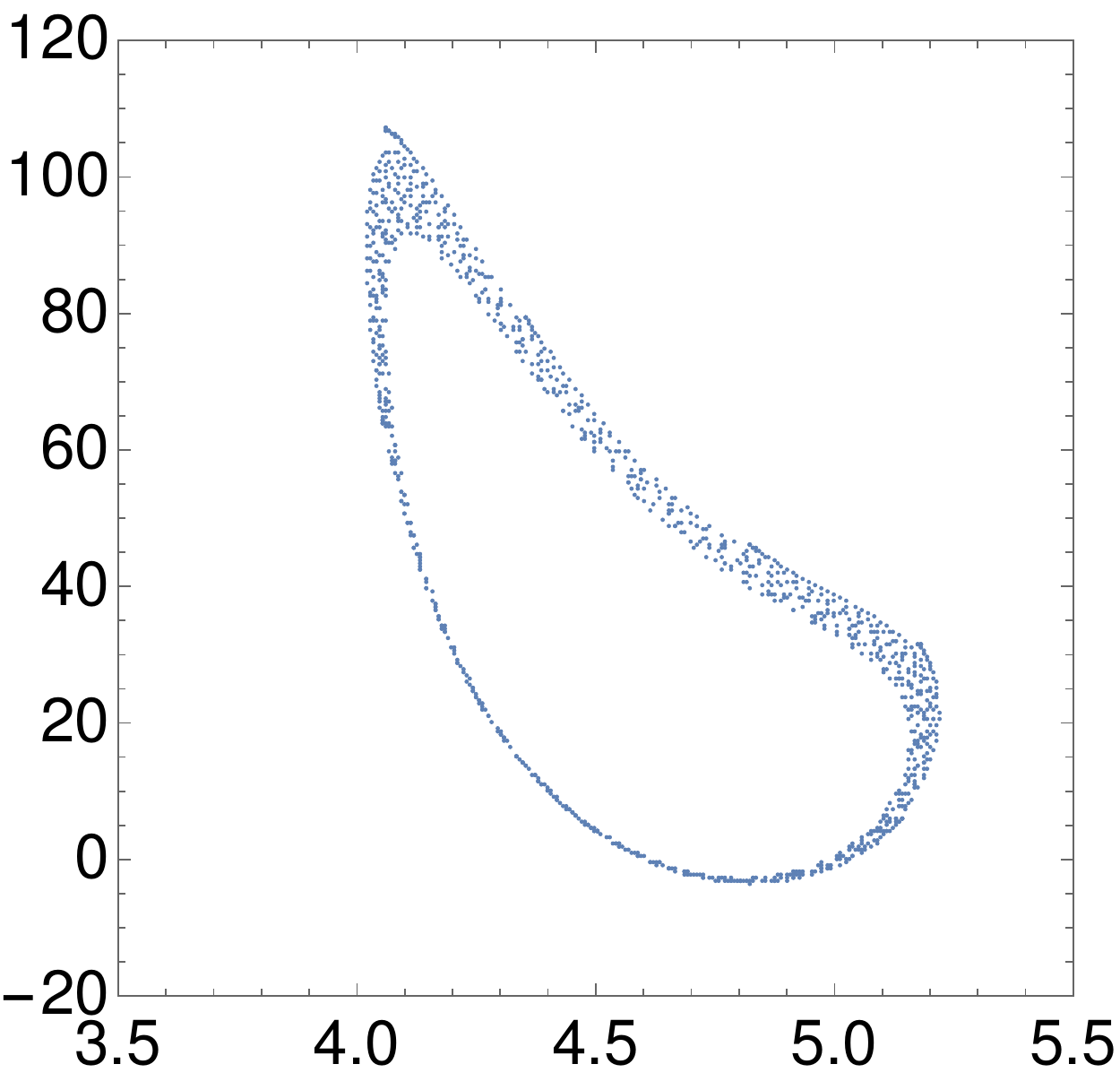}
		\caption{E=65}
	\end{subfigure}
	\begin{subfigure}[b]{0.4\linewidth}
		\includegraphics[width=\linewidth]{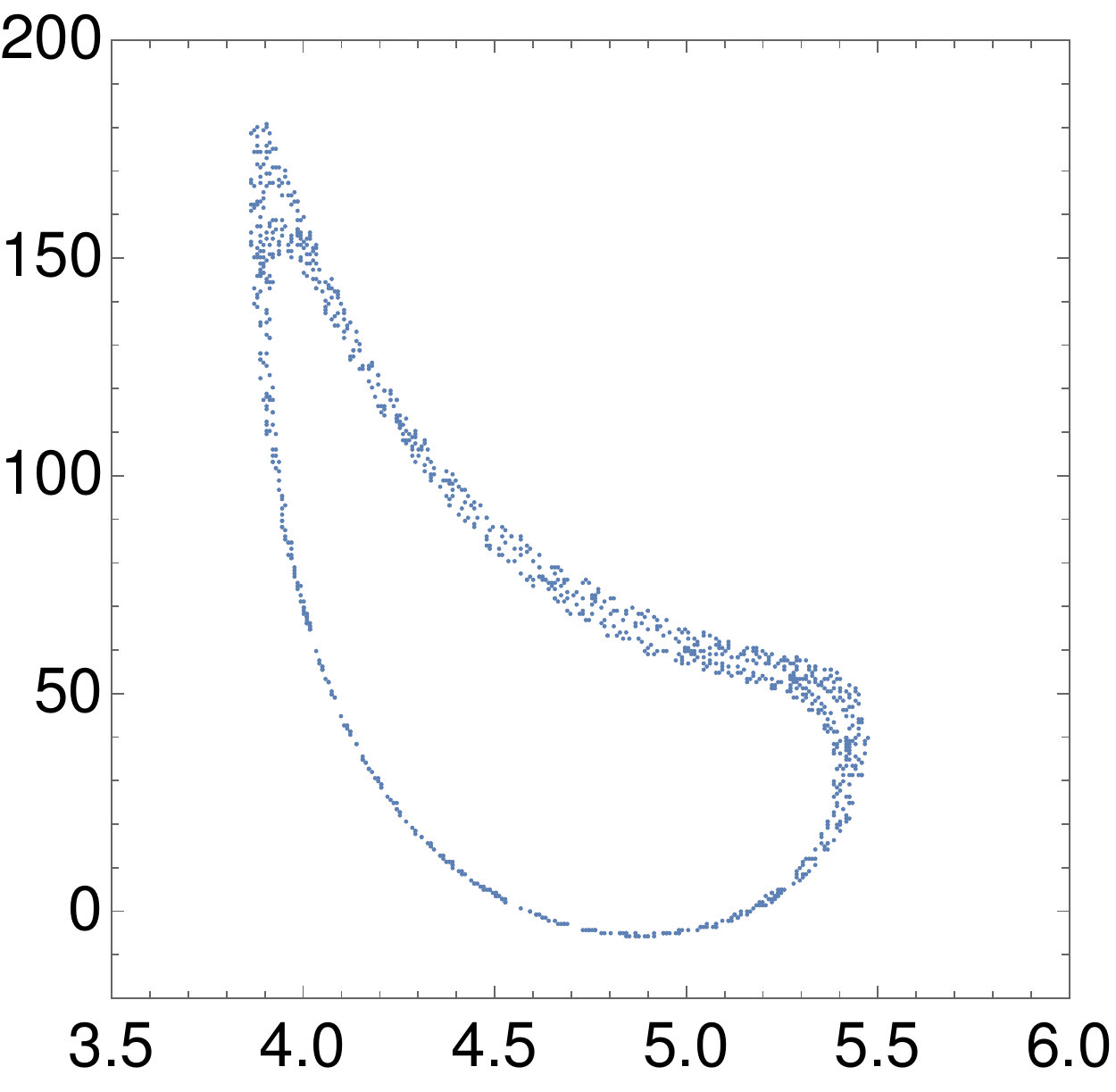}
		\caption{E=70}
	\end{subfigure}
	\begin{subfigure}[b]{0.4\linewidth}
		\includegraphics[width=\linewidth]{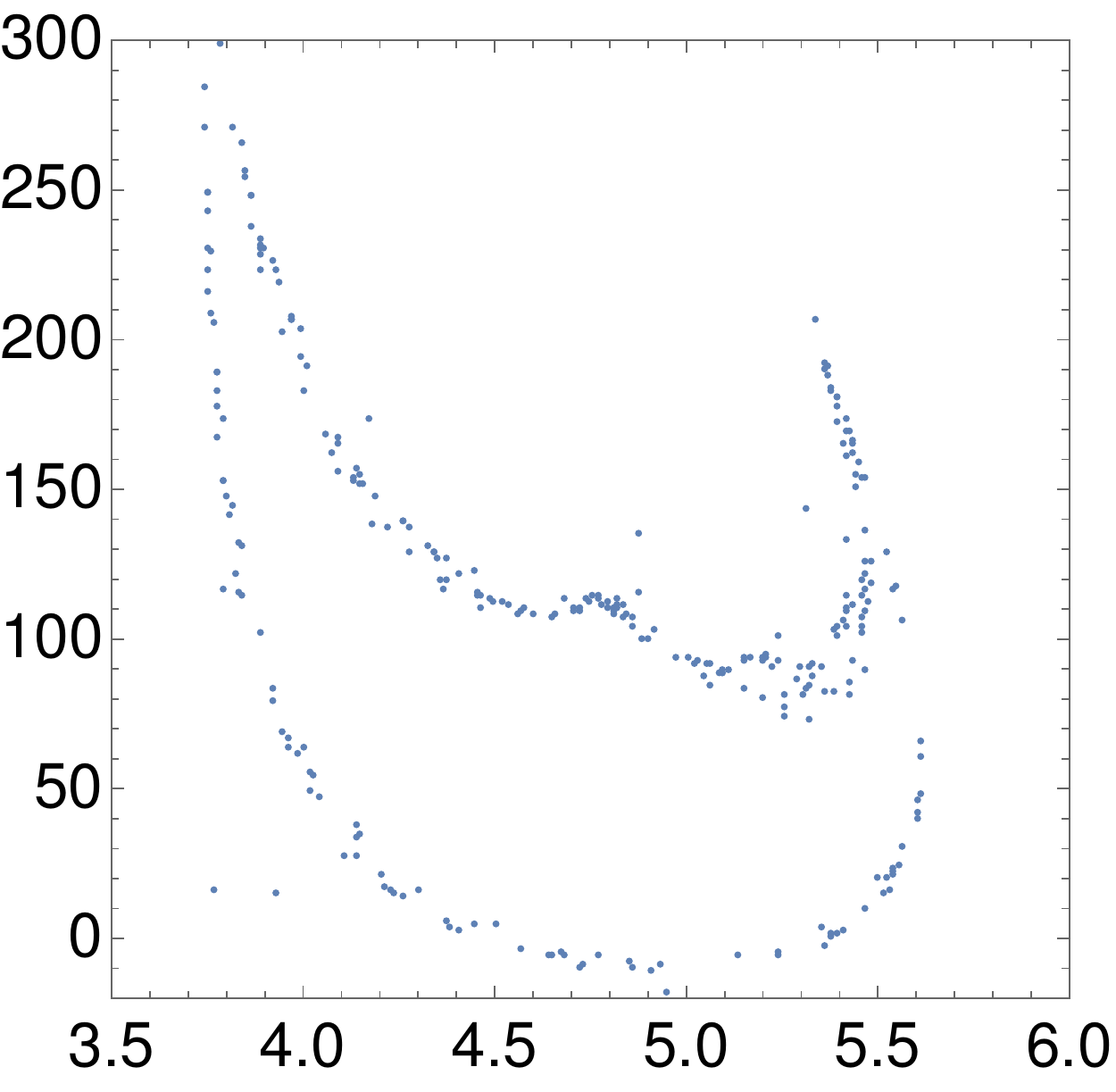}
		\caption{E=72}
	\end{subfigure}
	\caption{The Poincar\'e sections in the ($r,p_r$) plane with $\theta = 0$ and $p_\theta > 0$ at different energies for quantum-corrected Schwarzschild black hole. The horizontal and vertical axes in each of the graphs correspond to $r$ and $p_r$, respectively.}
	\label{fig:11}
\end{figure}
\par\noindent
In this case, the Poincar\'e sections in Fig. \ref{fig:11} are also plotted for $p_{\theta}>0$ and $\theta=0$ just like before. We have considered $M=1.0$, $K_r=100$, $K_\theta=25$, $r_c=4.5$, and $y_c=0$. In order to make sure that the particle resides near the horizon we have restricted the range of $r$ at $4.5<r<5.5$. Similar nature as before can be found in these plots. As we increase the value of the energy of the system $E=55,~60,~65,~70,~72$, the system gradually goes from the periodic state to the chaotic one. The appearance of the scattered points in the higher energy values, namely $E=70,~72$, is an indication of the chaotic fluctuations into the system. With the increase in the energy of the system, our massless test particle reaches nearer to the horizon and the more our system comes under the influence of horizon the more it becomes chaotic.  Some more examples of Poincar\'e sections for different initial conditions for this case are provided in Appendix \ref{App1}.

In the previous section, for  the particle motion in the background of the NC-deformed metric, we discussed the effect of changing values of the spring constants,  $K_r$ and $K_{\theta}$ (of the harmonic potential) on the Poincar\'e sections. Here also the same phenomena are observed as we change the values of $K_{r}$ and $K_{\theta}$. As we increase $K_r$ and $K_{\theta}$, the Poincar\'e sections start showing distortion in the higher energy regime. However, with  decreasing  values of $K_r$ and $K_{\theta}$, the Poincar\'e sections start showing chaos in the lower energy values as the effect of the horizon starts to dominate.

It is important to note that in the expression of the system energy $(\ref{energy qc})$, the first term  (containing the $f_{QC}$ term) is the quantum-corrected part if we compare it with the expression of the system energy in the standard Schwarzschild background \cite{Dalui:2018qqv}. Due to this quantum correction, the system starts showing chaotic fluctuations in the lower energy values in comparison to the case of a standard Schwarzschild one and this feature  is reflected in the  Poincar\'e sections profiles.


\begin{figure}[hb!]
	\centering
	\begin{subfigure}[b]{0.6\linewidth}
		\includegraphics[width=\linewidth]{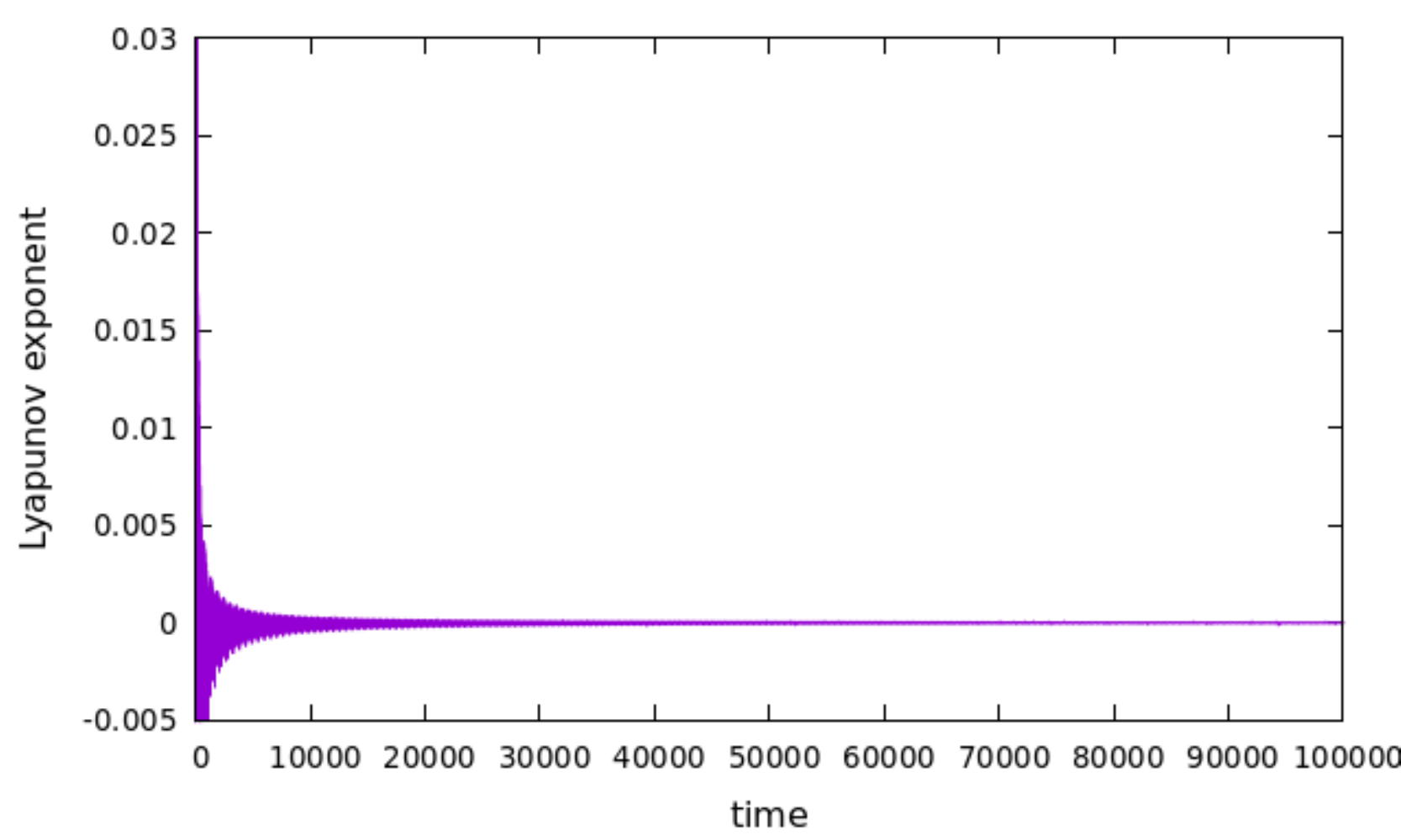} \caption{$E=60$}
	\end{subfigure}
	\begin{subfigure}[b]{0.6\linewidth}
		\includegraphics[width=\linewidth]{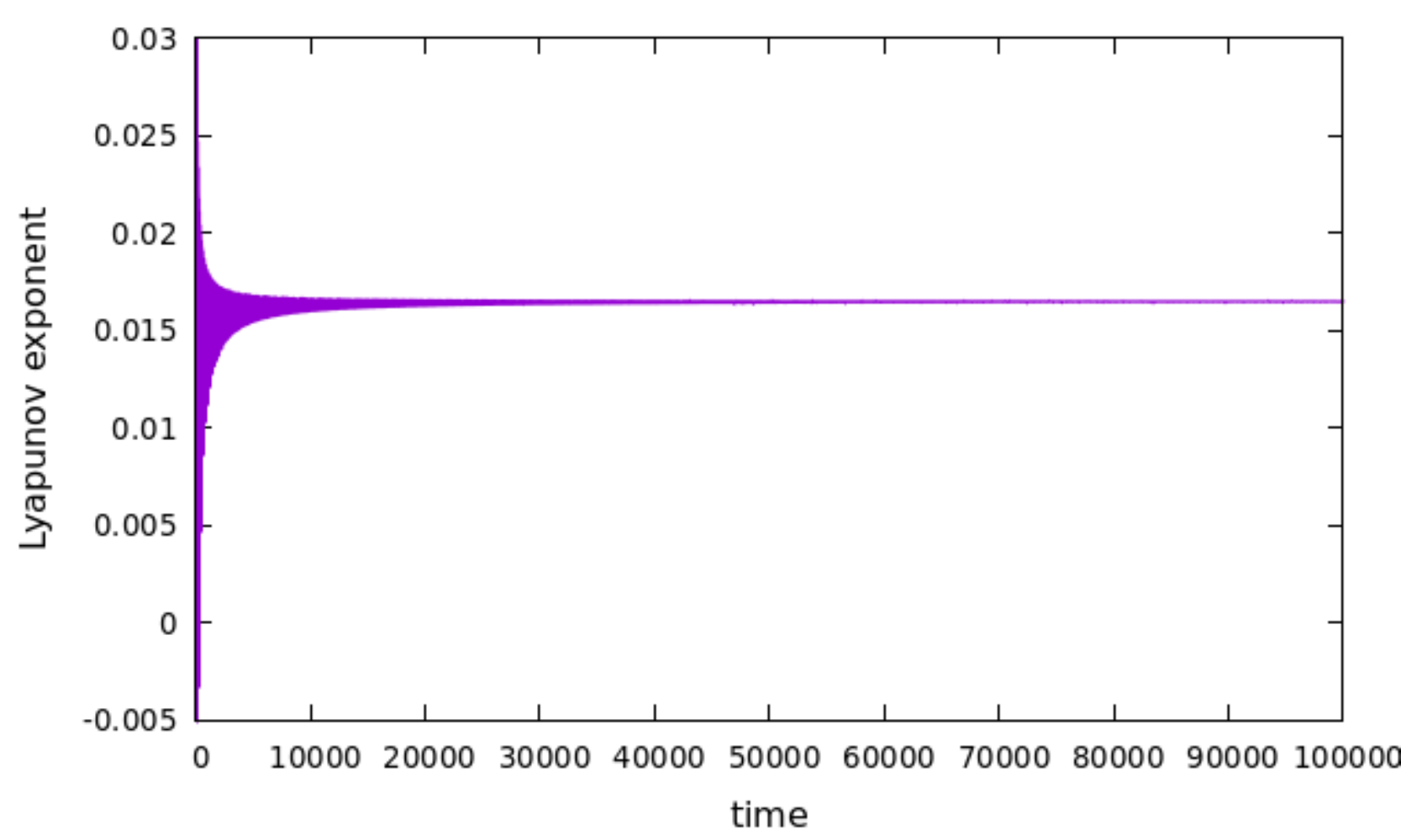} \caption{$E=65$}
	\end{subfigure}
	\begin{subfigure}[b]{0.6\linewidth}
		\includegraphics[width=\linewidth]{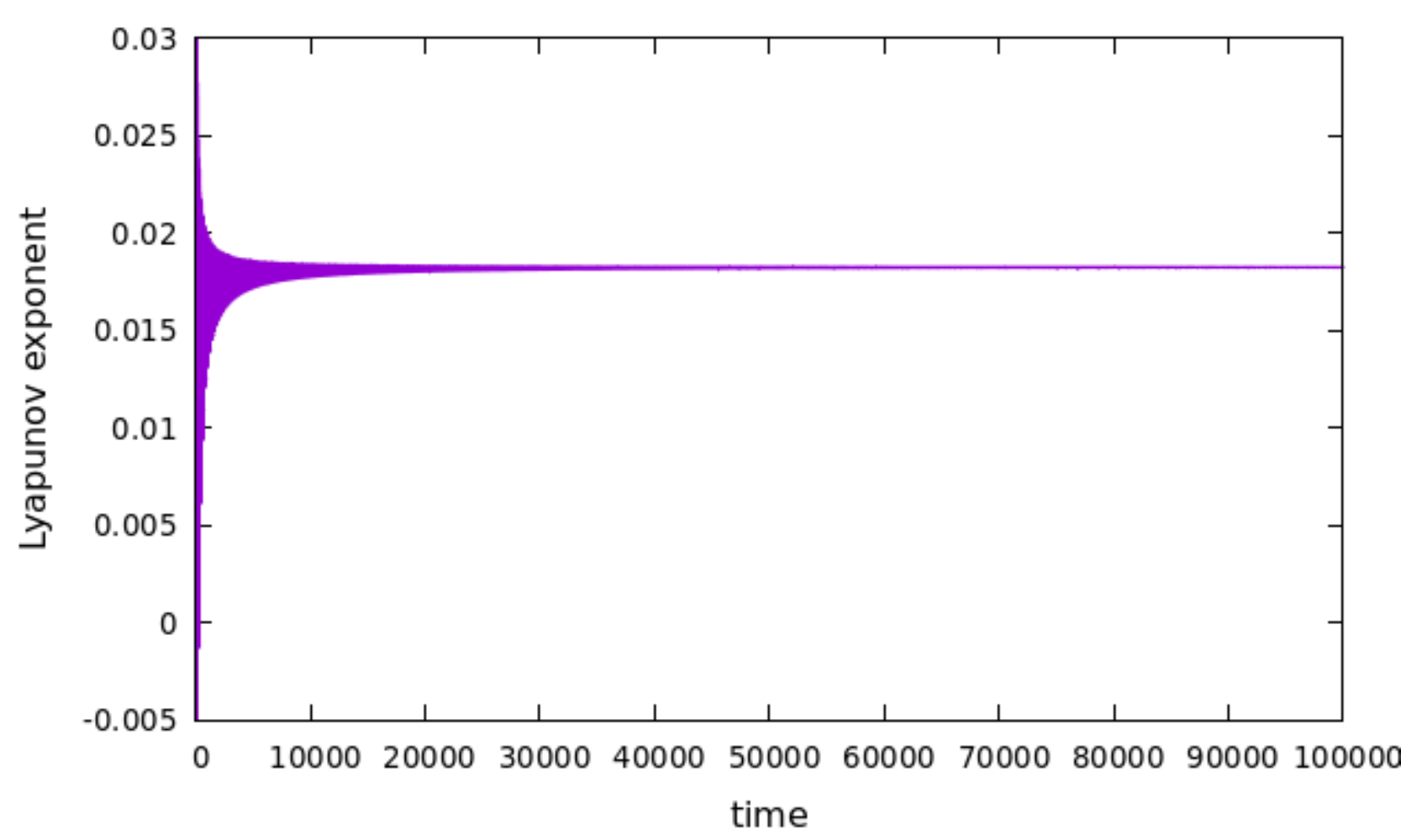} \caption{$E=70$}
	\end{subfigure}
	\begin{subfigure}[b]{0.6\linewidth}
		\includegraphics[width=\linewidth]{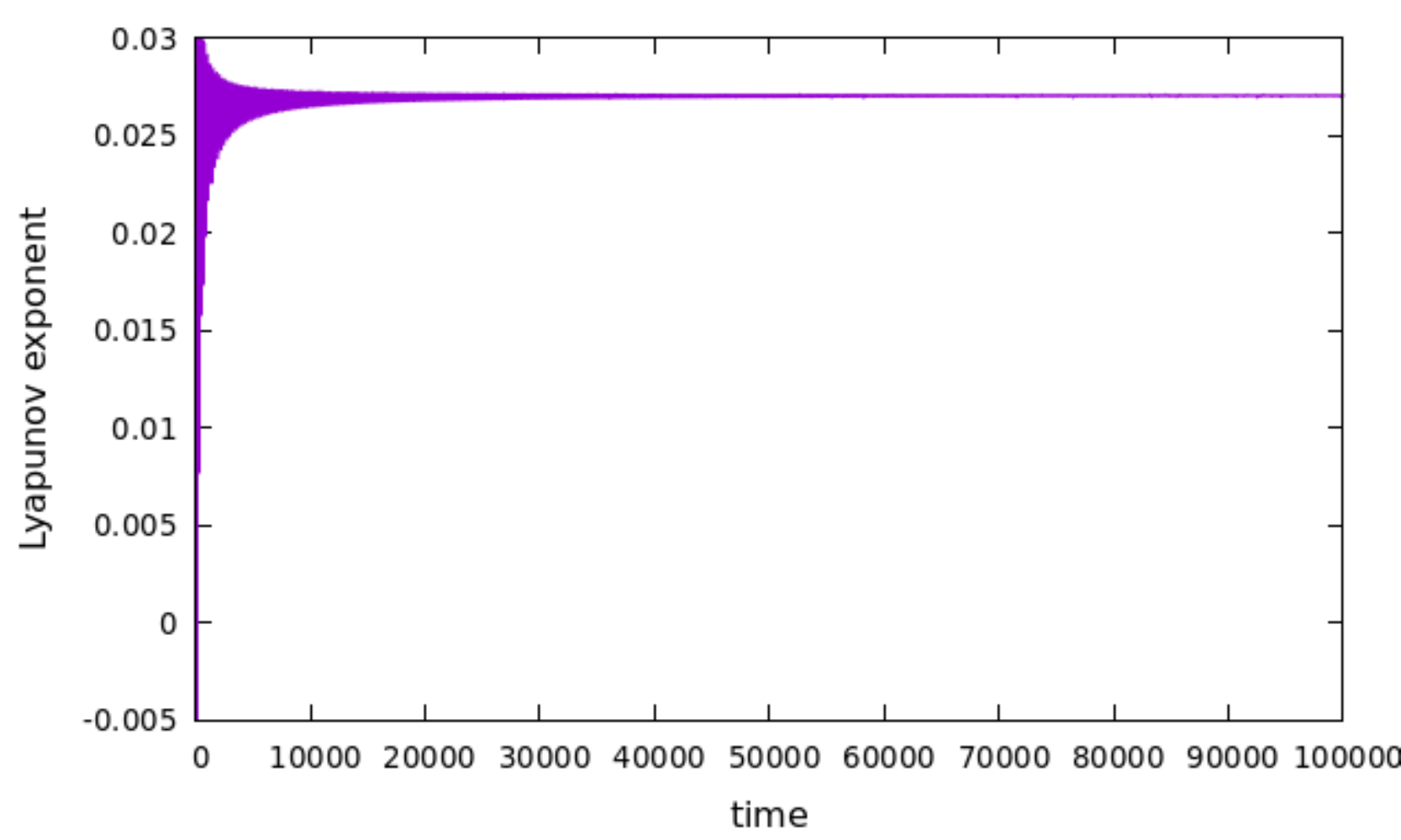} \caption{$E=72$}
	\end{subfigure}
	\caption{Largest Lyapunov exponents for different values of energy of the system $E$.}
	\label{fig:12}
\end{figure} 	 
%
%
%
%
%
%
\subsection{Lyapunov exponents for quantum corrected metric}
%
%
%
%
\par\noindent
Here, we also quantify the chaos in our system. From Fig. \ref{fig:12}, we can see that with the increase in the value of the energy of the system, the largest Lyapunov value of the system gets increased. As we increase the energy, the particle moves closer towards the horizon, subsequently  the influence of the horizon increases making   the system more chaotic. Moreover, it is worth to mention that the obtained value of the Lyapunov exponents for different values of energy of the system are lower than the upper bound \cite{Maldacena:2015waa}.

\section{Conclusion}
\par\noindent
Let us summarize the results obtained in the present work. The chaotic behavior of particle dynamics near Schwarzschild black hole has been under study for quite some time. In recent times, this research has received new impetus from deep results in a very different area, namely the existence of a universal upper bound of Lyapunov exponent for a finite-temperature quantum field theory. These two very distinct phenomena are intimately connected by the fact that black holes can be treated as thermodynamic systems with a Hawking temperature. Another area of recent interest in black hole physics is the quantum corrections incorporated in the metric, either arising from possible quantum gravity effects via noncommutative geometry, or generated through conventional quantum field theoretic effects. In the present work we have analysed in detail the effects of both metric extensions on the chaotic behavior of particles close to the black hole horizon.
\par
 Our results clearly show that in both cases, the metric extensions favour chaotic behavior, namely chaos is attained for relatively smaller particle energy. This is demonstrated numerically by exhibiting the breaking of the KAM torus in the Poincar\'e sections of the particle trajectories and also via explicit computation of the (positive) Lyapunov exponents of the trajectories. Also of interest is  that in all the cases considered here, the numerical values of the Lyapunov exponent are well within the universal saturation value.
%
%
%
%
\section{Acknowledgements}
\par\noindent
The authors would like to thank the anonymous referees for their constructive comments.\\
\appendix
\section{{\label{App1}}Poincar\'e sections for quantum-corrected Schwarzschild metric}
%
%
%
\par
\begin{figure}[h!]
		\centering
		\begin{subfigure}[b]{0.49\linewidth}
			\includegraphics[width=\linewidth]{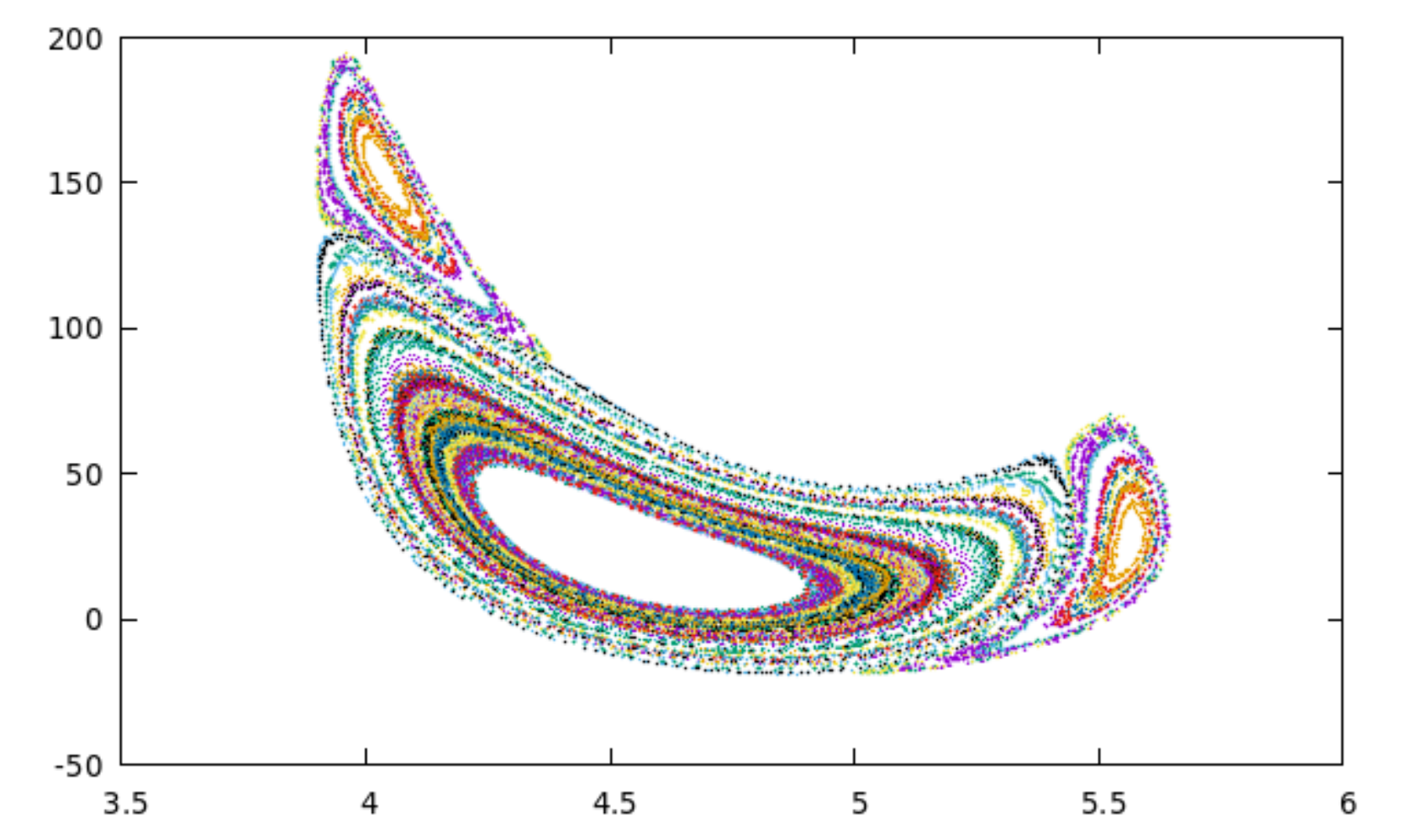}
			\caption{E=60}
		\end{subfigure}
		\begin{subfigure}[b]{0.49\linewidth}
			\includegraphics[width=\linewidth]{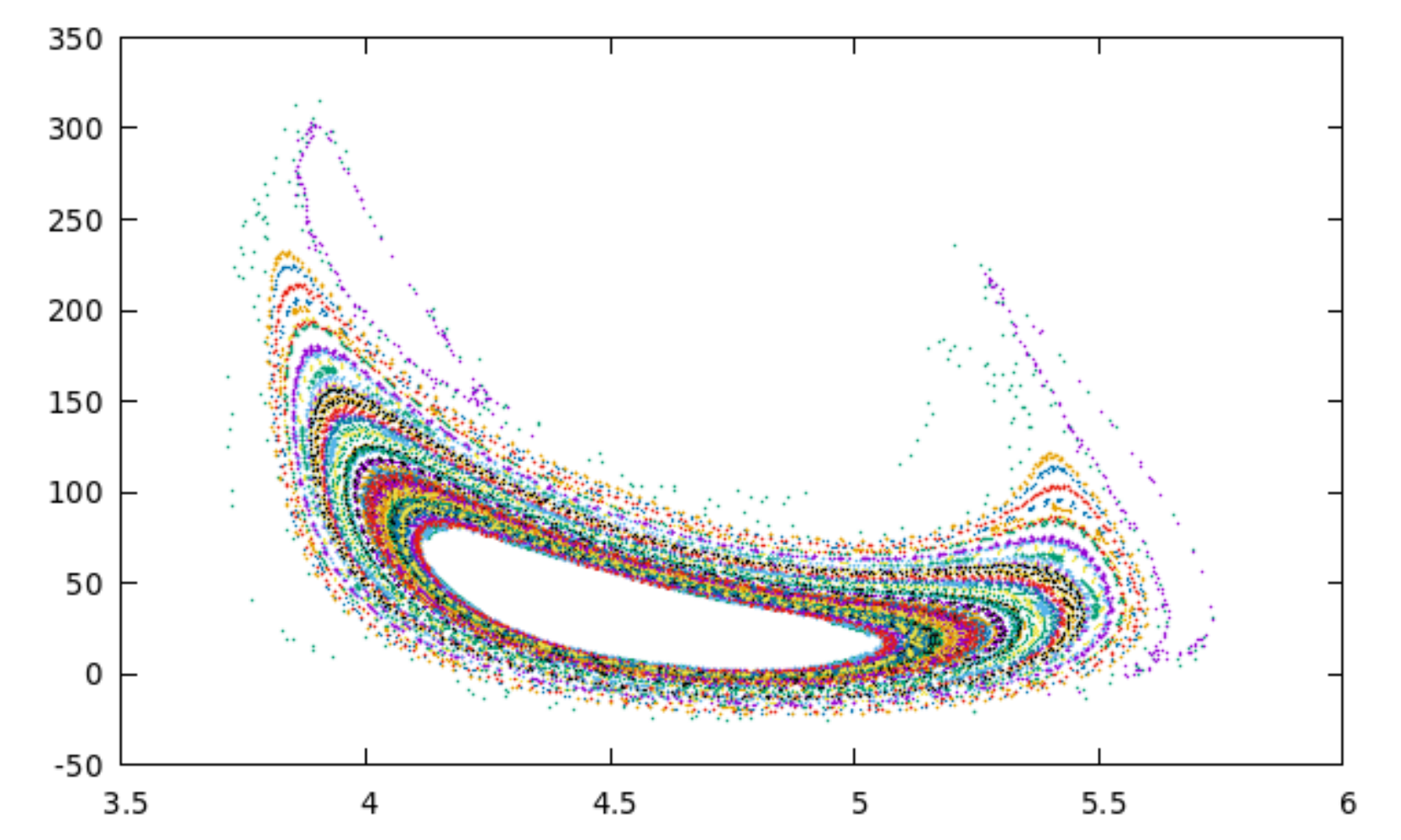}
			\caption{E=60}
		\end{subfigure}
		
		\begin{subfigure}[b]{0.49\linewidth}
			\includegraphics[width=\linewidth]{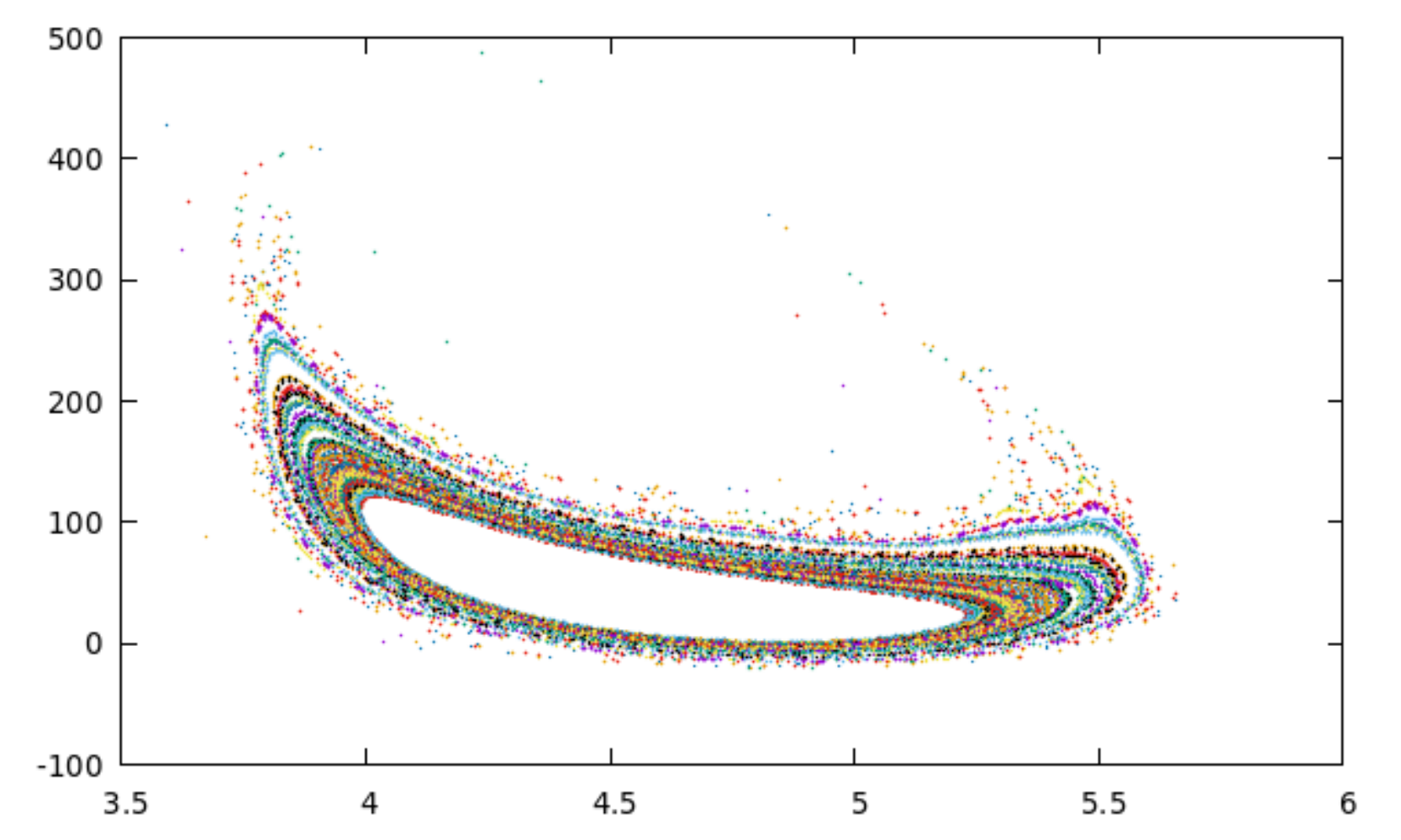}
			\caption{E=70}
		\end{subfigure}
		\begin{subfigure}[b]{0.49\linewidth}
			\includegraphics[width=\linewidth]{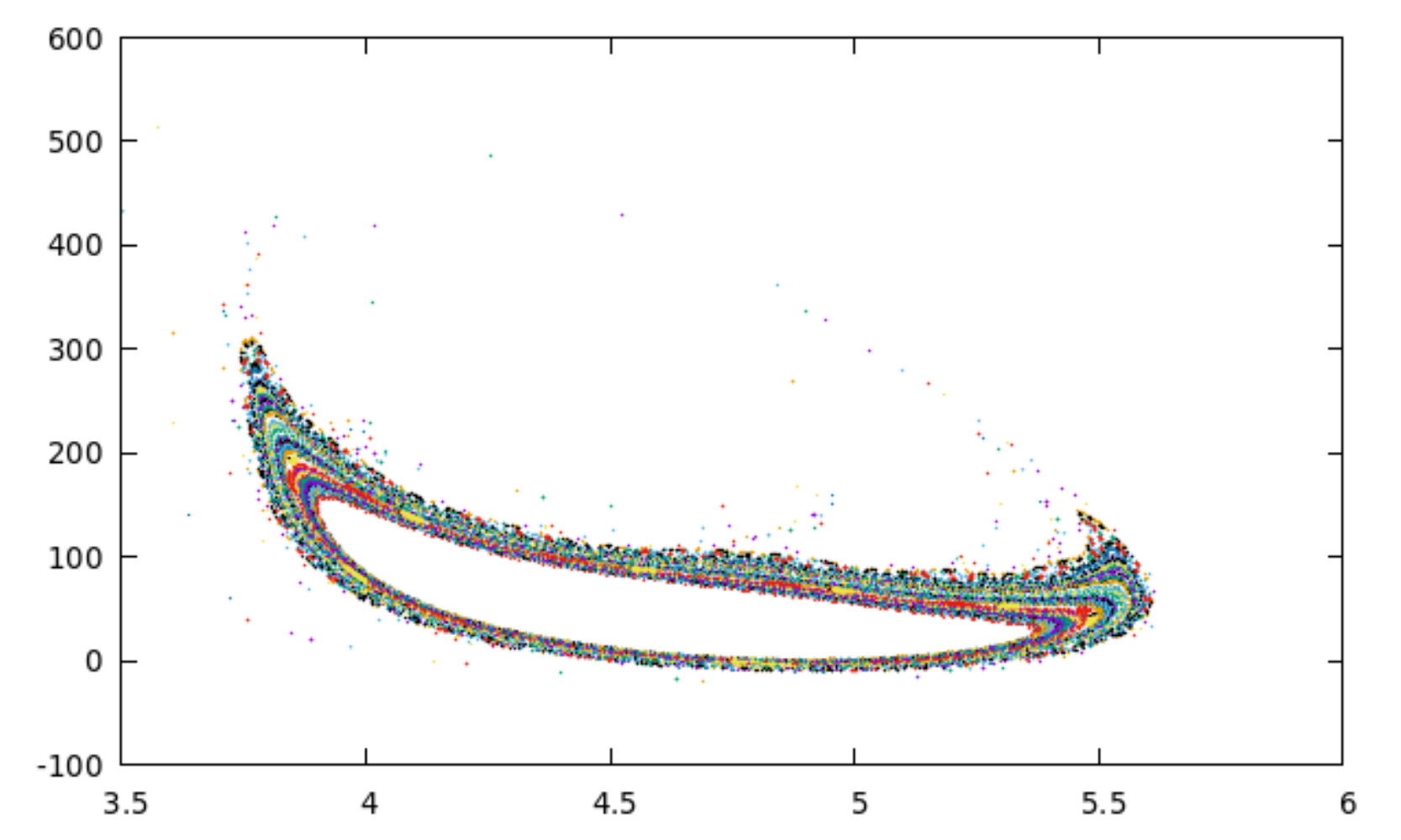}
			\caption{E=72}
		\end{subfigure}
		\caption{The Poincar\'e sections in the ($r,p_r$) plane with $\theta = 0$ and $p_\theta > 0$ at differernt energies for quantum-corrected Schwarzschild black hole. The horizontal and vertical axes in each of the graphs correspond to $r$ and $p_r$, respectively. Different coloured lines indicate the different initial conditions. }
		\label{fig:13}
\end{figure}

\noindent
The  plots in Fig. \ref{fig:13} represent the Poincar\'e sections for a particle motion near the QC Schwarzschild horizon. Here the dynamical equations of motion of the particle has been solved for different initial conditions of $r$ and $p_r$. Different colours in the figures represent different initial conditions. However, the initial conditions of $r$ are restricted in the range $4.5<r<5.5$. The scattered points in the Poincar\'e sections for the higher energy values show that chaos emerges with the increase in the value of the energy of the system $E$. 
%
%
%
%
\section{{\label{App2}}Orbits of the particle trapped in the harmonic potential in the near-horizon region}
%
%
%
%
\begin{figure}[H]
	\centering
		\begin{subfigure}[b]{0.49\linewidth}
		\includegraphics[width=\linewidth]{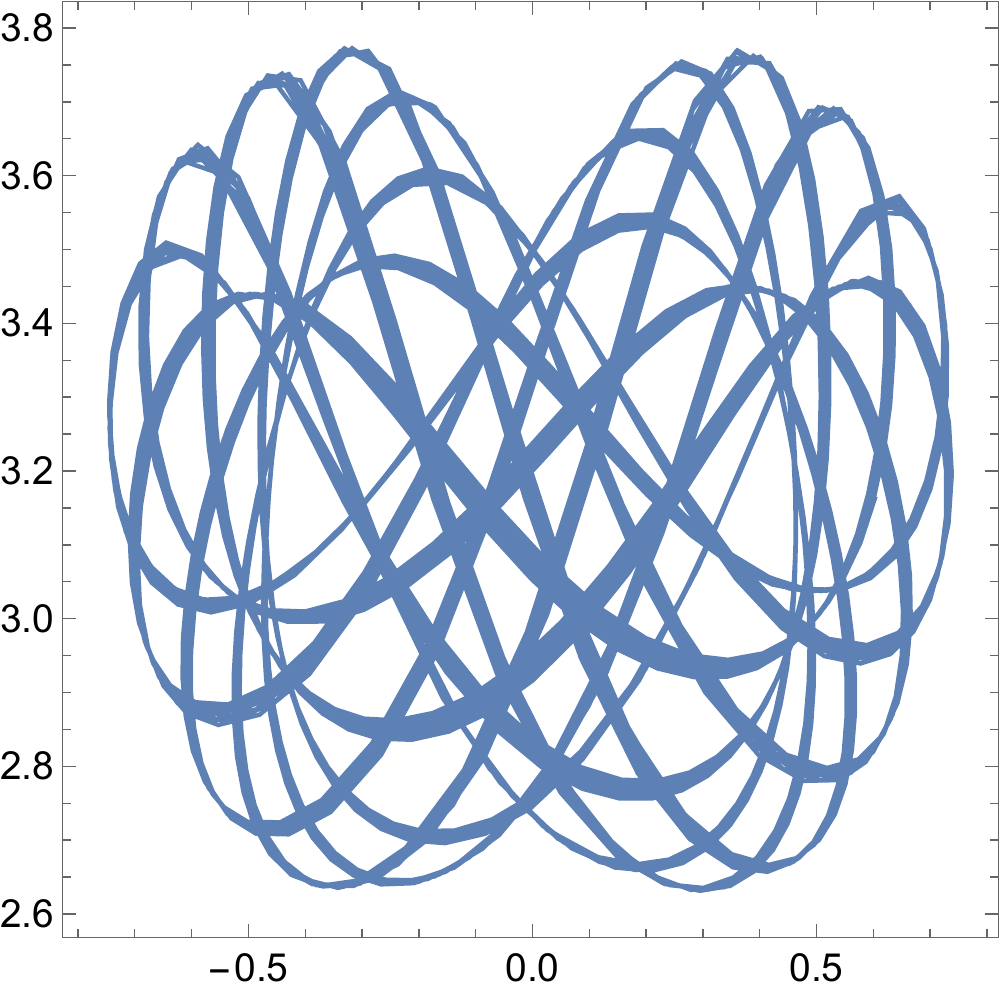}
		\caption{E=50}
	\end{subfigure}
	\begin{subfigure}[b]{0.49\linewidth}
		\includegraphics[width=\linewidth]{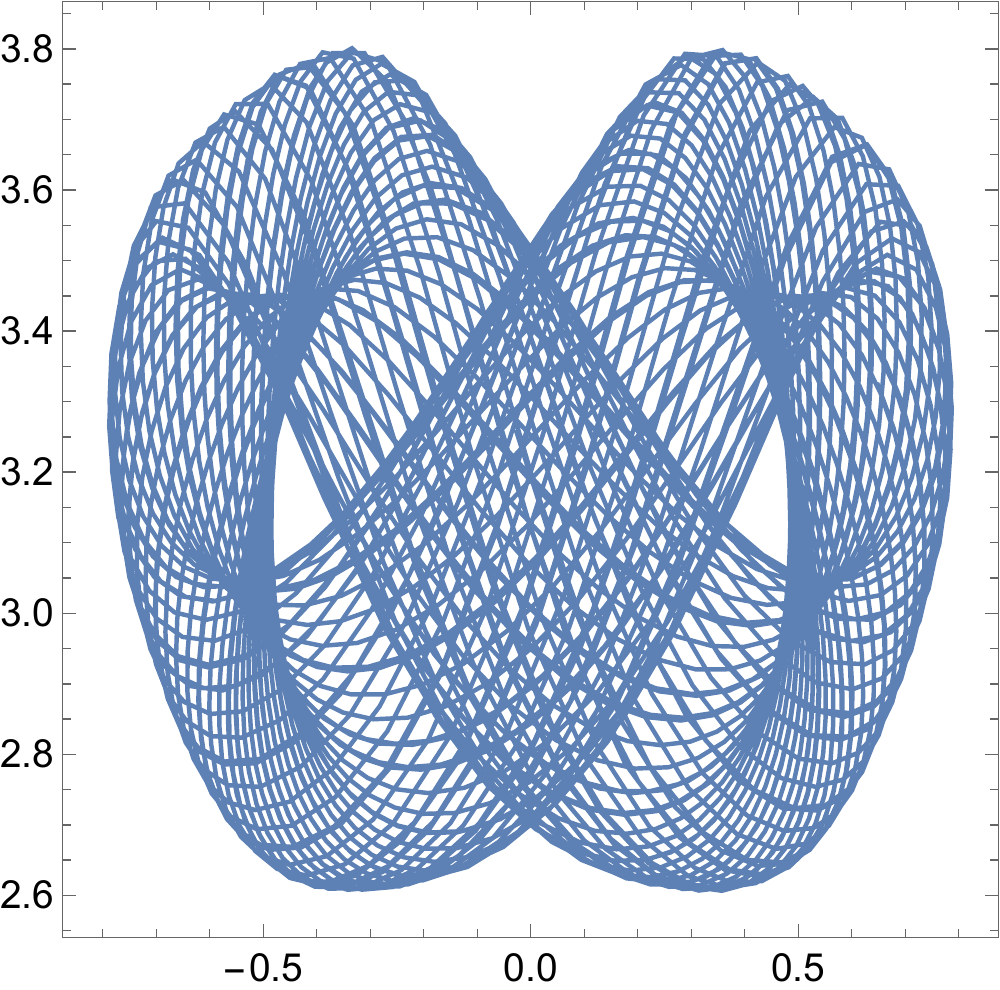}
		\caption{E=55}
	\end{subfigure}
	\begin{subfigure}[b]{0.49\linewidth}
		\includegraphics[width=\linewidth]{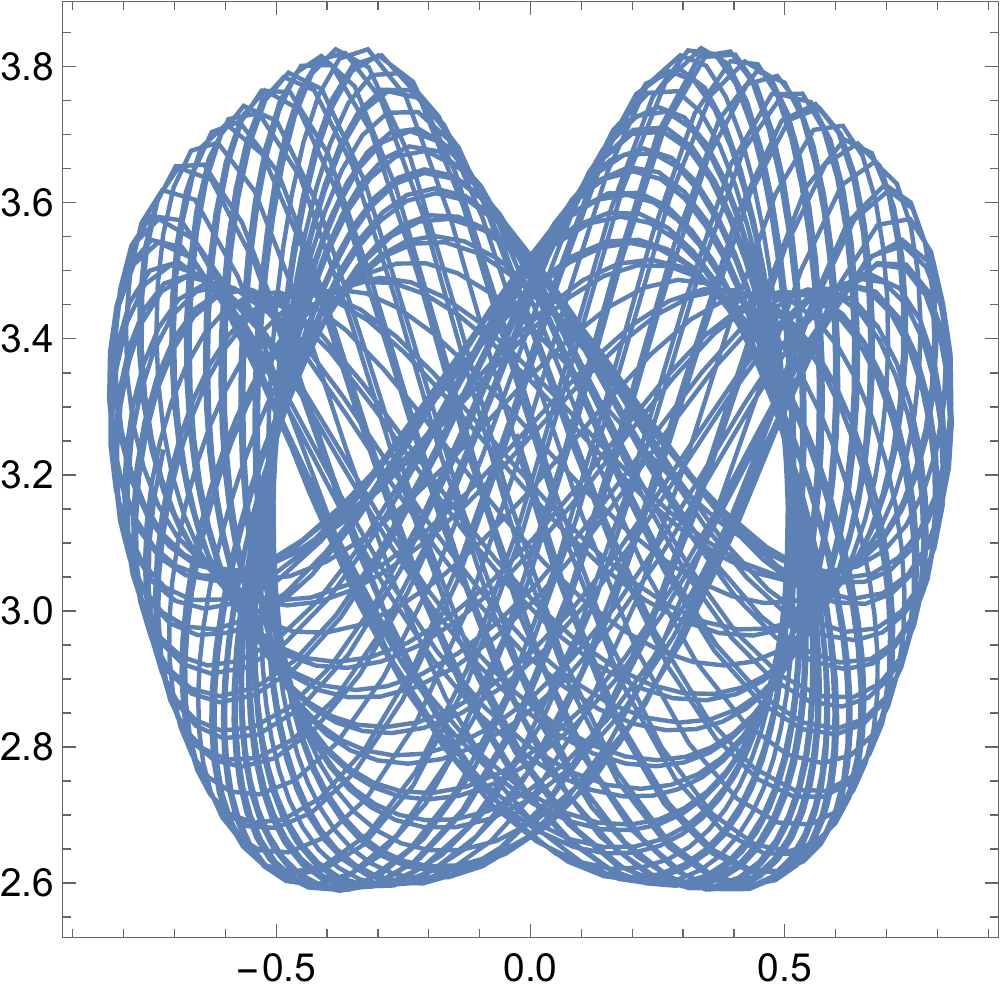}
		\caption{E=60}
	\end{subfigure}
	\begin{subfigure}[b]{0.49\linewidth}
		\includegraphics[width=\linewidth]{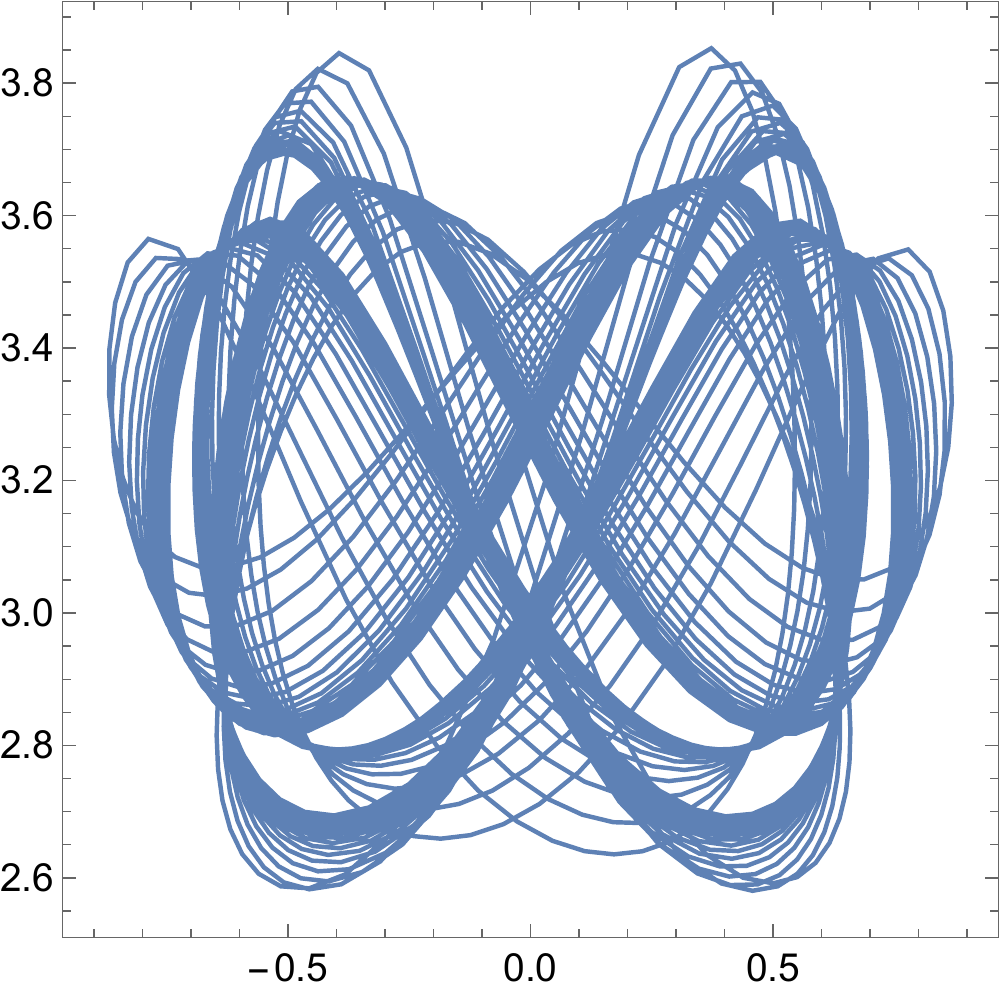}
		\caption{E=65}
	\end{subfigure}
	\caption{Orbits of the particle trapped in the harmonic potential in the near-horizon region in the background of (\ref{Nicolini metric}) for fixed $\theta_{nc}=0.16$ but for different values of the system energy $E$.}
	\label{fig:14}
\end{figure}
\begin{figure}[H]
	\centering
	\begin{subfigure}[b]{0.49\linewidth}
		\includegraphics[width=\linewidth]{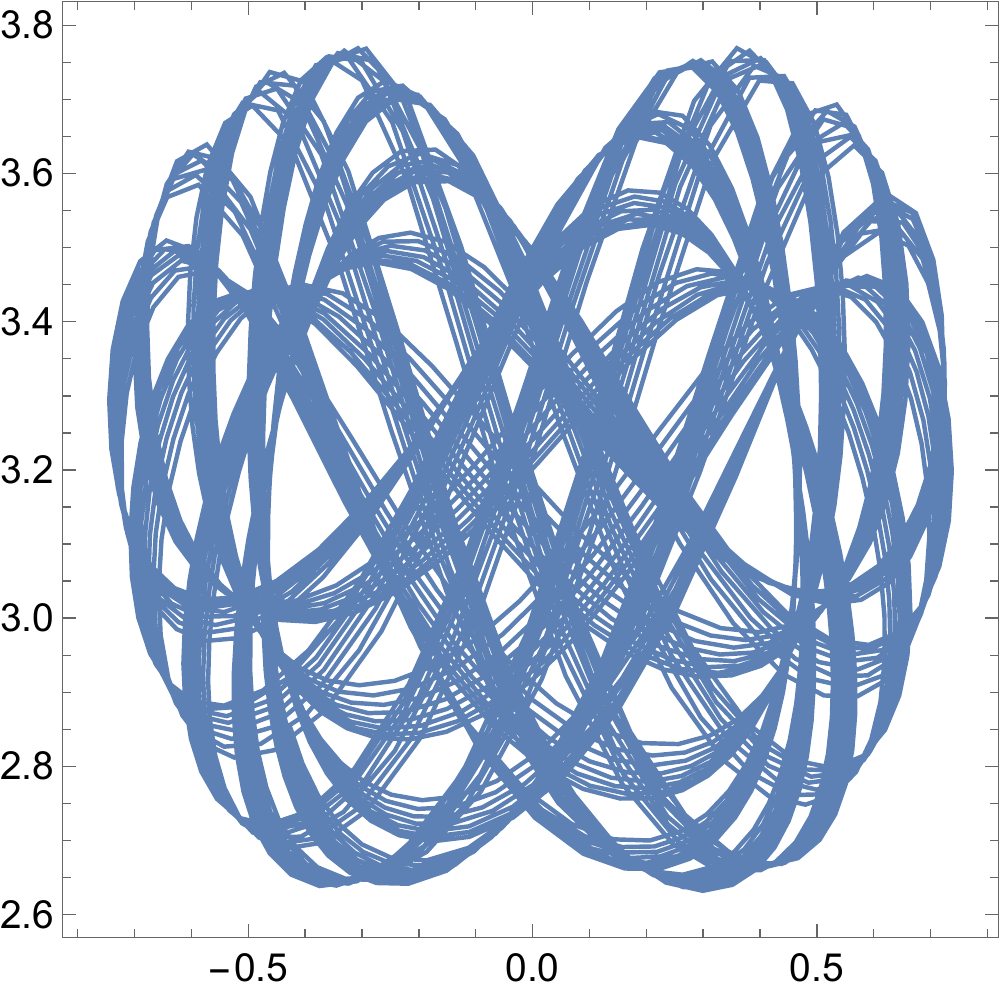}
		\caption{E=50}
	\end{subfigure}
	\begin{subfigure}[b]{0.49\linewidth}
		\includegraphics[width=\linewidth]{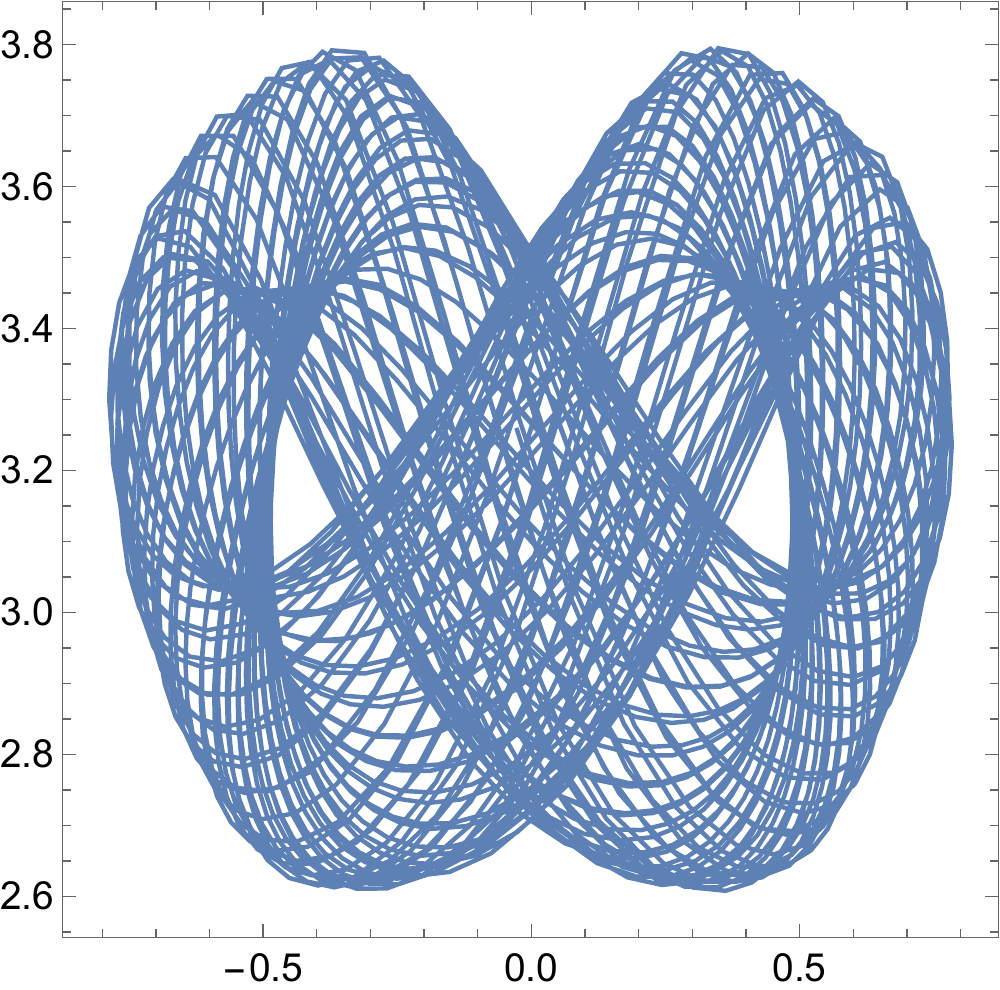}
		\caption{E=55}
	\end{subfigure}
	\begin{subfigure}[b]{0.49\linewidth}
		\includegraphics[width=\linewidth]{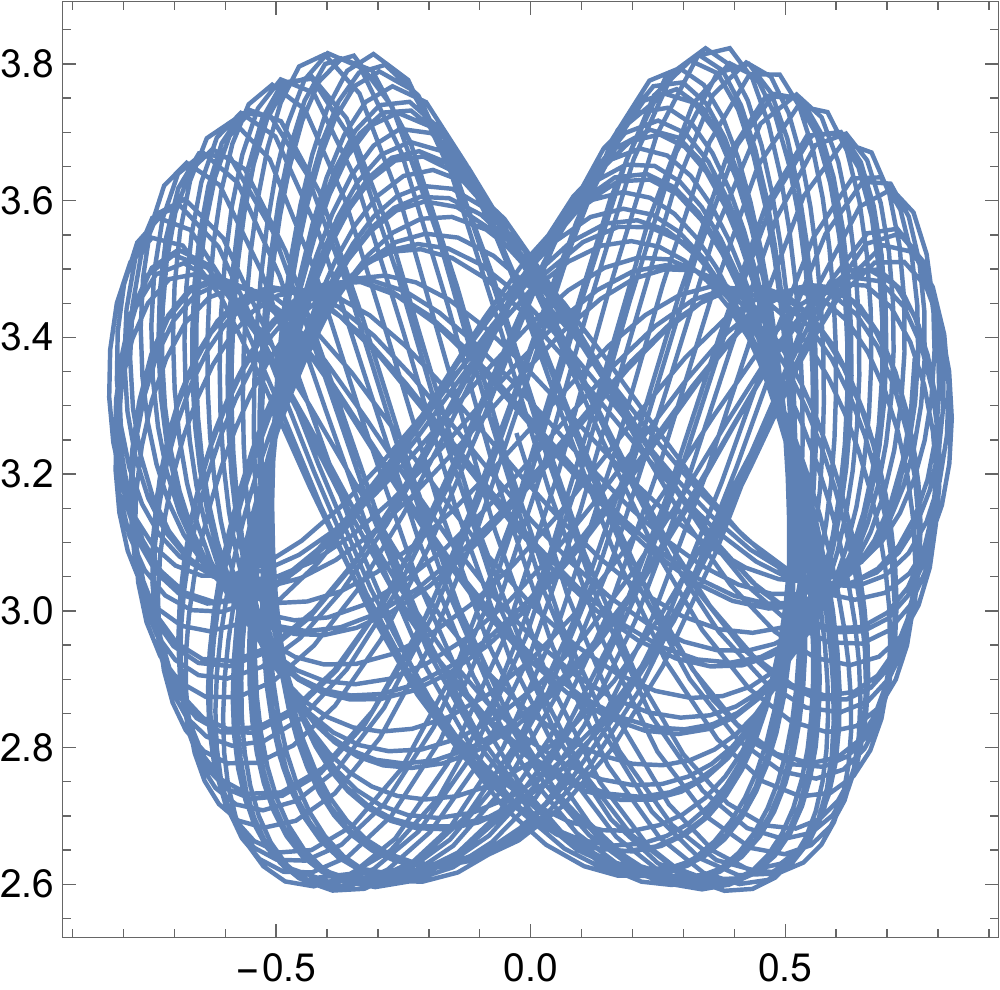}
		\caption{E=60}
	\end{subfigure}
	\begin{subfigure}[b]{0.49\linewidth}
		\includegraphics[width=\linewidth]{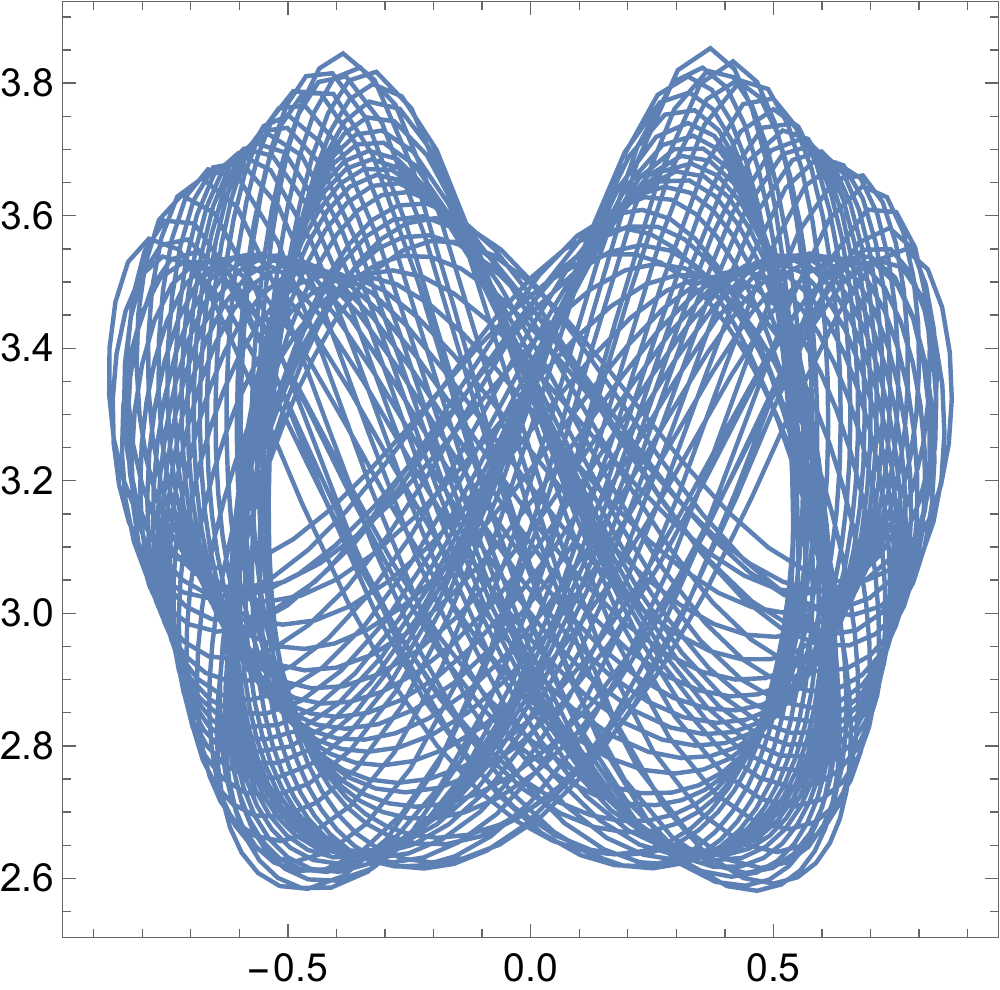}
		\caption{E=65}
	\end{subfigure}
	\caption{Orbits of the particle trapped in the harmonic potential in the near-horizon region in the background of (\ref{Nicolini metric}) for fixed $\theta_{nc}=0.2$ but for different values of the system energy $E$.}
	\label{fig:15}
\end{figure}
%
%
%
%
%
\section{{\label{App3}}Characteristic behaviour of Poincar\'e sections depending on the values of $K_r$ and $K_{\theta}$}
%
%
%
%
\par
\begin{figure}[H]
	\centering
	\begin{subfigure}[b]{0.49\linewidth}
		\includegraphics[width=\linewidth]{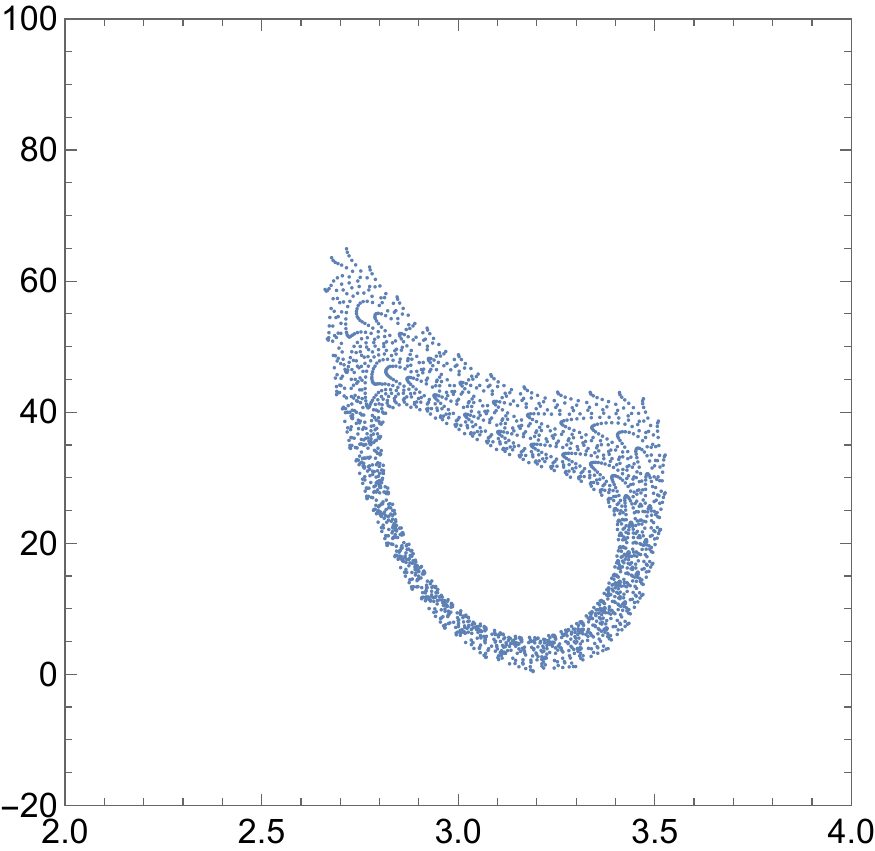}
		\caption{E=40}
	\end{subfigure}
	\begin{subfigure}[b]{0.49\linewidth}
		\includegraphics[width=\linewidth]{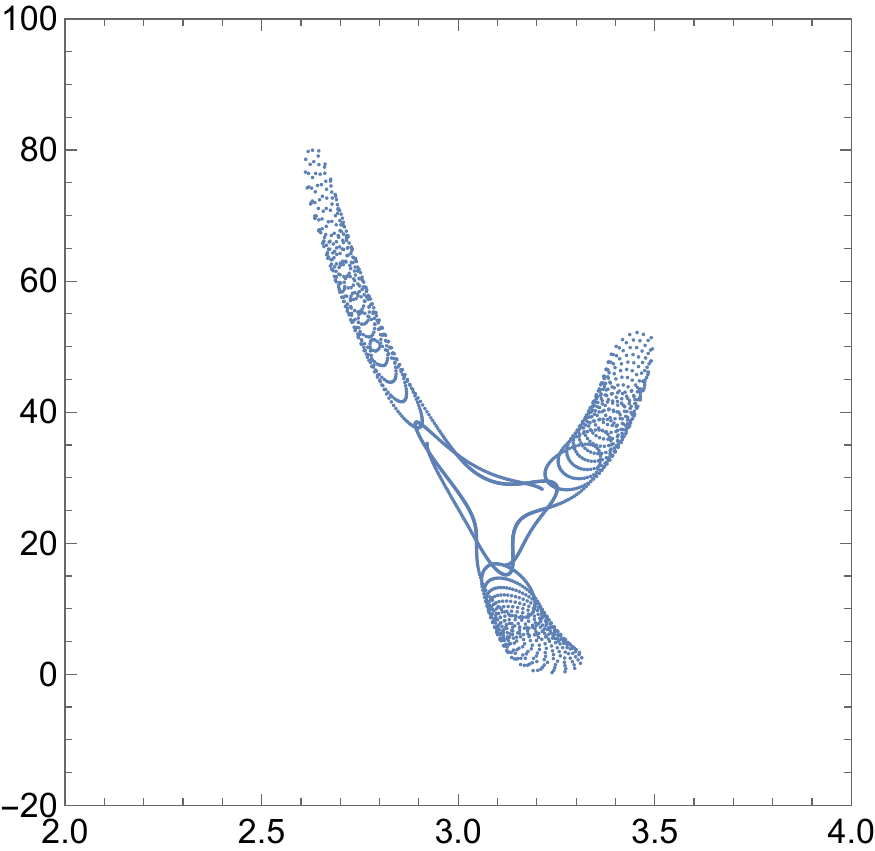}
		\caption{E=45}
	\end{subfigure}
	\begin{subfigure}[b]{0.49\linewidth}
		\includegraphics[width=\linewidth]{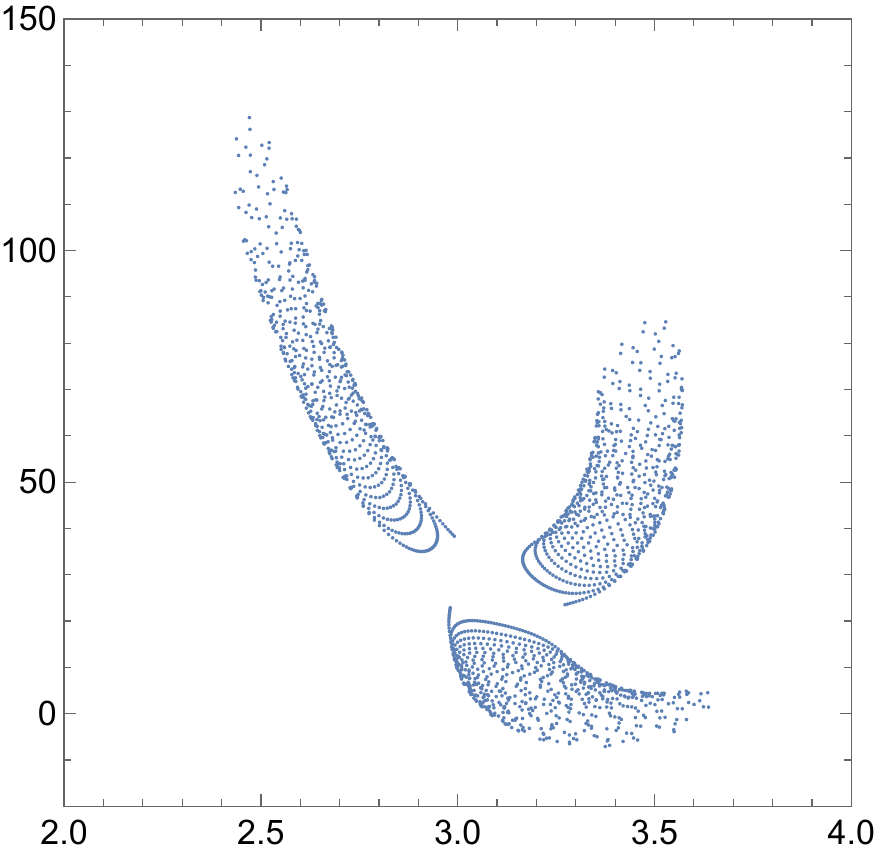}
		\caption{E=50}
	\end{subfigure}
	\begin{subfigure}[b]{0.49\linewidth}
		\includegraphics[width=\linewidth]{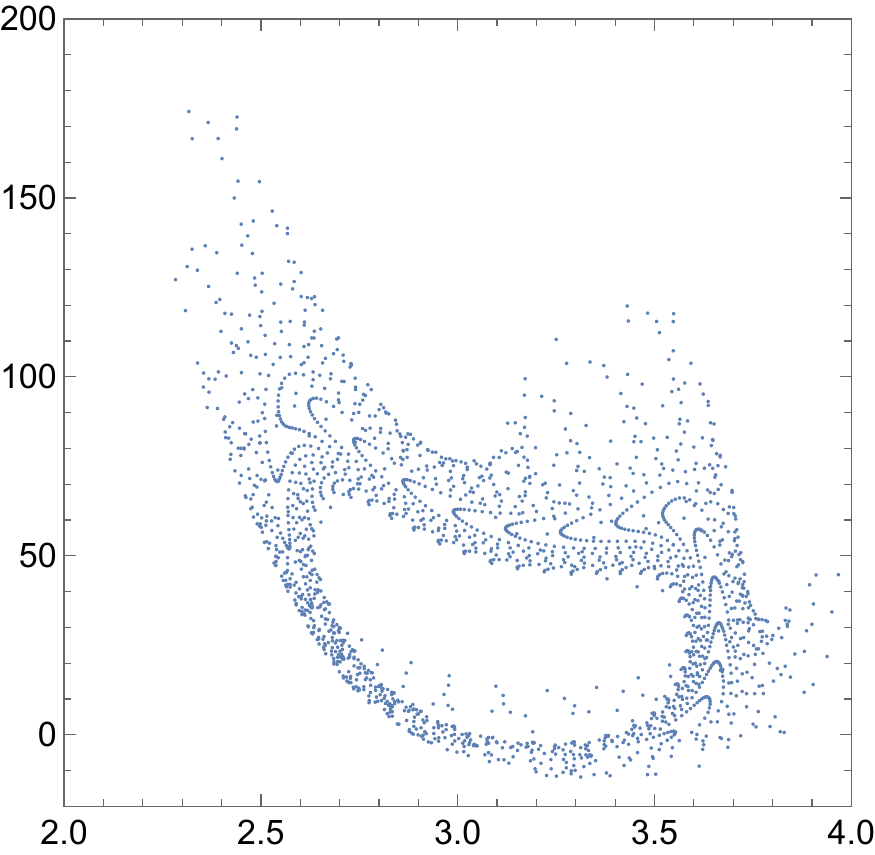}
		\caption{E=55}
	\end{subfigure}
	\caption{The Poincar\'e sections in the ($r,p_r$) plane with $\theta = 0$ and $p_\theta > 0$ at different system energies $E$ for the exact metric (\ref{Nicolini metric}) and for $\theta_{nc}=0.16$, $K_{r}=80$, and  $K_{\theta}=20$. The horizontal and vertical axes in each of the graphs correspond to $r$ and $p_r$, respectively.}
	\label{fig:16}
\end{figure}

\begin{figure}[hbt]
	\centering
	\begin{subfigure}[b]{0.49\linewidth}
		\includegraphics[width=\linewidth]{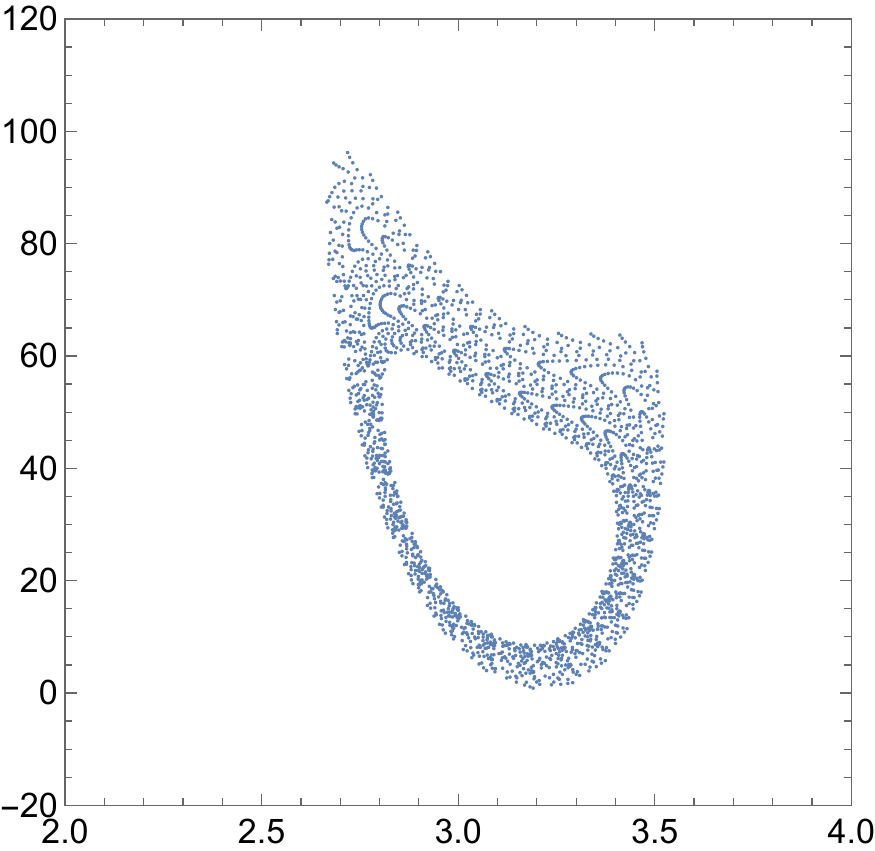}
		\caption{E=60}
	\end{subfigure}
	\begin{subfigure}[b]{0.49\linewidth}
		\includegraphics[width=\linewidth]{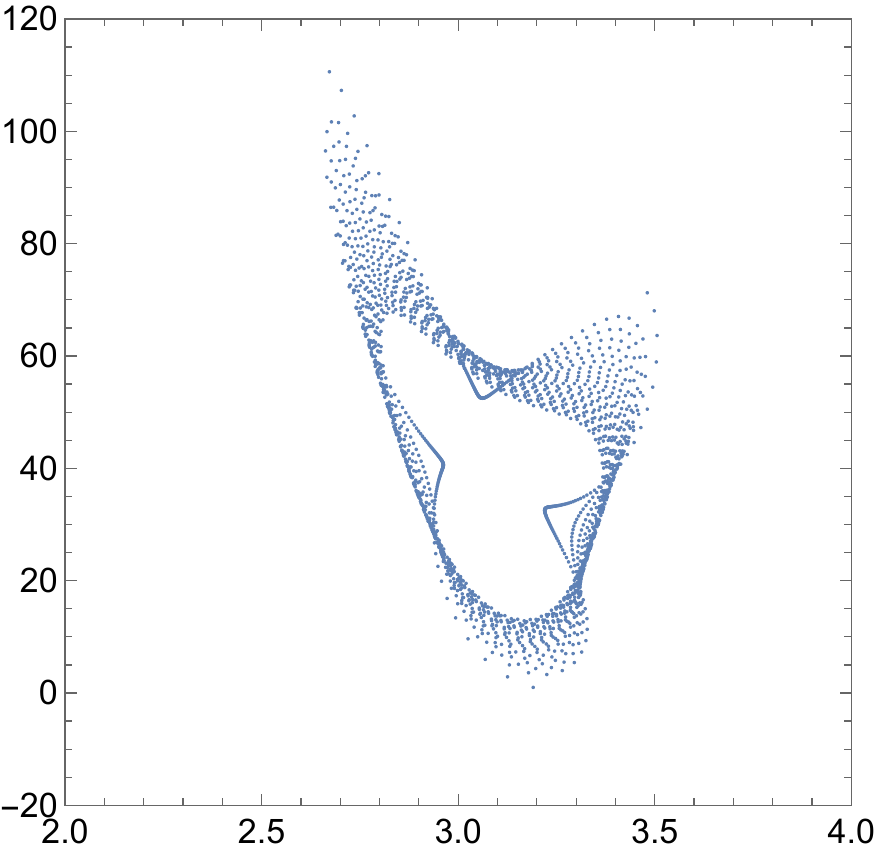}
		\caption{E=65}
	\end{subfigure}
	\begin{subfigure}[b]{0.49\linewidth}
		\includegraphics[width=\linewidth]{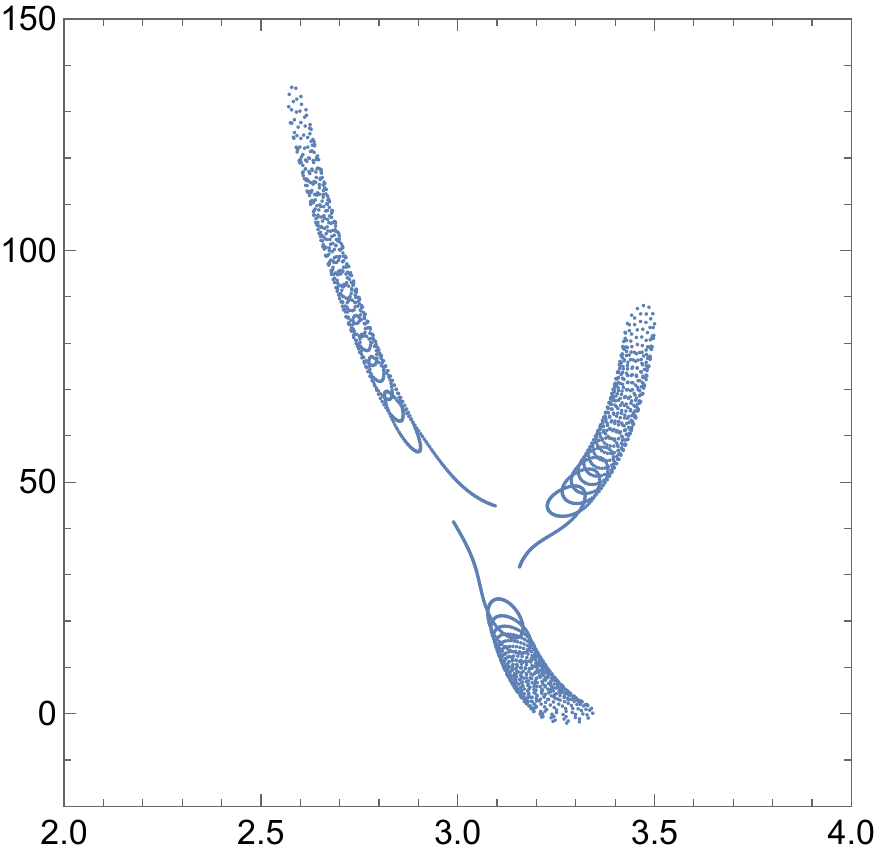}
		\caption{E=70}
	\end{subfigure}
	\begin{subfigure}[b]{0.49\linewidth}
		\includegraphics[width=\linewidth]{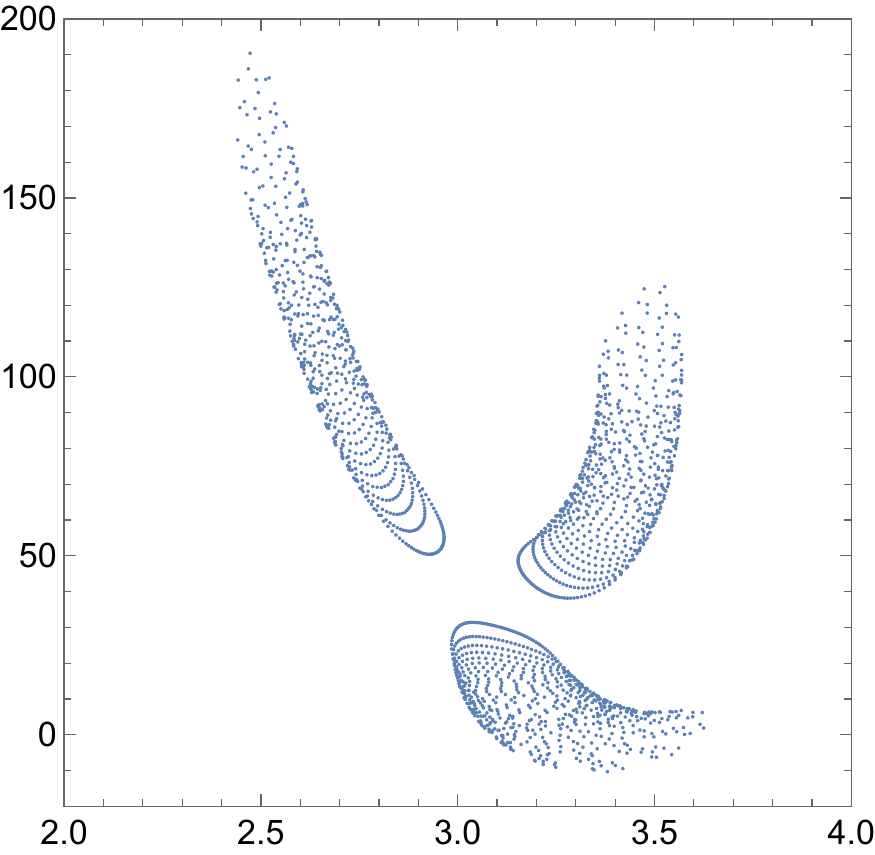}
		\caption{E=75}
	\end{subfigure}
	\begin{subfigure}[b]{0.49\linewidth}
	\includegraphics[width=\linewidth]{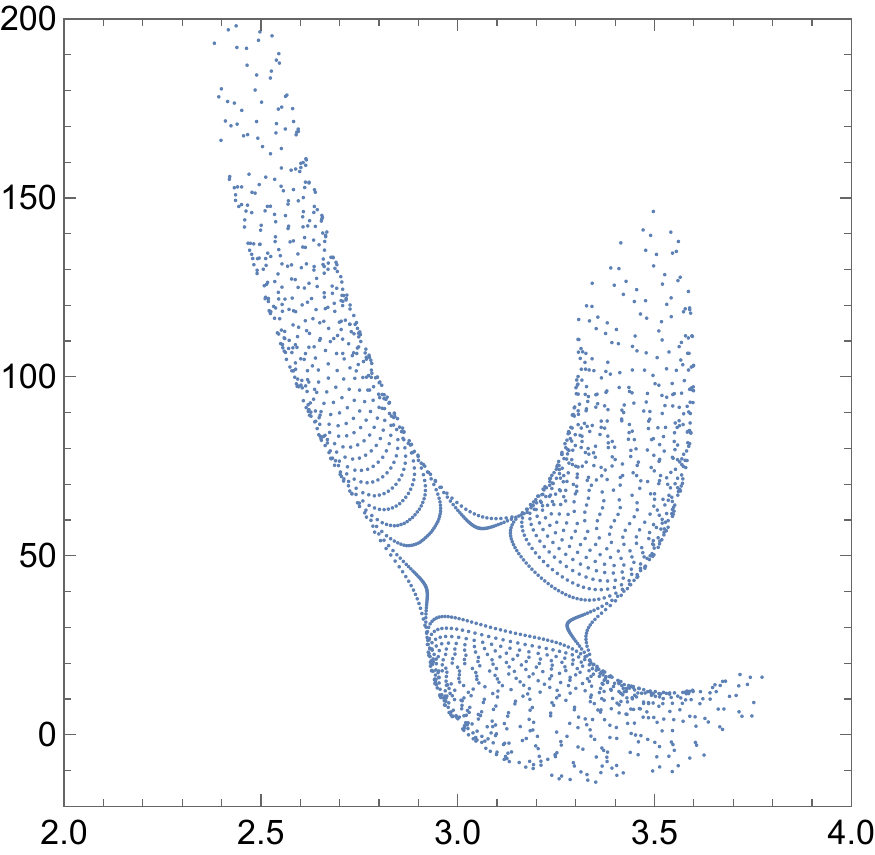}
	\caption{E=77.5}
	\end{subfigure}
	\begin{subfigure}[b]{0.49\linewidth}
	\includegraphics[width=\linewidth]{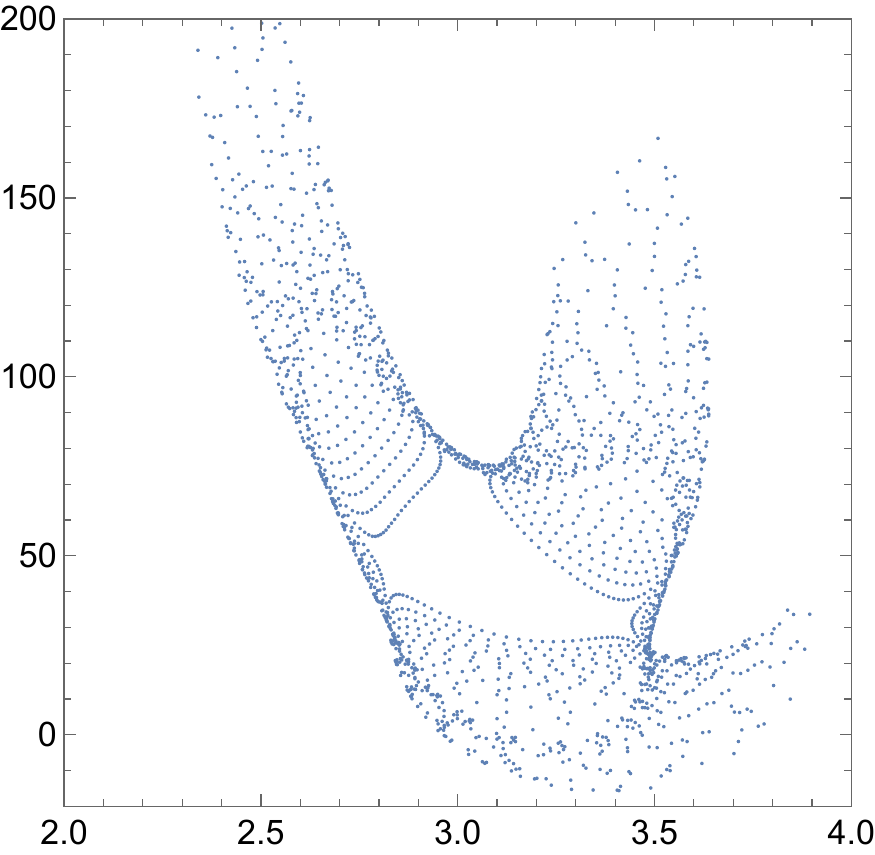}
	\caption{E=80}
	\end{subfigure}
	\caption{The Poincar\'e sections in the ($r,p_r$) plane with $\theta = 0$ and $p_\theta > 0$ at different system energies $E$ for the exact metric (\ref{Nicolini metric})  and for $\theta_{nc}=0.16$, $K_{r}=120$, and  $K_{\theta}=30$. The horizontal and vertical axes in each of the graphs correspond to $r$ and $p_r$, respectively.}
	\label{fig:17}
\end{figure}

\noindent
Here we have analysed the characteristic behaviour of Poincar\'e sections when we change the values of the spring constants, namely $K_{r}$ and $K_{\theta}$, of the harmonic potential. From Fig. \ref{fig:16}, it is evident that if we decrease the values of the spring constants, namely $K_{r}=80$ and $K_{\theta}=20$,  the Poincar\'e sections start getting distorted from  lower energy values, i.e., $E=45$ and $E=50$, with respect to the higher values of spring constants, i.e., $K_{r}=100$ and $K_{\theta}=25$. On the contrary, in Fig. \ref{fig:17} with the increment of the values of the spring constants, namely $K_{r}=120$ and $K_{\theta}=30$, the breaking of KAM torus start happening from higher energy values, i.e., $E=65$ and $E=70$.  
%
%
%
%
\section{{\label{App4}}Characteristic behaviour of Lyapunov exponent depending on the values of $K_r$ and $K_{\theta}$}
%
%
%
%
\par
\begin{figure}[H]
	\centering
	\begin{subfigure}[b]{0.49\linewidth}
		\includegraphics[width=\linewidth]{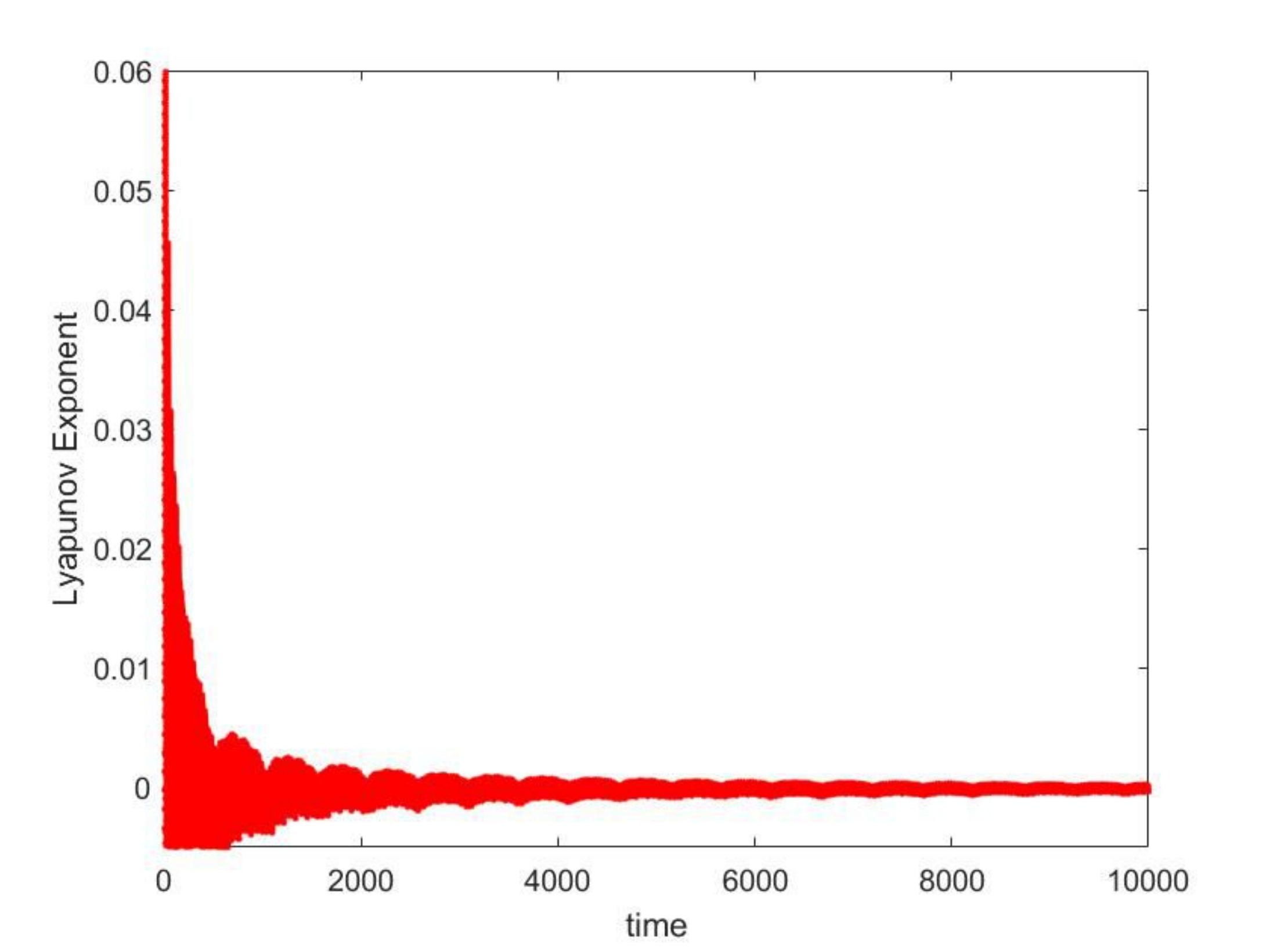}
		\caption{E=40}
	\end{subfigure}
	\begin{subfigure}[b]{0.49\linewidth}
		\includegraphics[width=\linewidth]{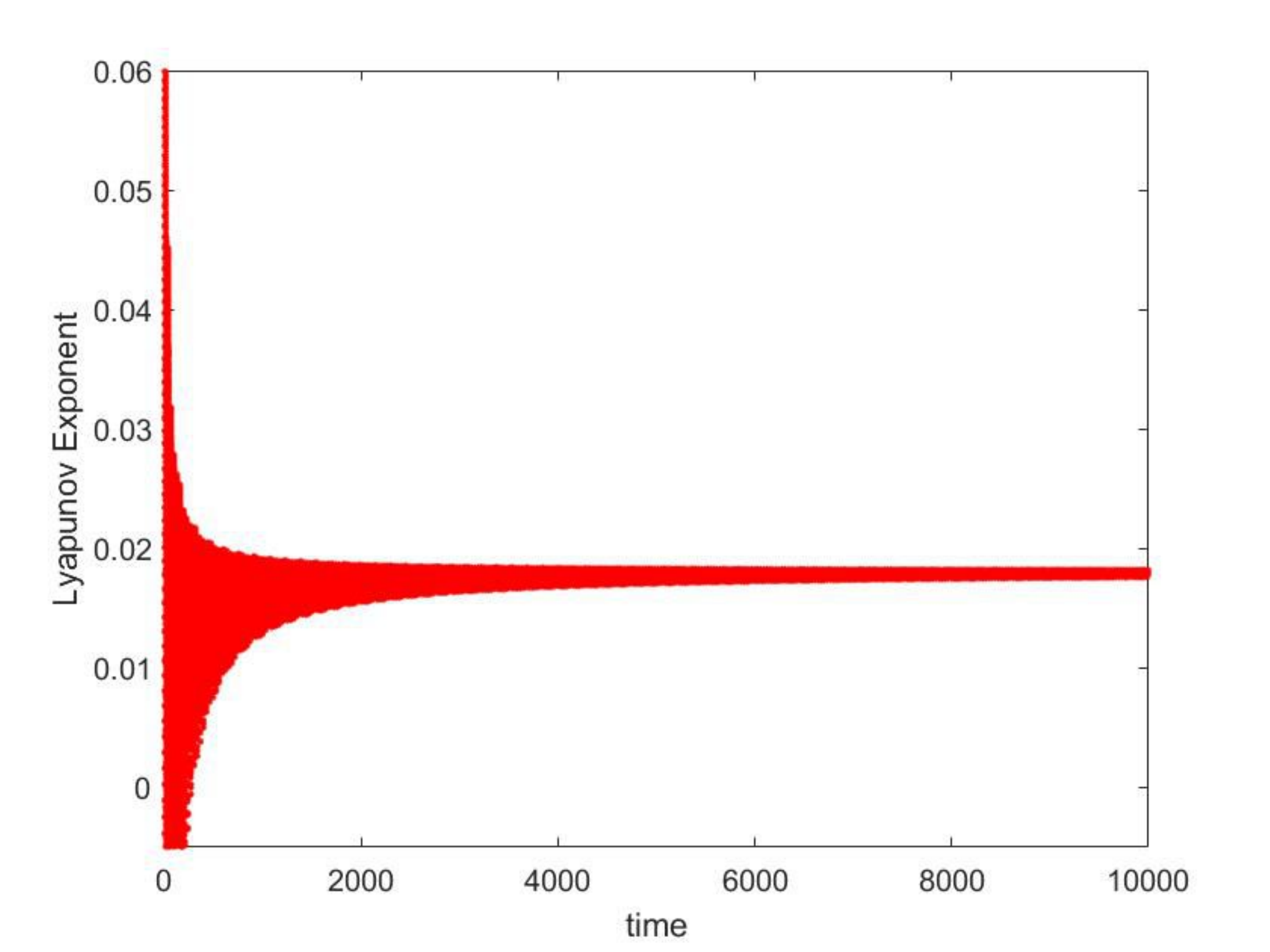}
		\caption{E=45}
	\end{subfigure}
	\begin{subfigure}[b]{0.49\linewidth}
		\includegraphics[width=\linewidth]{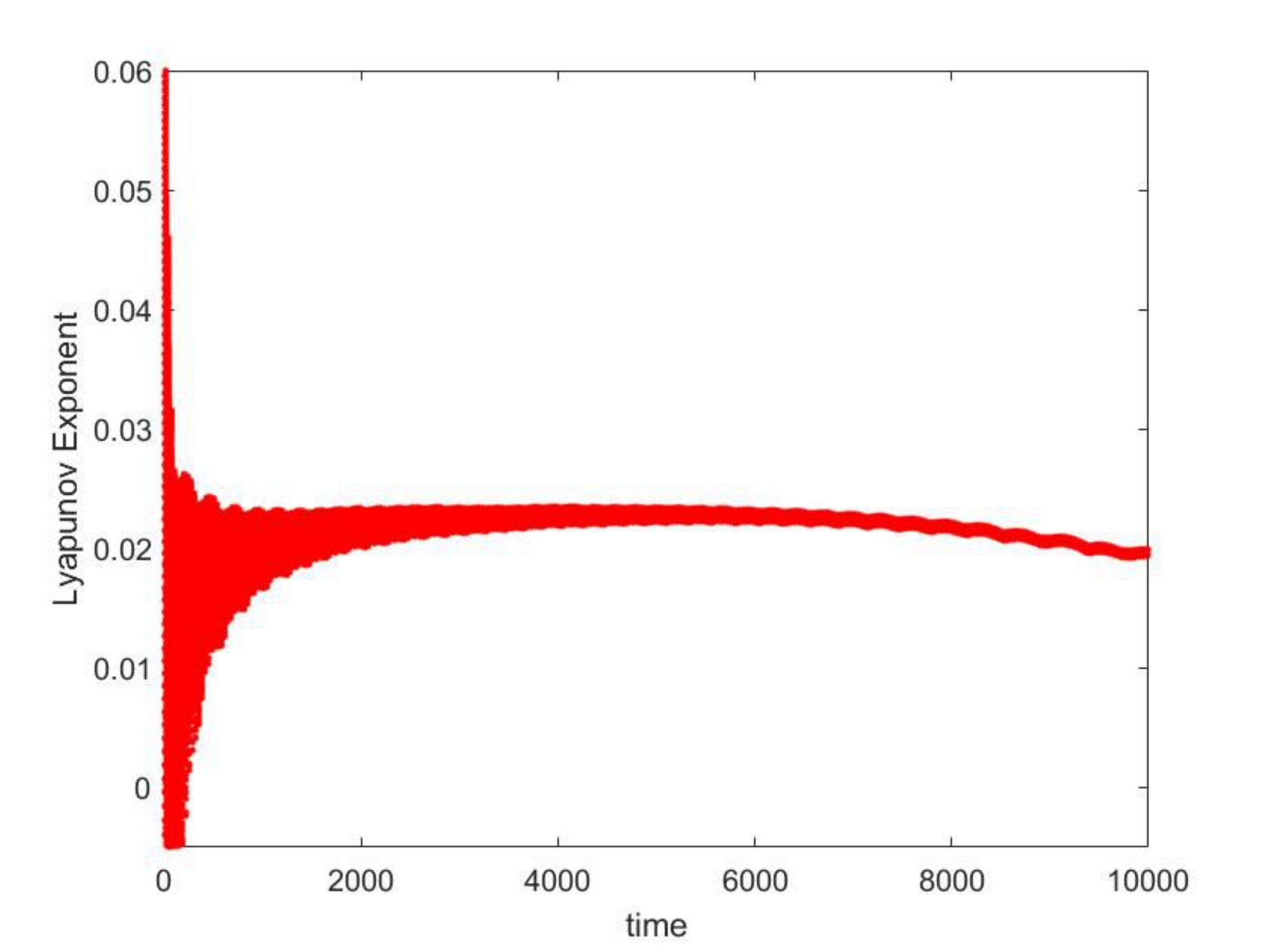}
		\caption{E=50}
	\end{subfigure}
	\begin{subfigure}[b]{0.49\linewidth}
		\includegraphics[width=\linewidth]{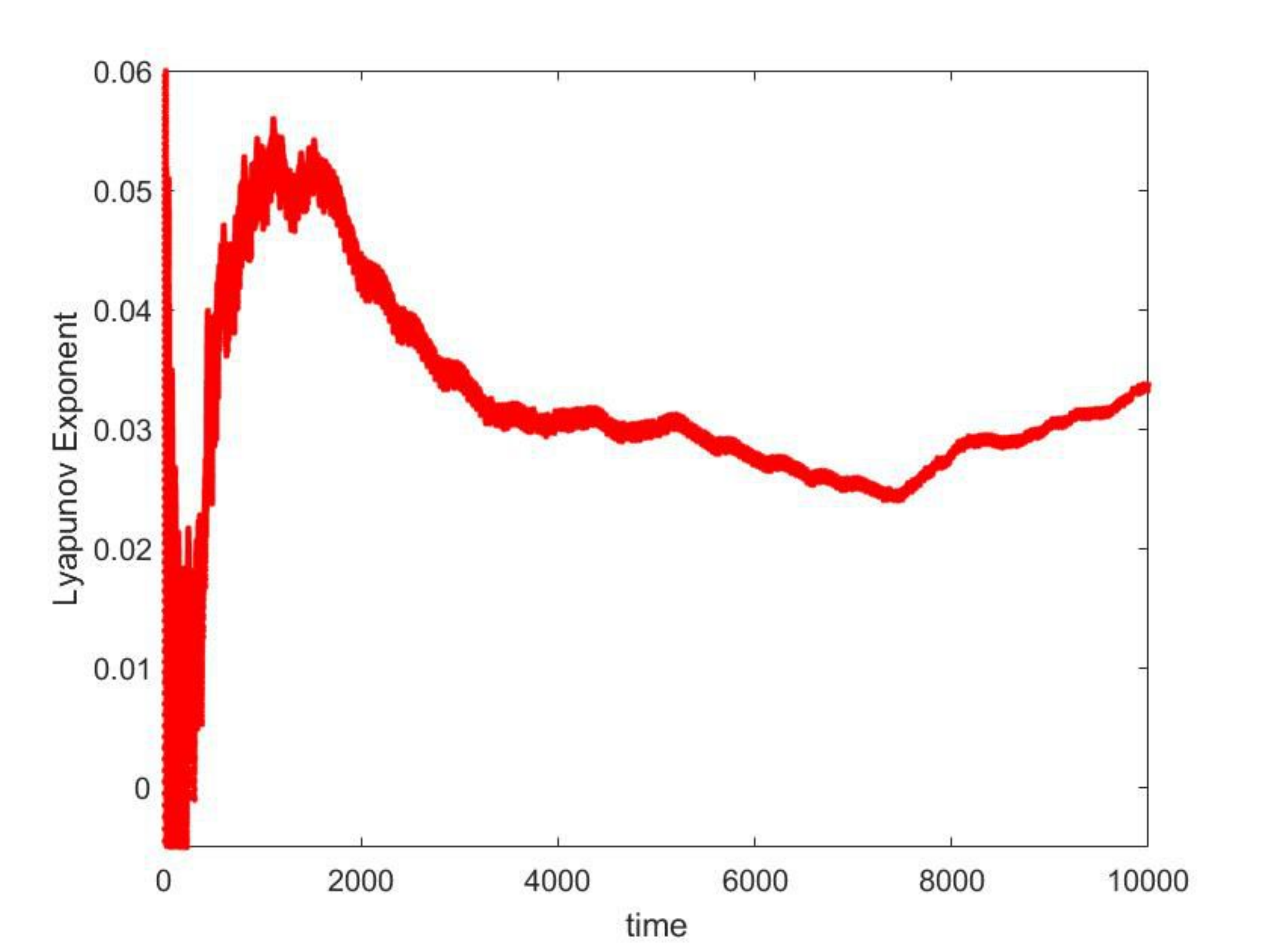}
		\caption{E=55}
	\end{subfigure}
	\caption{Largest Lyapunov exponents for different values of the system energy $E$, but for a particular value of $\theta_{nc}=0.2$, $K_{r}=80$, and $K_{\theta}=20$.}
	\label{fig:18}
\end{figure}

\begin{figure}[hbt]
	\centering
	\begin{subfigure}[b]{0.49\linewidth}
		\includegraphics[width=\linewidth]{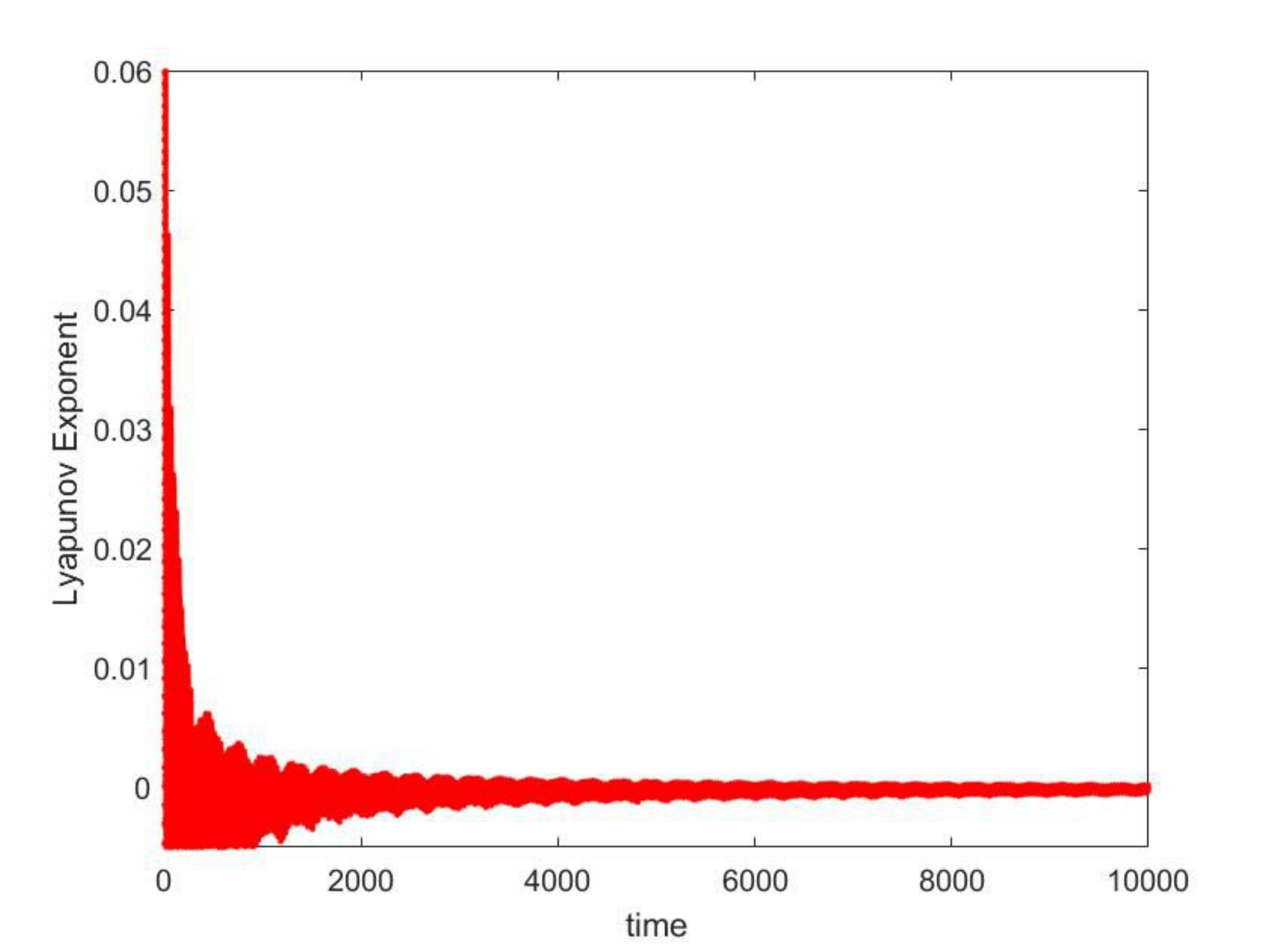}
		\caption{E=60}
	\end{subfigure}
	\begin{subfigure}[b]{0.49\linewidth}
		\includegraphics[width=\linewidth]{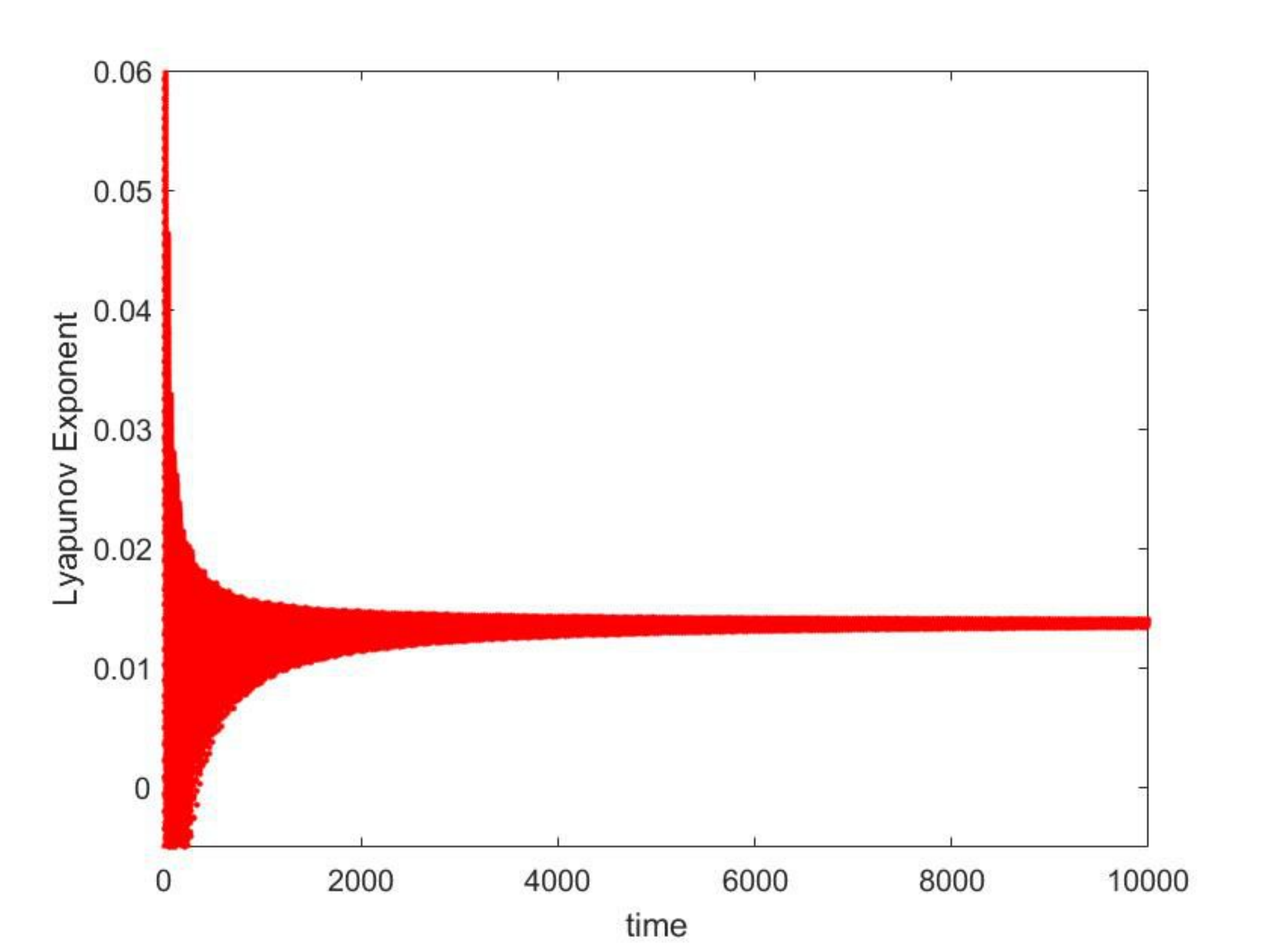}
		\caption{E=65}
	\end{subfigure}
	\begin{subfigure}[b]{0.49\linewidth}
		\includegraphics[width=\linewidth]{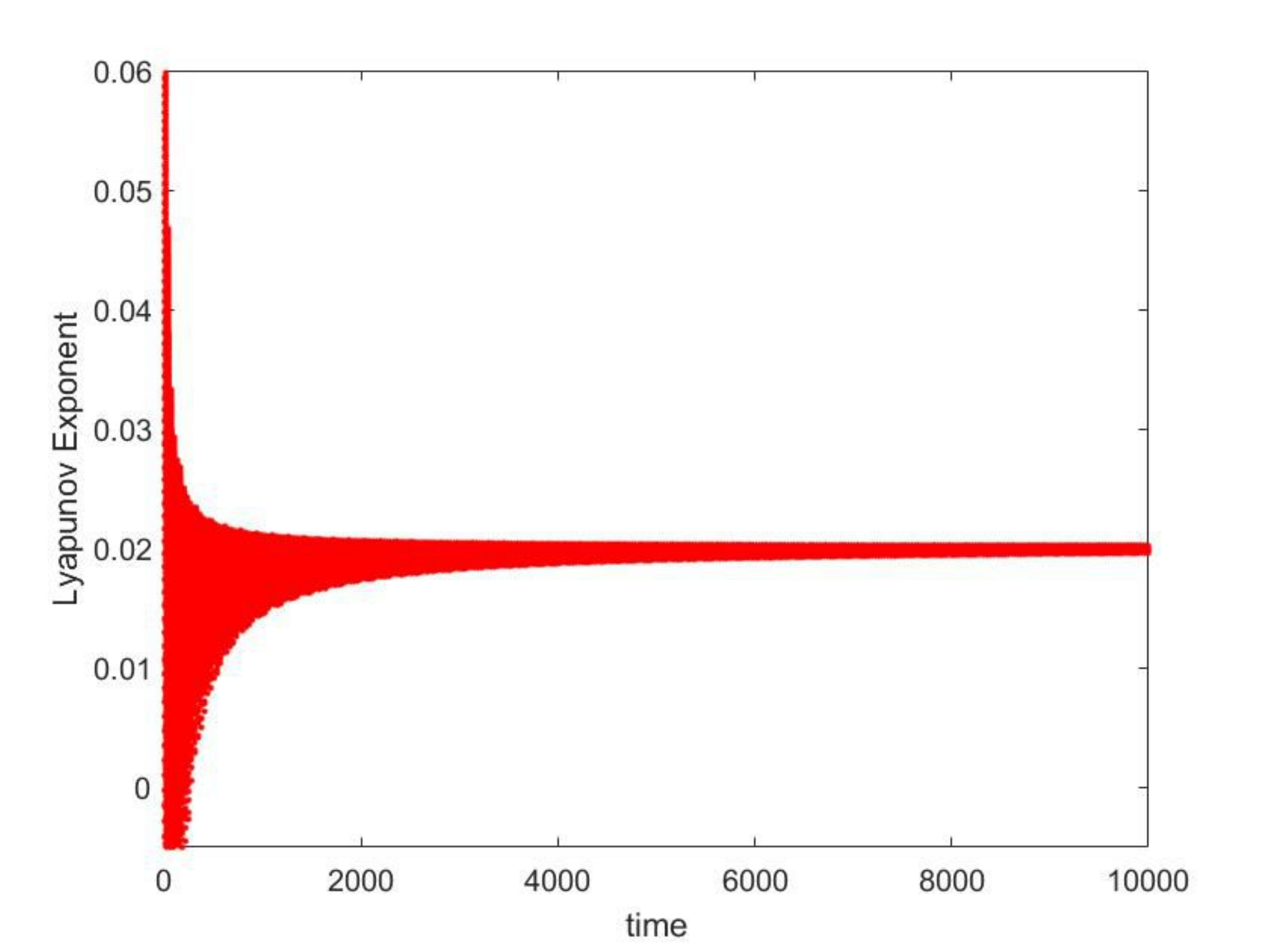}
		\caption{E=70}
	\end{subfigure}
	\begin{subfigure}[b]{0.49\linewidth}
		\includegraphics[width=\linewidth]{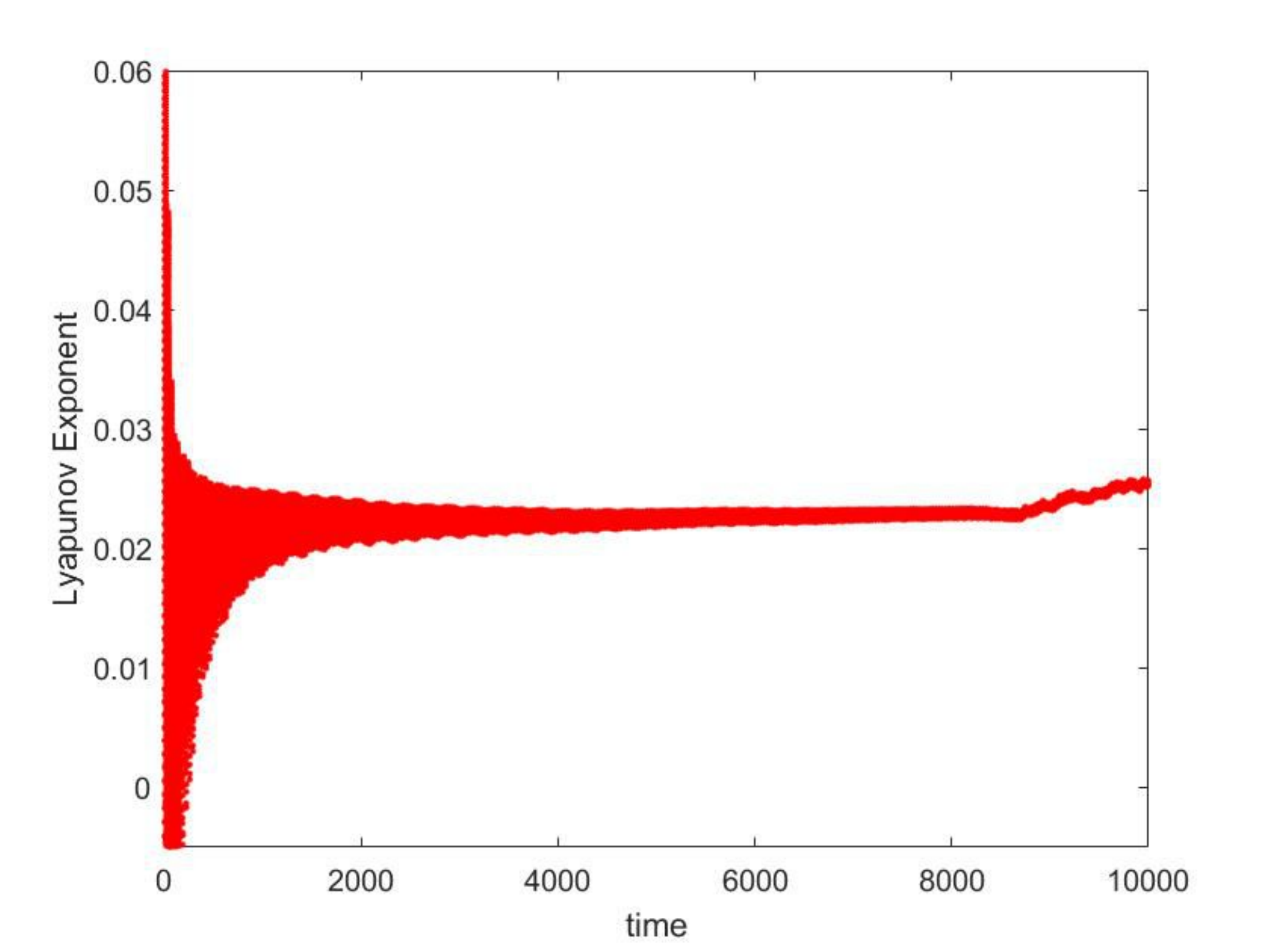}
		\caption{E=75}
	\end{subfigure}
	\begin{subfigure}[b]{0.49\linewidth}
		\includegraphics[width=\linewidth]{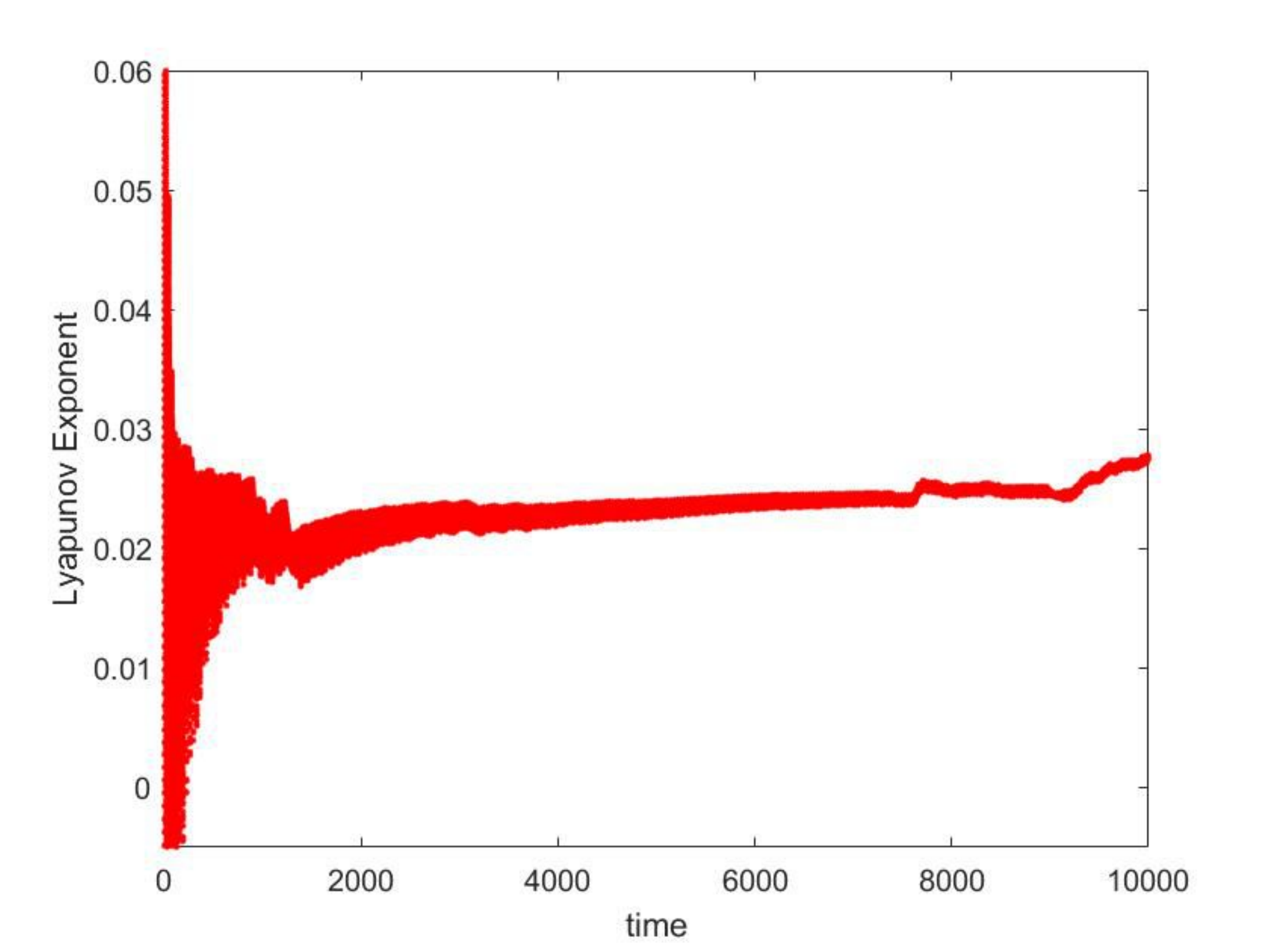}
		\caption{E=77.5}
	\end{subfigure}
	\begin{subfigure}[b]{0.49\linewidth}
		\includegraphics[width=\linewidth]{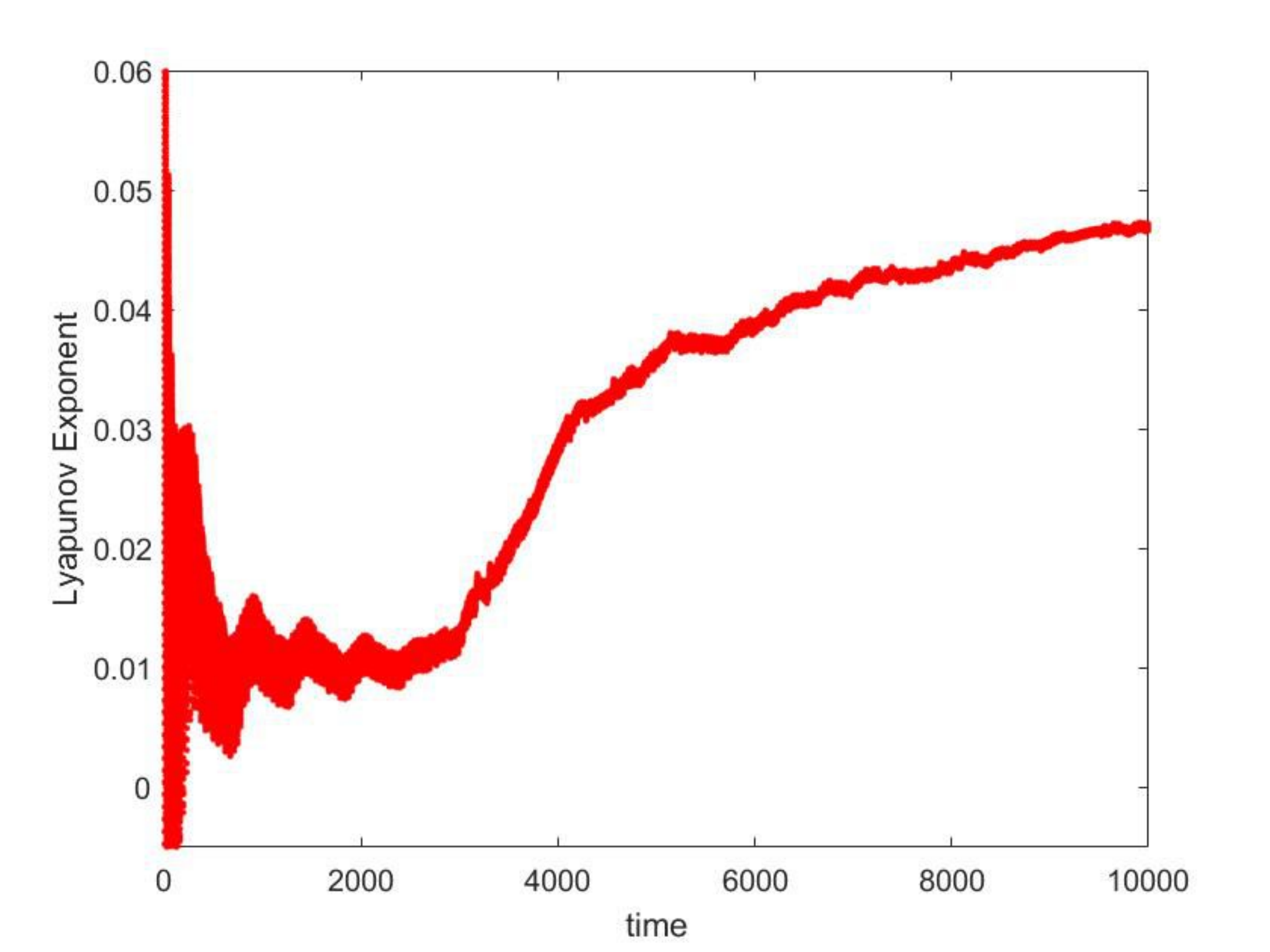}
		\caption{E=80}
	\end{subfigure}
\caption{Largest Lyapunov exponents for different values of the system energy $E$, but for a particular value of $\theta_{nc}=0.2$, $K_{r}=120$, and $K_{\theta}=30$.}
\label{fig:19}
\end{figure}

\noindent
Here we have analysed the characteristic behaviour of the largest Lyapunov exponent values with the changing of the spring constants, namely $K_{r}$ and $K_{\theta}$, of the harmonic potential. From Fig. \ref{fig:18}, we can easily see that if we decrease the values of the spring constants, i.e., $K_{r}=80$ and $K_{\theta}=20$, the largest Lyapunov exponent values attain higher positive values at lower energy values, i.e., $E=45$ and $E=50$, with respect to the case where the values of the spring constants were higher (see Fig. \ref{fig:9} for $K_{r}=100$ and $K_{\theta}=25$). On the contrary, in Fig. \ref{fig:19} we see that with the increment of the values of the spring constants, namely $K_{r}=120$ and $K_{\theta}=30$, the positive values of the largest Lyapunov exponents start coming in the higher energy values, i.e., $E=65$ and $E=70$).

\end{document}